\documentclass[accepted=2026-05-14,letterpaper,allowfontchangeintitle]{quantumarticle}
\pdfoutput=1
\usepackage[numbers,sort&compress]{natbib}

\usepackage{graphicx, color, graphpap}      
\usepackage{amsmath}
\usepackage{enumerate}
\usepackage{amssymb}
\usepackage{amsthm}
\usepackage{pstricks}
\usepackage{float}
\usepackage[colorlinks=true, citecolor=blue, linkcolor=red]{hyperref}
\usepackage[T1]{fontenc}
\usepackage{bbm}
\usepackage{dsfont}
\usepackage[linesnumbered,ruled,vlined]{algorithm2e}
\SetKwInput{kwInit}{Init}
\usepackage{mathtools}

\usepackage{tikz}
\usepackage{graphicx}
\usetikzlibrary{positioning}
\usepackage{graphicx}
\usepackage{xcolor}
\usepackage[caption=false]{subfig}
\usepackage[ruled,vlined]{algorithm2e}
\usepackage{pifont} 
\usepackage{thmtools}
\usepackage{thm-restate}


\renewcommand{\eqref}[1]{(\ref{#1})}

\newtheoremstyle{example}{\topsep}{\topsep}%
{}
{}
{\bfseries}
{:}
{   }
{\thmname{#1}\thmnumber{ #2}}
\theoremstyle{example}

\theoremstyle{definition}

\newcommand{\Jiuzhang}{\textit{Ji\u{u}zh\=ang}}
\newcommand{\Zuchongzhi}{\textit{Zuchongzhi}}

\newtheorem*{theorem*}{Theorem}

\def\orcid#1{\kern -0.4em\href{https://orcid.org/#1}{\includegraphics[keepaspectratio,width=0.7em]{orcid_logo.pdf}}}

\usepackage{lipsum}

\long\def\ca#1\cb{} 
\newcommand{\rcs}{\text{RCS}}
\newcommand{\gbs}{\text{GBS}}
\newcommand{\qsim}{QSim}

\begin{document}
\title{\LARGE A brief history of quantum vs classical computational advantage}

\author{Ryan LaRose}
\thanks{\href{mailto:rmlarose@msu.edu}{rmlarose@msu.edu}.}
\affiliation{Department of Computational Mathematics, Science, and Engineering, Michigan State University, East Lansing, MI 48823, USA}
\affiliation{Department of Electrical and Computer Engineering, Michigan State University, East Lansing, MI 48823, USA}
\affiliation{Department of Physics and Astronomy, Michigan State University, East Lansing, MI 48823, USA}
\affiliation{Center for Quantum Computing, Science, and Engineering, Michigan State University, East Lansing, MI 48823, USA}

\begin{abstract}
In this review article we summarize all experiments claiming quantum computational advantage to date. Our review highlights challenges, loopholes, and refutations appearing in subsequent work to provide a complete picture of the current statuses of these experiments. In addition, we also discuss theoretical computational advantage in example problems such as approximate optimization and recommendation systems. Finally, we review recent experiments in quantum error correction --- the biggest frontier to reach experimental quantum advantage in Shor's algorithm.
\end{abstract}

\maketitle


\section{Introduction}

Since the first experimental demonstration of quantum advantage on Oct 23 2019, several new experiments have been performed and previous experiments have been challenged or refuted. The purpose of this review article is to collect these experiments, challenges, and refutations in a single place to provide a complete picture of the current status of quantum vs classical computational advantage.

While computational advantage could refer to several aspects (e.g. lower energy consumption or higher accuracy), here we use it to mean lower time to solution as is common in the quantum computing literature~\cite{Preskill_2012}\footnote{Note that, after the first experiment, \textit{quantum advantage} is used preferentially to the previous term of \textit{quantum supremacy} although the technical meaning has not changed~\cite{Preskill_2019}. Note that there are several other terms used for the notion of quantum advantage --- e.g. \textit{beyond-classical computation}. Some authors prefer to use \textit{quantum advantage} only for ``useful'' computations as opposed to (probably) arbitrary computations designed around what quantum computers are good at. Still some other authors prefer \textit{quantum utility} for this concept. While these distinctions are useful, there is not yet standard terminology nor a standard definition for any term. In this article, by \textit{quantum advantage} we always mean a computation performed faster on a quantum computer than on a classical computer, irrespective of the utility of the problem.}. This captures the intuitive meaning of advantage as the answer to the following question: for a given  computational problem, which solves the problem faster --- the best classical algorithm on the best classical computer, or the best quantum algorithm on the best quantum computer? 
Two notes are in order. First, computational advantage is defined with respect to a particular problem, which may or may not have any practical relevance or utility.  Second, the \textit{best} computers, as well as the \textit{best} algorithms, change over time, both in the classical and quantum settings. Thus computational advantage is a moving target and can change hands between classical and quantum computers. The history of this changing status is the primary purpose of this review article. Table~\ref{tab:overview} provides a chronological list of all experiments claiming computational advantage to date along with their current statuses, and Sec.~\ref{sec:quantum-advantage-experiments} provides the narrative discussion.

In addition to this primary purpose, we also discuss \textit{theoretical} quantum vs classical computational advantage. In this setting, advantage refers to lower algorithmic complexity for a given computational problem, typically in the asymptotic setting, without regard to any computer or experiment implementing the algorithm. Even in this arena, computational advantage has shifted hands between quantum and classical as better classical algorithms, often inspired by the quantum algorithms, have been developed. Sec.~\ref{sec:quantum-advantage-theory} provides a discussion for the examples of recommendation systems, approximate optimization, and quantum chemistry.

Finally, since many quantum algorithms with hypothesized or proven computational advantage will require error correction to experimentally realize, we include a history of quantum error correction experiments, especially in recent years, as the final element to this review article. While error correction is typically considered separate and independent of computational advantage, some quantum algorithms with theoretical computational advantage can actually lose their speedup due to the overhead of error correction~\cite{Babbush_McClean_Newman_Gidney_Boixo_Neven_2021}. For this reason, and because quantum information is so inherently fragile, the two are intricately connected in quantum computing. Error correction can thus be considered the final frontier to experimentally demonstrating quantum computational advantage for problems such as Shor's algorithm for prime factorization~\cite{Shor_1997}. Table~\ref{tab:quantum_error_correction} provides a history of quantum error correction experiments, and Sec.~\ref{sec:qec-experiments} provides the narrative discussion.

\begin{table*}
    \centering
    \scriptsize
    \begin{tabular}{|p{0.085\linewidth}|c|c|c|l|l|c|c|c|} \hline
        Date        & Problem   & $n$   & $m$   & Group \& computer              & Computer type     & Ref.                                                                                  & Status                                                                                                                                                                                                    & Section   \\ \hline
        Oct 23 2019 & \rcs      & $53$  & $20$  & Google \textit{Sycamore}      & Superconducting   & \cite{Arute_Arya_Babbush_Bacon_Bardin_Barends_Biswas_Boixo_Brandao_Buell_}            & Refuted by~\cite{Zhao_Zhong_Pan_Chen_Fu_Su_Xie_Zhao_Zhang_Ouyang__2024}                                                                                                                                   & Sec.~\ref{sec:google-rcs-2019} \\
        Dec 03 2020 & \gbs      & $50$  & $100$ & USTC \Jiuzhang{}              & Photonic          & \cite{Zhong_Wang_Deng_Chen_Peng_Luo_Qin_Wu_Ding_Hu_2020}                              & Weakly refuted by~\cite{Oh_Liu_Alexeev_Fefferman_Jiang_2024} & Sec.~\ref{sec:ustc-gbs-1} \\
        Jun 28 2021 & \rcs      & $56$  & $20$  & USTC \Zuchongzhi{}            & Superconducting   & \cite{Wu_Bao_Cao_Chen_Chen_Chen_Chung_Deng_Du_Fan__2021}                              & Challenged by~\cite{Kalachev_Panteleev_Zhou_Yung_2021,Morvan_2024}                                                                                                                                        & Sec.~\ref{sec:ustc-rcs-1} \\
        Jun 29 2021 & \gbs      & $50$  & $144$ & USTC \Jiuzhang{} \textit{2.0} & Photonic          & \cite{Zhong_Deng_Qin_Wang_Chen_Peng_Luo_Wu_Gong_Su_2021}                              & Weakly refuted by~\cite{Oh_Liu_Alexeev_Fefferman_Jiang_2024} & Sec.~\ref{sec:ustc-gbs-2} \\
        Sep 08 2021 & \rcs      & $60$  & $24$  & USTC \Zuchongzhi{}            & Superconducting   & \cite{Zhu_Cao_Chen_Chen_Chen_Chung_Deng_Du_Fan_Gong_2022}                             & Challenged by~\cite{Liu_Guo_Liu_Yang_Song_Gao_Wang_Wu_Peng_Zhao_2021,Morvan_2024}                                                                                                                         &  Sec.~\ref{sec:ustc-rcs-2} \\
        Jun 01 2022 & \gbs      & $216$ & $216$ & Xanadu \textit{Borealis}      & Photonic          & \cite{Madsen_Laudenbach_Askarani_Rortais_Vincent_Bulmer_Miatto_Neuhaus_Helt_Collins}  & Weakly refuted by~\cite{Oh_Liu_Alexeev_Fefferman_Jiang_2024}                                                                                                                                                                                                 & Sec.~\ref{sec:xanadu-gbs-borealis}          \\
        Apr 21 2023 & \rcs      & $67$  & $32$  & Google \textit{Sycamore}      & Superconducting   & \cite{Morvan_2024}                                                                    & Unrefuted                                                                                                                                                                                                 & Sec.~\ref{sec:google-second-experiment} \\
        Apr 21 2023 & \rcs      & $70$  & $24$  & Google \textit{Sycamore}      & Superconducting   & \cite{Morvan_2024}                                                                    & Unrefuted                                                                                                                                                                                                 & Sec.~\ref{sec:google-second-experiment}          \\
        Apr 24 2023 & \gbs      & $50$  & $144$ & USTC \Jiuzhang{} \textit{3.0} & Photonic          & \cite{Deng_Gu_Liu_Gong_Su_Zhang_Tang_Jia_Xu_Chen_2023}                                & Weakly refuted by~\cite{Oh_Liu_Alexeev_Fefferman_Jiang_2024}                                                                                                                                                                                                 & Sec.~\ref{sec:ustc-gbs-jiuzhang-3} \\
        Jun 14 2023 & \qsim     & $127$ & $60$  & IBM \textit{Kyiv}                 & Superconducting     & \cite{Kim_Eddins_Anand_Wei_van_Rosenblatt_Nayfeh_Wu_Zaletel_Temme_2023}               & Refuted by~\cite{Tindall_Fishman_Stoudenmire_Sels_2023, refute_ibm_1,refute_ibm_2,refute_ibm_3,refute_ibm_4}                                                                                              &  Sec.~\ref{sec:ibm-qsim} \\
        Mar 01 2024 & \qsim     & $567$ & --    & D-Wave \textit{ADV1/2}                       & Annealing         & \cite{dwave2024}                                                                      & Unrefuted                                                                                                                                                                                                 & Sec.~\ref{sec:dwave-qsim}           \\ \hline
    \end{tabular}
    \caption{A chronological list of claims to quantum advantage and current statuses. Statuses show only the strongest challenge/refutations --- additional ones can be found in the main text. Problem abbreviations are random circuit sampling (RCS), Gaussian boson sampling (GBS), and quantum simulation (QSim). Variables $n$ and $m$ denote the size of the experiment: $n$ is the number of qubits and $m$ is the circuit depth for random circuit sampling and gate-model quantum simulation; $n$ is the number of photons and $m$ is the number of modes for Gaussian Boson sampling; and in quantum annealing, $n$ is the number of qubits but $m$ is not defined since the evolution is not decomposed into gates. \textit{Challenged} refers to literature improving the classical simulation and/or calling some aspect of the experiment into question; \textit{Weakly refuted} means a new algorithm has been developed which likely could classically simulate the experiment on a powerful enough classical computer that is achievable in the near future; \textit{Refuted} means a classical computer has simulated the experiment faster than the quantum computer.}
    \label{tab:overview}
\end{table*}

\section{Quantum advantage experiments} \label{sec:quantum-advantage-experiments}

Table~\ref{tab:overview} shows the complete list of articles claiming quantum computational advantage to date, ordered chronologically. As can be seen, so far three computational problems have been considered: Random Circuit Sampling (\rcs), Gaussian Boson Sampling (\gbs), and quantum simulation (\qsim). Each problem is defined when it appears in the chronological narrative. Briefly, random circuit sampling is the computational task of sampling bitstrings from a random quantum circuit; Gaussian boson sampling is the analogous problem for continuous variable quantum computers; and quantum simulation is the problem of (i) time-evolving a quantum state under a specified Hamiltonian then (ii) computing certain properties from the evolved state, e.g. expectation values of observables. Parameters $n$ and $m$ refer to the size of the experiment. In random circuit sampling and quantum simulation, $n$ is the number of qubits and $m$ is the depth of the random quantum circuit, also referred to as the number of layers or cycles. In Gaussian boson sampling, $n$ is the number of input photons and $m$ is the number of modes (locations) where photons can be sampled. Each experiment involves additional parameters --- such as the number of samples generated, the (estimated) fidelity, and the total runtime --- which are omitted from the table for brevity and discussed in the main text. The final column shows the current status of each experiment, for which we use the following terminology.

\paragraph*{Terminology} Here and throughout, we use \textit{quantum advantage}, \textit{quantum supremacy}, and \textit{beyond-classical computation} interchangeably to mean the experimental demonstration of a well-defined computational task performed faster on a quantum computer than is believed to be possible on any classical computer. A \textit{challenge} to a claim of quantum advantage is any work which demonstrates a significant improvement in the classical simulation time of a quantum advantage experiment. A \textit{weak refutation} is a paper with evidence, via an improved classical algorithm and/or numerical demonstration, that with reasonably expected future classical computers a  quantum advantage claim could be classically simulated. A \textit{(strong) refutation} is the classical simulation of a quantum advantage experiment faster than the quantum computer.

\subsection{General remarks on quantum advantage} \label{sec:general-remarks-on-quantum-advantage}

Before giving a detailed chronological history of experiments, challenges, and refutations, let us provide some important remarks and context. Arguably the biggest challenge experimental quantum computing is noise. In the context of quantum advantage experiments, this limits the size (both number of qubits and number of operations) of quantum circuits that can be executed. Additionally, noise can make classical simulation easier through clever exploitations~\cite{Gao_Duan_2018,Aharonov_Gao_Landau_Liu_Vazirani_2023,Schuster_Yin_Gao_Yao_2024,Nelson_Rajakumar_Hangleiter_Gullans_2024} --- particular strategies for this will be discussed in detail in the chronological history.

Broadly, noise can be characterized as low and high, where in the low (or white) noise regime the noise rate is inversely proportional to the system size~\cite{Dalzell_Jones_Brandao_2021}. While most quantum experiments are in this regime, there is also the high noise regime with constant noise rates. While noise can be characterized in many ways in this regime, one of the most prominent ways is unital vs non-unital noise --- a noise channel $\mathcal{E}$ being unital if it preserves identity $\mathcal{E}(I) = I$ and otherwise non-unital. Many open questions are present in this category. For example, it is an open question whether it is possible to efficiently sample from random circuits with non-unital noise at constant rates~\cite{Fefferman_Ghosh_Gullans_Kuroiwa_Sharma_2023}. Unital vs. non-unital noise has also been explored in the context of barren plateaus, or lack thereof, when optimizing variational quantum circuits~\cite{Mele_Angrisani_Ghosh_Khatri_Eisert_França_Quek_2024,Singkanipa_Lidar_2025}, and in particular classical simulation algorithms~\cite{Angrisani_Schmidhuber_Rudolph_Cerezo_Holmes_Huang_2025,Angrisani_Mele_Rudolph_Cerezo_Holmes_2025}.

A second important remark that we will see in the chronological history is that rigorous hardness proofs are not known for many quantum advantage experiments, and that claims to advantage are thus based on best-known algorithms at the time of publication. (This of course leads to challenges and refutations as better algorithms are developed.) Further, current experiments cannot be efficiently verified, but rather require hard classical computations. The holy grail is an experiment which is able to be performed on current or near-term quantum computers, is efficiently verifiable, and has a rigorous hardness/advantage proof~\cite{Aaronson_Zhang_2024}. In this arena many open questions remain, even with experiments that have been performed. For example, both the advantage and verification aspects of random circuits sampling with sublogarithmic depth are open questions. Recently in this vein, results on when random circuits anticoncentrate (and thus are hard to simulate classically)~\cite{Dalzell_Hunter}, form (approximate) unitary designs~\cite{Belkin_Allen_Ghosh_Kang_Lin_Sud_Chong_Fefferman_Clark_2024,Schuster_Haferkamp_Huang_2025}, and converge to the uniform distribution under noise~\cite{Deshpande_Niroula_Shtanko_Gorshkov_Fefferman_Gullans_2022} have been demonstrated.
For certain advantage proposals like instantaneous quantum polynomial (IQP) circuits, it is known that sampling from the output distribution can be classically hard in certain cases~\cite{Bremner_Montanaro_Shepherd_2016} --- including on a square lattice of qubits which is highly relevant to current quantum computers~\cite{Bremner_Montanaro_Shepherd_2017} --- and in general, assuming standard complexity assumptions~\cite{Bremner_Jozsa_Shepherd_2011}. The hardness of IQP circuits has been shown to hold with intermediate measurements~\cite{Jozsa_Ghosh_Strelchuk_2025}. On the other hand, it is known that the addition of small noise to sufficiently anticoncentrated IQP circuits leads to efficient classical simulation~\cite{Bremner_Montanaro_Shepherd_2017}. Noisy simulation results without assuming anticoncentration are known as well~\cite{Rajakumar_Watson_Liu_2025}, and state-of-the-art classical algorithms estimate 70-qubit IQP circuits are within reach for classical simulation on large clusters~\cite{Codsi_Wetering_2025}. While no experimental quantum advantage claims with IQP circuits have been performed, recent experiments with neutral atoms have prepared logical states in IQP like architectures~\cite{Bluvstein_Evered_Geim_Li_Zhou_Manovitz_Ebadi_Cain_Kalinowski_Hangleiter_2024}.

Finally, we note that there are many more proposals for demonstrating quantum advantage, for example with (short-)time dynamics~\cite{BermejoVega_2018,Haferkamp_Hangleiter_Bouland_Fefferman_Eisert_Bermejo2020}, Gibbs sampling~\cite{Rajakumar_Watson_2024,Bergamaschi_Chen_Liu_2024}, and projected designs/analog simulators~\cite{Cotler_Mark_Huang_Hernandez_Choi_Shaw_Endres_Choi_2023,Choi_Shaw_Madjarov_Xie_Finkelstein_Covey_Cotler_Mark_Huang_Kale_2023}, among others --- our focus in what follows is on experimental realizations thus far.

\subsection{The first demonstration of quantum advantage: Google's 2019 experiment in random circuit sampling on the \textit{Sycamore} quantum computer}

\subsubsection{Description of the experiment} \label{sec:google-rcs-2019}

On Oct 23 2019, Google Quantum AI announced the first experimental demonstration of quantum advantage~\cite{Arute_Arya_Babbush_Bacon_Bardin_Barends_Biswas_Boixo_Brandao_Buell_}. The paper, bypassing the conventional arXiv submission prior to peer review and first appearing in \textit{Nature} after peer review, showed computational advantage for the problem of sampling from random quantum circuits with $n = 53$ qubits and depth $d = 20$ on the \textit{Sycamore} quantum computer. One million bitstrings were sampled in 200 seconds with estimated fidelity of $0.2\%$. (See the next section for an explanation of fidelity and other parameters described here.) Referring to the best-known classical simulation algorithms at the time, and assuming a state-of-the-art classical supercomputer (Summit), the authors estimated it would take ten thousand years to classically simulate the experiment. Thus computational advantage was demonstrated for the problem of random circuit sampling.

\begin{center}
    \hyperref[tab:overview]{\textit{Back to Table I}}
\end{center}

\subsubsection{Background on random circuit sampling}

Random circuit sampling is the computational problem of sampling bitstrings from an $n$-qubit, depth $d$ quantum circuit. This quantum circuit is designed to be hard to classically simulate relative to known classical algorithms. For example, quantum circuits with low entanglement such as short-depth circuits in one dimension can be efficiently classically simulated~\cite{Vidal_2003}, quantum circuits with only Clifford gates can be efficiently classically simulated~\cite{Aaronson_Gottesman_2004}, and quantum circuits with peaked output distributions can be efficiently classically simulated~\cite{Bravyi_Gosset_Liu_2024}. In constructing the random circuit, two-qubit gates are applied, usually in a two-dimensional pattern, to rapidly increase entanglement, and single-qubit gates drawn randomly from an ensemble are inserted afterwards. This pattern of single- and two-qubit gates generally defines a \textit{cycle} or \textit{layer}, and the depth is the number of cycles. The random circuit $U$ (and fixed input state $|0\rangle \equiv |0\rangle ^{\otimes n}$) defines a probability distribution 
\begin{equation}
    p(z) := | \langle z| U | 0\rangle | ^2
\end{equation}
and is designed to follow a Porter-Thomas (exponential) distribution. The computational problem is to produce samples from the distribution $p(z)$. Prior to the experiment, the problem of random circuit sampling and its theory had been developed in several papers~\cite{Neill_2017,Harrow_Montanaro_2017,Neill_Roushan_Kechedzhi_Boixo_Isakov_Smelyanskiy_Megrant_Chiaro_Dunsworth_Arya_2018,Bouland_Fefferman_Nirkhe_Vazirani_2019}, and the question of when quantum advantage would be achieved was highly anticipated and speculated on~\cite{Markov_Fatima_Isakov_Boixo_2018}.

Experimentally, quantum computers are noisy. To ensure the device is producing signal instead of pure noise in the experiment, a metric known as the (linear) cross entropy benchmark (XEB) fidelity is computed. Letting $z_i \in \{0, 1\}^n$ be the sampled bitstrings from the device, the (linear) XEB fidelity is
\begin{equation} \label{eqn:xeb-fidelity}
    \mathcal{F}_{\text{XEB}} := 2^n \mathbb{E} [ p(z_i) ] - 1
\end{equation}
where $n$ is the number of qubits and $p(z_i)$ is the probability of sampling bitstring $z_i$. On a perfect (noiseless) quantum computer, $\mathcal{F}_{\text{XEB}} = 1$ for random circuits following the Porter-Thomas (exponential) distribution. On a completely noisy quantum computer which causes the output distribution to be completely flat, $\mathcal{F}_{\text{XEB}} = 0$. Thus $0 \le \mathcal{F}_{\text{XEB}} \le 1$, and the value of the fidelity gives an estimate to the quality of the quantum computer. Note importantly that $p(z_i)$ must be computed classically, thus $\mathcal{F}_\text{XEB}$ is intractable to evaluate in the supremacy or beyond-classical regime and must be estimated. As noted, the the authors of~\cite{Arute_Arya_Babbush_Bacon_Bardin_Barends_Biswas_Boixo_Brandao_Buell_} estimate an XEB fidelity of $\mathcal{F}_{\text{XEB}} = 0.002$. This estimate is done by evaluating the fidelity for small circuits which can be classically simulated, as well as evaluating the fidelity of large circuits that are modified to be easy to classically simulate, e.g. by removing some or all two-qubit gates across a bipartition of the circuit. A key reason why linear cross entropy was chosen is because it tracks fidelity in certain regimes and because it can be computed efficiently from samples whereas other metrics like fidelity cannot~\cite{Ware_Deshpande_Hangleiter_Niroula_Fefferman_Gorshkov_Gullans_2023}. 

Although the random circuit sampling problem is classically hard and quantumly easy, it is not something (obviously) useful since the samples are drawn from random quantum dynamics. However, the idea to use the random circuit sampling to generate \textit{certified} randomness quickly emerged and is being formalized~\cite{Bassirian_Bouland_Fefferman_Gunn_Tal_2021, Aaronson_Hung_2023}. Thus, although not originally designed for this purpose, the random circuit sampling problem may have applications in certified random number generation.

\subsubsection{Significance of the experiment}

Prior to the experiment, as mentioned, whether quantum advantage could be achieved, and when it could be achieved, were highly discussed open questions~\cite{Preskill_2012,Preskill_2019,Markov_Fatima_Isakov_Boixo_2018}. Although it is questionable whether the computational problem solved by \textit{Sycamore} has practical value, the experiment has remarkable aspects and profound implications for both the theory of quantum physics and the theory of computer science. Physically, as noted in a talk by John Martinis on Nov 01 2019 at Caltech just after the results were announced, this experiment can be considered a test of quantum mechanics at a scale which has never been achievable before. Indeed, the experiment probes a Hilbert space of size roughly $10^{16}$, whereas the previous largest tests of quantum mechanics took place in Hilbert spaces of size roughly $10^{3}$. The fact that the experiment behaved as expected at this scale is a remarkable testament to the theory of quantum mechanics. In computer science, the experiment provides evidence that the extended Church-Turing thesis --- at the core of the theory of quantum computational complexity since the 1980s~\cite{Deutsch_Penrose_1997} --- may be violated. This thesis posits that a probabilistic Turing machine (i.e., classical computer) can efficiently simulate any physical process (i.e., computation) with polynomial overhead.  If a quantum computer is truly performing a computation that cannot be simulated classically, the extended Church-Turing thesis would be violated. As noted in~\cite{Harrow_Montanaro_2017}, this has both practical and and philosophical implications, demonstrating that quantum mechanics challenges our models of information and computation built on classical physics. For example, Bell's inequality shows quantum correlations (entanglement) are stronger than classical correlations and refutes local hidden variable theories. Quantum advantage experiments can thus be thought of as computational analogues to Bell experiments, as noted in~\cite{Harrow_Montanaro_2017}. Finally, the extremely good performance of the \textit{Sycamore} computer ---  average simultaneous single-qubit error of 0.16\% and average simultaneous two-qubit error of 0.62\%~\cite{Arute_Arya_Babbush_Bacon_Bardin_Barends_Biswas_Boixo_Brandao_Buell_} --- are remarkable achievements in experimental physics and engineering. These comments generally apply equally to subsequent quantum advantage experiments but are only stated here for the first experiment.

\subsubsection{Challenges to the experiment} \label{sec:challenges-to-google-rcs-2019}

While the experiment was announced when the paper was published in \textit{Nature} on Oct 23, 2019, the first challenge to the experiment came two days before on Oct 21, 2019\footnote{As developments were rather rapid and chronology is important, here and throughout we use the first posting (typically to arXiv) for the date.}~\cite{Pednault_Gunnels_Nannicini_Horesh_Wisnieff_2019}. This seemingly bizarre fact is because the paper was accidentally leaked on a NASA pre-print server on Sep 14, 2019 during peer review~\cite{IBM_challenge_leak, supremacy_leak_nasa_internet_archive}. This challenge built on previous work~\cite{Pednault_Gunnels_Nannicini_Horesh_Magerlein_Solomonik_Draeger_Holland_Wisnieff_2020} and claimed that with memory modifications it would only take Summit a matter of days, rather than thousands of years, to simulate the experiment using the simple Schr\"odinger algorithm in which the entire wavefunction is stored and updated gate-by-gate. However this is a (disputed) claim and not an actual computation, and no actual computation has been performed since the challenge appeared five years ago. So this challenge appears to be unsubstantiated, however it is of historical interest as the first challenge to a claim of quantum advantage, humorously coming two days before the claim of quantum advantage was made.

The first chronologically intelligible challenge to Google's 2019 experiment appeared on Feb 5, 2020~\cite{Gray_Kourtis_2021}, estimating a 10,000 times speedup over the 10,000 year estimate of~\cite{Arute_Arya_Babbush_Bacon_Bardin_Barends_Biswas_Boixo_Brandao_Buell_}. This speedup is achieved through new algorithms to find good contraction paths in tensor networks, formulating the problem of simulating a quantum circuit as contracting a tensor network~\cite{Markov_Shi_2008}. The best algorithm for contracting tensor networks from Google's 2019 experiment, roughly speaking, uses a divisive approach to partition a vertex in the contraction graph into $k$ partitions ($k$ being a hyperparameter) then emplys a greedy or optimal algorithm to construct the contraction tree for each subgraph. This approach, known as hypergraph partitioning
, is shown to produce an orders of magnitude better contraction cost compared to five other state-of-the-art algorithms when applied to Google's 2019 experiment. This approach is also combined with slicing to parallelize tensor network contraction. To mimic the fidelity $f$ of Google's experiment, it is proposed to sample from the maximally mixed state $I / 2^n$ with probability $1 - f$ and sample from $|\psi\rangle$ (i.e., contract the tensor network) with fidelity $f$. Referencing a similar recent tensor network contraction on Summit, the authors estimate that Google's experiment could be performed in 195 days on Summit, assuming a $68\%$ FLOPS efficiency.

The first challenge to Google's 2019 experiment utilizing noise to make the classical simulation easier occurred on Feb 18, 2020~\cite{Zhou_Stoudenmire_Waintal_2020}. As mentioned, the fidelity of Google's experiment was estimated at $0.2\%$. The authors of~\cite{Zhou_Stoudenmire_Waintal_2020} ask what random circuits are classically simulable with low fidelity. Their technique uses the well-known time-evolving block decimation (TEBD) algorithm~\cite{Vidal_2003} with truncation to emulate loss of fidelity. In this technique, the wavefunction is represented by a matrix product state (MPS) and gates are applied by contracting the tensors into the MPS. For two-qubit operations, a singular value decomposition is performed after contracting the operation into the MPS. A degree of approximation is achieved by truncating the singular value (Schmidt) spectrum, either by keeping a maximum number of singular values (Schmidt rank) or discarding singular values below a pre-determined threshold. Arguing that singular value truncation is a proxy for physical noise in quantum computers, and arguing that overall fidelity due to approximation is multiplicative in each gate fidelity, the authors of~\cite{Zhou_Stoudenmire_Waintal_2020} are able to simulate two-dimensional random circuits on $n = 54$ qubits with depth $20$, the same parameters as in Google's 2019 experiment~\cite{Arute_Arya_Babbush_Bacon_Bardin_Barends_Biswas_Boixo_Brandao_Buell_}. However, the random circuits in~\cite{Zhou_Stoudenmire_Waintal_2020} are not identical to Google's experiment since they use controlled-$Z$ (CZ) gates and different random rotations on single qubits. The different two-qubit gate is notable because CZ has Schmidt rank two whereas the $\sqrt{\text{iSWAP}}$ used in Google's experiment has Schmidt rank four. Although the entanglement dynamics are not completely determined by the Schmidt rank, it is generally expected that higher Schmidt rank corresponds to harder classical simulation. Using a maximum MPS bond dimension of $\chi = 320$ to produce a final fidelity of around $0.2\%$ as in Google's experiment, the total classical simulation on a single core took less than $48$ hours and used $4.5$ GB of memory. When using a two-qubit gate with Schmidt rank four, as in Google's 2019 experiment, the authors are unable to perform the classical simulation with fidelity as large as Google's 2019 experiment. They speculate that a parallelized version with a few hundred cores and a few terabytes of memory could reach comparable fidelity, but note that such a calculation has not been attempted. While this work raises many interesting questions, and is very notable as the first to exploit approximate classical simulation to challenge Google's 2019 experiment, it is generally not considered to refute Google's 2019 experiment because of differences in the circuits and differences in the definition of fidelity.

On May 14 2020, another approach using optimized tensor network contraction~\cite{Huang_Zhang_Newman_Cai_Gao_Tian_Wu_Xu_Yu_Yuan_2020} was put forth to challenge the 10,000 year estimate. Specifically, for depth $14$ circuits with $53$ qubits, the authors collect three million samples with $1\%$ fidelity in $264$ seconds. In addition to significantly improving the 1.1 year estimate of~\cite{Arute_Arya_Babbush_Bacon_Bardin_Barends_Biswas_Boixo_Brandao_Buell_}, this time is actually two times faster than Sycamore. For depth $20$, it was estimated to take twelve hours to sample bitstrings at $0.2\%$ fidelity. However, this estimate is assuming $100\%$ FLOPS efficiency, when in practice on depth $14$ circuits only a low FLOPS efficiency of around $15\%$ was reached. The main contribution in this approach is optimization of the tensor network contraction through several veins. First, the authors identify a ``stem'' of the contraction tree which dominates the cost, and optimize this stem through several recently introduced techniques --- notably hypergraph partitioning~\cite{Gray_Kourtis_2021} and dynamic slicing~\cite{Chen_Zhang_Huang_Newman_Shi_2018} --- to find a good contraction path. As in~\cite{Villalonga_Boixo_Nelson_Henze_Rieffel_Biswas_Mandra_2019,Markov_Fatima_Isakov_Boixo_2018}, six time-like edges are left open in the tensor network to compute a batch of amplitudes and sample bitstrings via frugal rejection sampling. Although there are many improvements introduced in this work and the depth $14$ results are impressive, the depth $20$ results in $12$ hours would have required substantial engineering effort and improved hardware to be done in practice, thus we do not consider this work to be a weak refutation of Google's 2019 experiment.

On Mar 04 2021, a challenge to Google's 2019 experiment was announced by two researchers from the Chinese Academy of Sciences~\cite{Pan_Zhang_2021}. In this work, the authors are able to sample one million \textit{correlated} bitstrings from the $53$ qubit depth $20$ Sycamore circuits in \textit{five days}. The reason why bitstrings are correlated can be understood from the method. As in previous works, the authors use tensor network contraction to simulate the circuit. Whereas previous works have kept a small number (typically six) of qubits open to compute a batch of amplitudes rather than a single amplitude, a key contribution of this work is to select a large number of open qubits, significantly increasing the batch size. As noted by the authors, keeping more qubits open increases the complexity of the contraction, and finding an optimal set of qubits to keep open is a hard combinatorial optimization problem. To deal with these challenges, the authors first find a good contraction order and then select the set of open qubits based on this order. This contraction order is optimized to find a structure --- called a ``big-head structure'' by the authors --- which encapsulates the bottleneck of the contraction. The overall tensor network is then partitioned into the \textit{head} and \textit{tail} sub tensor networks, where qubits in the head are open and qubits in the tail are closed, which are further partitioned and contracted independently. The final step is to contract the resulting head and tail tensors (vectors) via a dot product. Because the key idea of this approach is to maximize the number of open qubits which occur in the ``head'' tensor network, this algorithm is dubbed the ``big head algorithm.'' It is important to note that the authors use dynamic slicing to contract the head and tail tensor networks, building on previous work~\cite{Gray_Kourtis_2021}. This results in $2^{23}$ subtasks, each of which have $2^{30}$ space complexity and are put into $32$ GB of GPU memory. For sampling different bitstrings, the algorithm can be perfectly parallelized; the authors use a cluster of $48$ NVIDIA V100 GPUs and $12$ NVIDIA A100 GPUs to sample two million bitstrings in about five days. In practice, the head tensor network contains $21$ open qubits and the tail tensor network contains $32$ fixed (closed) indices. This means that each contraction yields a set of $2^21$ bitstrings, of which $32$ bits are fixed. Thus, the sampled bitstrings are correlated. This bitstrings are then post-selected to reduce correlations and obtain a linear XEB fidelity of $0.739$, higher than that achieved by Sycamore in Google's 2019 experiment. Thus, less than eighteen months after Google's announcement of quantum supremacy, the estimated classical runtime of $10,000$ years was reduced to five days, albeit with caveats such as the fact that bitstrings were correlated.

On Oct 27 2021, a scheme to sample one bitstring from Google's 2019 experiment in 304 seconds on the Sunway supercomputer was presented~\cite{Yong_Liu_Xin_Liu_Fang_Li_Fu_Yang_Song_Zhao_2021}. Estimating a runtime of 7,610 years on Sunway at 100\% compute efficiency to do the ``naive'' tensor network contraction, the authors build on previously discussed work and utilize a PEPS-based contraction algorithm optimized for both the circuits and the underlying Sunway architecture. Sampling is done by frugal rejection sampling in which 512 amplitudes are computed in a batch (i.e., ten qubit wires are left open in the contraction). The authors use 41,932,800 cores on Sunway, each having 32 GB memory and a peak performance of 4.7 TFlops, in their largest experiment. Circuit sizes of up to $n = 100$ qubits and $m = 42$ cycles are considered, however the definitions of random circuits here are not identical with those used experimentally, differing both in the definition of depth as well as the type of two-qubit gate (which drastically affects complexity). For example, this work references and builds off~\cite{Guo_Liu_Xiong_Xue_Fu_Huang_Qiang_Xu_Liu_Zheng_2019}, consisting of several of the same authors, in which $n = 64$ and depth $m = 25$ quantum circuits were simulated on a personal computer, numbers which would have already refuted Google's 2019 experiment if they corresponded directly to the definitions used in the experiment. In the published version of the paper in ACM, a video of a talk given by an author shows that with further improvements it is projected that one uncorrelated sample could be drawn in 60.4 seconds. Recall that \textit{Sycamore} sampled one million bitstrings in 200 seconds.

On Nov 01 2021, a single uncorrelated bitstring from the $n = 52$, $m = 20$ random circuits in Google's 2019 experiment is sampled on the Sunway supercomputer in 400 seconds in single-precision arithmetic and 276 seconds in mixed-precision arithmetic~\cite{Liu_Guo_Liu_Yang_Song_Gao_Wang_Wu_Peng_Zhao_2021}. This work applies the ``big head'' contraction algorithm of~\cite{Pan_Zhang_2021} combined with the previous optimizations for Sunway (by a similar group of authors) in~\cite{Yong_Liu_Xin_Liu_Fang_Li_Fu_Yang_Song_Zhao_2021}, again using 41,932,800 cores but this time with experimental parameters (i.e., $n$, $m$, and the two-qubit gates + overall structure of the random circuits) matching those of Google's 2019 experiment. In this work six qubit wires are left open to compute a batch of $64$ amplitudes per contraction, then frugal sampling is performed to extract uncorrelated bitstrings. In addition to the $m = 20$ cycle experiment, the authors also consider the $m = 12$ and $m = 14$ cycle experiments, showing they can draw bitstrings in 18 and 82 seconds, respectively.

\subsubsection{Refutations of the experiment} \label{sec:weak-refutation-of-google-rcs-2019}

On Nov 04 2021, two years and twelve days after Google's announcement of quantum advantage, the experiment was (weakly) refuted~\cite{Pan_Chen_Zhang_2022}. This work, posted by the same authors as~\cite{Pan_Zhang_2021} in addition to one other co-author, generated one million uncorrelated bitstrings from the Sycamore circuits with fidelity $0.0037$ in fifteen hours. This computation was performed on a cluster of $512$ GPUS, and, while slower than Sycamore which sampled one million bitstrings in $200$ seconds, it is estimated and reasonably believed that on a larger exa-scale cluster the entire classical simulation could be done in a few dozens of seconds, \textit{faster} than Sycamore. For this reason we take this to be a weak refutation of Google's 2019 experiment. The key new idea in this work is removing a subset of edges to contract in the original tensor network to obtain an approximate tensor network. This approximation can be viewed as the analogue of noise in the quantum processor. Specifically, breaking an edge is equivalent to a Pauli error $E := (I + Z) / 2 = |0\rangle \langle 0|$, the projector onto the $|0\rangle$ state, effectively replacing the local state with a $|0\rangle$ state. As each path in the error channel $E$ contributes weight $1/2$, each broken edge halves the fidelity. In practice, eight edges (or four two-qubit gates) are removed to get a fidelity of $2^{-8} \approx 0.0039$, which combined with other approximations yield the final fidelity estimate of $0.0037$. These other approximations come from exploiting the singular value spectrum of the two-qubit gates in special cases. The approximate tensor network is then contracted using the big head algorithm~\cite{Pan_Zhang_2021}, which again builds on previous work by, e.g., incorporating dynamic slicing~\cite{Gray_Kourtis_2021}. In this work, however, six qubits are left open in the head tensor network, and importance sampling is used to remove correlations in the sampled bitstrings. Ultimately, $2^{20} \approx 10^{6}$ uncorrelated bitstrings were computed in just over fifteen hours. The code, contraction path(s), and sampled bitstrings are openly available~\cite{Pan_2023}. We remark that Ref.~\cite{Kalachev_Panteleev_Zhou_Yung_2021} is also notable for achieving similar runtime (15 hours) just two months later. 

On Jun 27 2024, Zhao \textit{et al.} claimed the first \textit{strong} refutation of Google's 2019 experiment, using a cluster of 1432 GPUs to compute three million samples in 86.4 seconds, \textit{faster} than the time it took \textit{Sycamore} to sample one million bitstrings (200 seconds). The algorithm is a culmination of previous techniques (which have been previously described) including the ``big head'' algorithm, dynamic slicing, and optimizing the stem of the contraction tree. Markov chain Monte Carlo is used to sample, then bitstrings are post-selected to maximize the XEB fidelity. In particular, ten qubits are left open in the ``head'' of the tensor network and $3 \cdot 10^6$ samples are drawn. Out of a total $2^{24}$ subtasks (from slicing), only 0.03\% of these are contracted to reach an XEB value of 0.002 which is what was achieved by Sycamore. Distributed across 1432 GPUs the entire simulation takes 86.4 seconds. To the author's knowledge, this work constitutes the first strong refutation of of Google's 2019 random circuit sampling experiment. It is interesting to note that the authors also consider power consumption of both experiments. Noting that previous challenges to Google's 2019 experiment have used roughly three orders of magnitude more energy than Sycamore (which is taken to be 4.3 kWh for cooling), the authors note that their entire simulation requires 13.7 kWh. Thus, although still larger, this is significantly closer to energy consumption of \textit{Sycamore}. In a comment on the arXiv version of this paper, it is noted that the work was completed in Aug 2023 and that a 50x improvement has since been achieved, however this has yet to be published. Finally, it is worth remarking that as classical techniques have steadily improved culminating in this refutation, quantum computers have likewise steadily improved resulting in larger random circuit sampling experiments which have not been refuted (see Tab.~\ref{tab:overview} and the subsequent discussion).

It is notable that Google's decision to openly share data from the experiment~\cite{Arute_Arya_Babbush_Bacon_Bardin_Barends_Biswas_Boixo_Brandao_Buell_} --- the exact circuits and sampled bitstrings are made available --- has significantly catalyzed these developments and advanced science. It is also important to note that, as can be seen in the preceding discussion, there is a clear thread of ideas being used and built upon from the first challenge to the final refutation. Without the infrastructure or the incentive to publish scientific articles freely online --- in this case the arXiv --- the developments described here would have taken significantly longer or may not have happened at all.

\subsubsection{Loopholes in the cross entropy benchmark fidelity} \label{sec:loopholes-rcs}

In the previous section, the classical simulators input quantum circuits and output bitstrings in the same way that a quantum computer would, and the XEB fidelity was used as a benchmark of the quality of the bitstrings. A different strategy is to focus on the definition of XEB fidelity itself and devise a classical algorithm which takes advantage of any loopholes in this metric. The first such work towards this end was presented in~\cite{Barak_Chou_Gao_2020} in which the authors try to use locality to ``spoof'' the cross entropy benchmark. In this work, a randomized classical algorithm is given to yield $\mathcal{F}_\text{XEB} = 1 / \text{poly}(n)$ (on average) in running time $2^{O(m^c)}$ where the depth of the Haar random circuit is logarithmic in the number of qubits $m = O (\log n)$ and $c$ is a constant. Note that random circuit sampling experiments are designed to scale the depth as $m \sim \sqrt{n}$. However, as noted by the authors, prior to their work only results with constant $m = O(1)$ were known. As mentioned, the key idea of this spoofing strategy is to exploit locality. In particular, the algorithm works by identifying a small number of qubits with disjoint light cones to compute the output distribution of. Bits are sampled from these qubits according to the computed probability distributions, then bits are sampled uniformly at random for the remaining qubits. For a light cone size of $L$ the time to find disjoint qubits can be done in $\text{poly}(n)$ time, while computing the marginal distributions requires time $\text{poly}(n, 2^L)$. The main question is thus the locality of the XEB fidelity. In follow-up experimental work~\cite{Morvan_2024} (Sec.~\ref{sec:google-second-experiment}), this question is experimentally probed to find that global correlations are the biggest contributing factor to XEB fideltiy.

The validity of XEB was also called to question in~\cite{Gao_Kalinowski_Chou_Lukin_Barak_Choi_2024}. In this work, an efficient classical algorithm which achieves high XEB values \textit{without} simulating the full circuit is presented. This algorithm simply partitions the circuit into $l$ parts of size $\lceil n / l \rceil $ qubits which can be simulated directly in time exponential in $l$. It is argued that these partitioned circuits will achieve relatively high XEB values, and several ways to improve the XEB based on post-processing sampled bitstrings are given. For Google's experiment, the circuits are partitioned into roughly equal halfs of $l = n / 2$ qubits, and this subsystem is simulated on an NVIDIA Tesla V100 GPU with 32GB memory in about one second, achieving an XEB of $2.7 \times 10^{-4}$, a value that is 12.3\% of the experimentally achieved value (around 0.2\%). Additionally, a general discussion of vulnerabilities of XEB as a metric is given. In particular, examples where the XEB fidelity and ``traditional'' fidelity $F := \langle \psi | \rho | \psi \rangle$ (where $\rho$ is the noisy state and $|\psi\rangle$ is the target state) diverge are given. The authors also show numerically that the ratio of XEB to ``traditional'' fidelity actually increases in $n$ with their algorithm faster than on a noisy quantum processor with fixed error rate. However, while this approach works for local circuits, there is evidence that it has limitations --- for example, the classical spoofer can suffer from exponentially large variance for shallow Brownian circuits and circuits with all-to-all connectivity~\cite{Bentsen_Fefferman_Ghosh_Gullans_Liu_2024}.

\subsection{Two Gaussian Boson sampling experiments on the \Jiuzhang{} quantum computer} \label{sec:ustc-gbs-jiuzhang-1-and-2}

\subsubsection{Description of the first experiment} \label{sec:ustc-gbs-1}

On Dec 03 2020, researchers from USTC reported an $n = 50$, $m = 100$ Gaussian boson sampling experiment with threshold detectors measuring up to $N_c = 76$ clicks~\cite{Zhong_Wang_Deng_Chen_Peng_Luo_Qin_Wu_Ding_Hu_2020}. The authors estimate a quantum computational advantage of $10^{14}$ --- i.e., it would take the best-known classical algorithm running on the TaihuLight supercomputer $10^{14}$ times as long as the quantum computer. As the Gaussian boson sampler \Jiuzhang{} took 200 seconds to collect all samples, this corresponds to an estimate of 2.5 billion years to classically simulate the experiment.

The experiment uses 25 two-mode squeezed states, equivalent to $n = 50$ single-mode squeezed states, as input. Both the spatial and polarization degrees of freedom are used to implement a random $100 \times 100$ unitary transformation acting on the basis $|H\rangle_1, |V\rangle_1, ..., |H\rangle_{50}, |V\rangle_{50}$. Here, $H$ denotes horizontal polarization, $V$ denotes vertical polarization, and subscripts index the mode. The unitary transformation is implemented by 300 beam splitters and 75 mirrors. The $m = 100$ output modes are detected by 100 threshold detectors with an average efficiency of 81\%. Up to $N_c = 76$ photons are detected in the experiment, and the total data collection time of the experiment is 200 seconds.

Just as in random circuit sampling (Sec.~\ref{sec:google-rcs-2019}), the experimental apparatus is noisy and needs to be benchmarked or validated to show it is behaving as intended. Also similar to random circuit sampling, this is tricky because the experiment is designed to be beyond classical so that the output cannot directly be verified. Nonetheless, several measures are taken towards this end. First, the $100 \times 100$ unitary matrix is measured experimentally and computed analytically and shown to be in agreement. The experimentally measured matrix is also shown to be (Haar) random which is required for the computational hardness. Next, a smaller experiment with $n = 6$ single-mode squeezed states and an output photon number of two is performed. In this regime, the experiment is small enough to be classically simulated. The fidelity $F := \sum_i \sqrt{p_i q_i}$ and total variation distance $D := \sum_i \left| p_i - q_i \right| / 2$ are used to compare the ideal and experimental distributions $p$ and $q$. For these small experiments. the average fidelity is $0.990$ and the average total variation distances is $0.103$. (For a perfect boson sampler, the fidelity would be one and the total variation distance would be zero.) Finally, the authors consider several scenarios under which the large-scale experiment would be classically simulable, and show these scenarios produce outcomes which do not agree with the experimental results. For example, the authors consider the scenario in which the input light was distinguishable, as well as the case in which the boson sampler was just doing uniform random sampling, and show strong deviations from the actual experiment.

Last, we note importantly that this experimental configuration is not programmable, meaning that performing a different experiment would require physically changing hardware components.

\begin{center}
    \hyperref[tab:overview]{\textit{Back to Table I}}
\end{center}

\subsubsection{Description of the second experiment} \label{sec:ustc-gbs-2}

Approximately six months after the first experiment, on Jun 29 2021 the same team from USTC published an improved version of their Gaussian boson sampling experiment~\cite{Zhong_Deng_Qin_Wang_Chen_Peng_Luo_Wu_Gong_Su_2021}. In this improved version, a Gaussian boson sampling experiment with threshold detection is performed on the photonic quantum computer \textit{Jiuzhang 2.0} with $n = 50$, $m = 144$, and up to $N_c = 113$ clicks. This largest number of clicks corresponds to sampling from a Hilbert space with dimension around $10^{43}$. It is estimated that the same computational task would take $10^{24}$ times longer using brute-force simulation on classical supercomputers. The total sampling time in the experiment is 200 seconds. The \textit{Jiuzhang 2.0} computer is also partially programmable --- phases can be changed via the addition of adjustable electric delay lines. This is an improvement over the first experiment which was not programmable, however the majority of the hardware is still fixed and this does not represent a programmable computer.

This experiment introduces the new concept of stimulated emission of squeezed states to address the difficulty of generating large arrays of squeezed states suitable for Gaussian boson sampling. Using this, 25 two-mode squeezed states, corresponding to $n = 50$ single-mode squeezed states, are sent into an $m = 144$ mode interferometer implementing a random unitary transformation. After, the output photons are detected with 144 single-photon detectors with an average efficiency of 83\%.

As in the first experiment, the authors consider several methods which could spoof their experiment --- such as using thermal states (due to excessive photon) loss or performing uniform random sampling --- and show they deviate from their experimental results. This is also done for smaller experiments in which the results can be classically simulated, providing some verification that the Gaussian boson sampler is performing as intended in the beyond-classical regime.

\begin{center}
    \hyperref[tab:overview]{\textit{Back to Table I}}
\end{center}

\subsubsection{Background on (Gaussian) boson sampling}

Boson sampling~\cite{Aaronson_Arkhipov_2011} is the quantum analogue of a Galton's board --- spherical balls dropped through a series of equally spaced pegs to illustrate, in the limit of many balls, a normal distribution. Here, the probability of finding a ball in a particular location is given by the permanent of the matrix $A = [a_{ij}]$ where $a_{ij}$ is the probability of ball $i$ landing at location $j$.
%
%
In the classical case, $a_{ij} \ge 0$. This leads to an efficient (classical) algorithm to approximate the permanent~\cite{Jerrum_Sinclair_Vigoda_2004}. The Boson sampling problem is identical to this problem, however the matrix $A$ has complex elements. In this case it turns out that unless $\text{P}^{\#\text{P}} = \text{BPP}^{\text{NP}}$ and the polynomial hierarchy collapses to the third level, the exact Boson sampling problem is not efficiently solvable by a classical computer~\cite[Theorem 1]{Aaronson_Arkhipov_2011}. Typically photons are used experimentally, and we will use photon and boson interchangeably. Experimentally, a setup analogous to a Galton's board is constructed in which $n$ identical photons can be found in $m$ \textit{modes} (locations). The experiment can be described by an $m \times m$  unitary transformation, which is constructed via a sequence of smaller unitary transformations that are realized experimentally by optical elements, e.g. beamsplitters and phaseshifters. Let $U$ be the $m \times m$ unitary implemented by the experiment, and let $A$ be the $m \times n$ isometry defined by the first $n$ columns of $U$. Then, the probability of sampling $z = z_1 \cdots z_m$ where $\sum_i z_i = n$ is given by
\begin{equation}
    p(z) = \frac{| \text{Per} \, A_z |^2 }{z_1 ! \cdots z_m !}
\end{equation}
where $A_z$ is the square submatrix of $A$ with $z_i$ copies of row $i$~\cite{Neville_Sparrow_Clifford_Johnston_Birchall_Montanaro_Laing_2017}. The permanent of a square $n \times n$ matrix $M$ is given by $\sum_{\sigma \in S_n} \prod_{i = 1}^{n} a_{i\sigma(i)}$ where $S_n$ is the symmetric group of $n$ elements --- i.e., the set of all permutations of $n$ elements. As in random circuit sampling, the computational problem of boson sampling is to produce samples from the distribution $p(z)$. Note that boson sampling is not a universal model for quantum computation, however when combined with adaptive measurements the model is universal~\cite{Knill_Laflamme_Milburn_2001}. 

Sources of single photons (primarily spontaneous parametric down conversion) are non-deterministic, creating experimental challenges to scaling boson sampling. Because of this several variants have been proposed. In Gaussian boson sampling (GBS)~\cite{Hamilton_Kruse_Sansoni_Barkhofen_Silberhorn_Jex_2017}, a particular Gaussian state known as a single-mode squeezed state is used as input to the array of optical elements. The output is measurements in the photon number basis, as in boson sampling. The difference is that now the probability of sampling depends on the \textit{hafnian} of a matrix characterizing the state, rather than the permanent. The hafnian is another matrix function related to the permanent by $\text{Perm}(A) = \text{Haf} \left[ \begin{matrix}
    0 & A \\
    A^T & 0 \\
\end{matrix} \right] $ and computing it is also in the class \#P, even in the approximate case (under some assumptions)~\cite{Hamilton_Kruse_Sansoni_Barkhofen_Silberhorn_Jex_2017}. Due to experimental advantages, Gaussian boson sampling has thus far been the preferred experimental route to demonstrating quantum advantage via boson sampling.

An output sample from a boson sampling experiment depends on the type of measurement apparatus, or detector, that is used. Certain detectors count the number of photons in each mode, termed \textit{photon number resolving detectors}. In this case, an output sample is a string of non-negative numbers $z_1, ... z_m$ where $z_i$ is the number of photons detected in mode $i$. Photon number resolving detectors are currently based on superconducting technology and must be operated at cryogenic temperatures. An alternative, proposed in~\cite{Quesada_Arrazola_Killoran_2018}, is to use \textit{threshold detectors} which distinguish between two cases: zero photons, or one or more photons. The case of one or more photons is referred to as a \textit{click}. Thus in this case the output is a bitstring $z \in \{0, 1\}^m$, and the number of ones in the bitstring $z$ is the number of clicks in the experiment. Threshold detectors can be operated at room temperature, providing an experimental advantage over photon umber resolving detectors. The best-known classical algorithm for classically simulating Gaussian boson sampling with threshold detectors scales as $O \left(m^2 2^{N_c} \right)$ where $N_c$ is the number of clicks (and $m$ as before is the number of modes)~\cite{Quesada_Arrazola_Killoran_2018}.

After the theoretical proposal, several small, proof-of-principle boson sampling experiments with increasing complexity had been performed~\cite{Tillmann_Dakić_Heilmann_Nolte_Szameit_Walther_2013,Spring_Metcalf_Humphreys_Kolthammer_Jin_Barbieri_Datta_Thomas_Peter_Langford_Kundys_2013,Spagnolo_Vitelli_Bentivegna_Brod_Crespi_Flamini_Giacomini_Milani_Ramponi_Mataloni_2014,Loredo_Broome_Hilaire_Gazzano_Sagnes_Lemaitre_Almeida_Senellart_White_2017,Crespi_Osellame_Ramponi_Brod_Galvão_Spagnolo_Vitelli_Maiorino_Mataloni_Sciarrino_2013,Carolan_Meinecke_Shadbolt_Russell_Ismail_Wörhoff_Rudolph_Thompson_OBrien_Matthews_2014,Carolan_Harrold_Sparrow_Martín_Lopez_Russell_Silverstone_Shadbolt_Matsuda_Oguma_Itoh_2015,Broome_Fedrizzi_Rahimi_Keshari_Dove_Aaronson_Ralph_White_2013,Bentivegna_Spagnolo_Vitelli_Flamini_Viggianiello_Latmiral_Mataloni_Brod_Galvão_Crespi_2015}. Prior to the first claim of quantum advantage in boson sampling, the state of the art was an Oct 22 2019 paper in which results from an $n = 20$ photon and $m = 60$ mode Gaussian boson sampling experiment with up to $N_c = 14$ detected photons were presented~\cite{Wang_Qin_Ding_Chen_Chen_You_He_Jiang_Wang_You_2019}.

\subsubsection{The status of classically simulating Gaussian boson sampling prior to quantum advantage experiments}

Prior to these experiments claiming quantum advantage, Neville \textit{et al.} in 2017 developed and benchmarked classical algorithms for boson sampling to set the goalposts for quantum advantage~\cite{Neville_Sparrow_Clifford_Johnston_Birchall_Montanaro_Laing_2017}. In this work, a small cluster was used to classically compute $250$ samples from an $n = 30$, $m = 900$ boson sampling problem in five hours. From this benchmark of their algorithm, as well as from experimental parameters of recent boson sampling experiments, the authors estimate it would require around $n = 80$ photons with $m = n ^ 2$ modes to reach quantum advantage. 

After this work, in 2017 Clifford and Clifford presented an exact classical algorithm for boson sampling which improves the brute-force sampling algorithm~\cite{Clifford_Clifford_2018}. For an $n$ boson, $m$ mode problem, the brute-force algorithm requires computing ${m + n - 1 \choose n}$ permanents of $n \times n$ matrices, giving a runtime of $O ( {m + n - 1 \choose n} n 2 ^ n)$. As noted by the authors, in the case that $m \ge n ^ 2$ this becomes $\Theta ( 2^n e^n  (m / n)^n \sqrt{n} )$ as shown in~\cite{Aaronson_Arkhipov_2011}. The authors present a new algorithm for exact sampling which takes time $O(n 2 ^ n + m n ^ 2)$ and $O(m)$ space. This algorithm is inspired by hierarchical sampling in Bayesian computation uses several innovations, for example introducing permutations which do not change the permanent and make it easier to compute, using conditional sampling, and using the Laplace expansion for permanents. Ultimately, the algorithm presents a large ${m + n - 1 \choose n}$ speedup for the boson sampling problem and has small constant factors making it amenable to implementation. Indeed the algorithm has been implemented in R and is freely available~\cite{Clifford_Clifford_boson_sampling_r_package}. The authors estimate it would require $n = 50$ photons with $m = n^2$ modes to achieve quantum advantage in boson sampling. 

In 2018, two algorithms for computing the permanent of $n \times n$ matrices with complexity $O(n^2 2^n) $ were implemented on the Tianhe-2 supercomputer as a classical benchmark of boson sampling~\cite{Wu_Liu_Zhang_Jin_Wang_Wang_Yang_2018}. The two algorithms are the Ryser algorithm~\cite{Ryser_1963} implementing the formula
\begin{equation*}
    \text{Per} \, A = (-1)^n \sum_{S \subset \{1, 2, ..., n\}} (-1)^{|S|} \prod_{i = 1}^{n} \sum_{j \in S} a_{ij}
\end{equation*}
and the so-called BBFB algorithm \cite{Balasubramanian_1980, Bax_1998} implementing the formula
\begin{equation*}
    2^{n - 1} \text{Per} \, A = \sum_\delta \prod_{k = 1}^{n} \delta_k \prod_{i = 1}^{n} \sum_{j = 1}^{n} \delta_j a_{ij}
\end{equation*}
where $\delta = \{\delta_1, ... \delta_n\}$ with $\delta_1 = 1$ and $\delta_i = \pm 1$ for $i > 1$. These algorithms are parallelized across up to 312,000 CPU cores, and the result is that one sample from an $n = 50$ boson sampling experiment (with $m = n^2$) can be drawn in around 100 minutes.

In 2018, Björklund \textit{et al.} devised an algorithm to compute the hafnian of a complex matrix in time $O(n^3 2^{n / 2})$ and benchmarked it on the Titan supercomputer~\cite{Björklund_Gupt_Quesada_2019}. Recall that computing the permanent corresponds to boson sampling and computing the hafnian corresponds to Gaussian boson sampling. Although the developed algorithm is perfectly parallel, the authors estimate that computing the hafnian of a $100 \times 100$ matrix would require the 288,000 CPUs of Titan roughly 1.5 months, assuming perfect parallelization over distributed nodes. The largest computed benchmarks were for $50 \times 50$ matrices which took a few hundred seconds on a single node and a few seconds on around 100 nodes, both using 16 threads per node.

Also in 2018, a similar group of authors in Ref.~\cite{Gupt_Arrazola_Quesada_Bromley_2020} implemented the proposed sampling algorithm from the previously-described original paper proposing Gaussian boson sampling with threshold detectors~\cite{Quesada_Arrazola_Killoran_2018}. Using a similar parallel scheme on the Titan supercomputer, this time with up to 240,000 nodes, a classical simulation of Gaussian boson sampling with $m = 800$ modes and $N_c = 20$ clicks is implemented, producing a sample in around two hours. This benchmarking indicates that Gaussian boson sampling with threshold detectors could achieve quantum computational advantage with $m = 2 N_c ^2$ modes and $N_c > 22$ clicks.

In 2019, Qi \textit{et al.} considered noisy processes in Gaussian boson sampling and used this to reduce the classical simulation time~\cite{Qi_2020}. In particular, a model of photon loss during state preparation is considered and it is shown that lossy Gaussian boson sampling can be efficiently simulated classically when the average number of detected photons is quadratically related to the number of input photons. This result establishes a condition which Gaussian boson sampling experiments must satisfy in order to claim computational hardness. Interestingly, the authors also show that \textit{exact} lossy Gaussian boson sampling is likely still hard. Specifically, it is shown that if lossy Gaussian boson sampling can be efficiently simulated classically, then the polynomial hierarchy collapses to the third level.

Later in 2019, Wu \textit{et al.} developed an alternative algorithm and sampled from an $n = 18$, $m = n ^ 2 = 324$ Gaussian boson sampling experiment on a laptop in twenty hours~\cite{Wu_Cheng_Jia_Zhang_Yung_Sun_2020}. This algorithm assumes single mode squeeze states as input and its complexity depends on the squeezing parameter. The algorithm is only benchmarked against brute force sampling and not rigorously compared to other relevant work.

In September 2020, just a few months before the first experiment claiming quantum advantage was posted, authors including those on the experimental paper~\cite{Zhong_Wang_Deng_Chen_Peng_Luo_Qin_Wu_Ding_Hu_2020} benchmarked Gaussian boson sampling with threshold detection on the Sunway TiahuLight cluster~\cite{Li_Gan_Chen_Chen_Lu_Lu_Pan_Fu_Yang_2022}. An $n = 50$ Gaussian boson sampling experiment is distributed across 32,286 nodes (157,144 processors) of Sunway TaihuLight (which contains a total of 40,960 nodes). Using a Markov Chain Monte Carlo sampling method, computing one sample took from a $N_c = 50$ click detection took about 20 hours in 128 bit precision and about two days for 256 bit precision. This large-scale classical computation set the boundary for quantum advantage which was to appear just a few months later.

Finally, in October 2020, a paper by Queseda \textit{et al.} presents a quadratic speedup for clasically simulating Gaussian boson sampling~\cite{Quesada_2022}. The algorithm would be corrected in subsequent version of the manuscript in August 2021, however the main claim of quadratic speedup remains. The algorithm introduces partial heterodyne measurements so that only pure state probabilities need to be calculated. Implementing the algorithm on a small cluster of 96 CPUs, the authors are able to compute a pure state probability in around an hour for a $50 \times 50$ matrix, corresponding to a Gaussian boson sampling experiment with $N_c = 50$ detection events.

\subsubsection{Challenges to the experiments}

On May 20, 2021, Shi and Byrnes provided a new result which reduces the complexity of classically simulating Gaussian boson sampling with both threshold detectors and photon number resolving detectors when input photons are distinguishable~\cite{Shi_Byrnes_2022}. In particular, the authors define the \textit{indistinguishability efficiency} $\eta_{\text{ind}}$ as the probability that an input indistinguishable photon does not become distinguishable. In other words, $\eta_{\text{ind}} = 1$ corresponds to all indistinguishable photons, while $\eta_{\text{ind}} = 0$ corresponds to all distinguishable photons. In the complexity of generating a sample with $N_c$ photons (clicks) $O(\text{poly}(N_c) 2 ^{N_c / 2})$, the algorithms provide a speedup in the exponential term from $2^{N_c / 2}$ to $2^{N_c \eta_{\text{ind}} / 2}$. This encapsulates the result that Gaussian boson sampling with distinguishable photons, corresponding to $\eta_{\text{ind}} = 0$, is efficiently simulable classically, as well as the speedups obtained for \textit{partial} (in)distinguishability $0 < \eta_{\text{ind}} < 1$. No large-scale numerical simulations are performed and no estimated speedup over either Gaussian boson sampling experiment is provided. We remark that the indistuingishability in Gaussian boson sampling was also considered by Renema in 2019~\cite{Renema_2020}.

On Aug 03 2021, a new work exploiting collisions in Gaussian boson sampling was presented, estimating a $10^{9}$ reduction in the time to classically simulate the \Jiuzhang{} experiments~\cite{Bulmer_Bell_Chadwick_Jones_Moise_Rigazzi_Thorbecke_Haus_Vaerenbergh_Patel_2022}. A collision in boson sampling occurs when two or more photons are measured in the same mode. The computational hardness results for boson sampling assume photon number resolving detectors, but the experiments used threshold detectors. As shown in~\cite{Quesada_Arrazola_Killoran_2018}, the hardness results do not change for threshold detectors \textit{as long as the probability of collisions is sufficiently small}. However, the authors note that collision probability in the \Jiuzhang{} experiments is non-negligible, also noted in~\cite{Shi_Byrnes_2022}, and develop algorithms for experiments in which collisions are likely. Intuitively, collisions allow one to group photons into pairs such that the number of terms when computing hafnians is reduced. Letting $n_i$ be the number of photons measured in mode $i$, the lower bound for the number of terms when grouping is $\prod_i \sqrt{n_i + 1}$, as opposed to $2^{ \sum_i n_i / 2}$ terms without grouping. In the case of threshold detectors with $N_c$ clicks, the authors algorithm reduces the $2^{N_c}$ complexity to $2^{N_c / 2}$. The algorithms are implemented on a cluster with 1024 nodes, and the authors are able to classically simulate Gaussian boson sampling experiments with threshold detectors with up to $m = 100$ modes and $N_c = 60$ clicks. In the case of photon number resolving detectors, the authors are able to simulate a $92$ photon sample, but at significantly increased runtime.

On Sep 09 2021, work by Kaposi \textit{et al.} presented a polynomial speedup in Gaussian boson sampling with threshold detectors~\cite{Kaposi_Kolarovszki_Kozsik_Zimborás_Rakyta_2022}. While not commenting directly on the claimed computational advantage of either experiment, the authors show a 237x speedup over the best-known classical algorithm for problem sizes larger than around $N_c = 20$. We note that this algorithm is implemented in an open source Python package~\cite{Quesada_Arrazola_2020}. This speedup is obtained by careful reuse of calculations during the evaluation of the Torontonian matrix function. As fitted empirically, this reduces the complexity of the best-known algorithm from $O(N_c^{2.7355} 2^{N_c / 2})$ to $O(N_c^{1.0695} 2^{N_c / 2})$, i.e., a speedup of approximately $N_c^{1.666}$, where $N_c$ is the number of detected photons (clicks).

\subsubsection{Loopholes in Gaussian boson sampling} \label{sec:loopholes-gbs}

On Sep 23 2021, researchers from Google published a classical algorithm which provides better total variation distance than the experiment and which is quadratic in the number of modes~\cite{Villalonga_Niu_Li_Neven_Platt_Smelyanskiy_Boixo_2022}. The algorithm exploits a key difference between Gaussian boson sampling and random circuit sampling --- in random circuit sampling, marginal distributions for any subset of qubits are (exponentially) close to the uniform distribution, however in Gaussian boson sampling marginal distributions are not generally close to uniform and in fact can be efficiently approximated. The intuition for this algorithm is to sample from a distribution that approximates all single-mode and two-mode marginals. While the cost of computing a probability is still exponential in the number of clicks, restricting to subsets with one or two modes makes the cost upper bounded by the size of these subsets, which is chosen to be small. Thus marginal distributions over a small number of modes are easy to compute in (Gaussian) boson sampling. The authors use a Boltzmann machine and Gibbs sampling to produce ``mockup samples''. The runtime of the algorithm scales as $O(m^2 L)$ where $m$ is the number of modes and $L$ is the number of samples. Using this algorithm, the authors are able to compute millions of bitstrings per minute on a single workstation.  In an updated version of the original experimental paper~\cite{Zhong_Wang_Deng_Chen_Peng_Luo_Qin_Wu_Ding_Hu_2020}, an added note emphasizes that collision modes dominate in the experiment, and that the actual photon number is approximately twice the number of clicked detectors. It is therefore claimed that the algorithm of~\cite{Villalonga_Niu_Li_Neven_Platt_Smelyanskiy_Boixo_2022} cannot readily be used to speed up their experiment.

As a final note, we remark that Ref.~\cite{Deshpande_Mehta_Vincent_Quesada_Hinsche_Ioannou_Madsen_Lavoie_Qi_Eisert_2022} attempts to close several potential loopholes in Gaussian boson sampling and introduces a new architecture --- termed ``high-dimensional Gaussian boson sampling'' --- which is programmable and which may experimentally outperform previous Gaussian boson sampling experiments.

\subsection{Two random circuit sampling experiments on the \Zuchongzhi{} quantum computer} \label{sec:ustc-rcs-zuchongzhi-1-and-2}

\subsubsection{Description of the first experiment} \label{sec:ustc-rcs-1}

On Jun 28 2021, the USTC group reported a random circuit sampling experiment with $n = 56$ qubits and $m = 20$ cycles~\cite{Wu_Bao_Cao_Chen_Chen_Chen_Chung_Deng_Du_Fan__2021}.  Approximately $19$ million bitstrings are sampled in 1.2 hours at an estimated fidelity~\eqref{eqn:xeb-fidelity} of $0.0662\%$. The authors estimate that the classical computational cost of simulating their experiment is two to three orders of magnitude higher than Google's 2019 random circuit sampling experiment~\cite{Arute_Arya_Babbush_Bacon_Bardin_Barends_Biswas_Boixo_Brandao_Buell_} (Sec.~\ref{sec:google-rcs-2019}). In particular, building on recent advances in tensor network simulation of random quantum circuits (Sec.~\ref{sec:challenges-to-google-rcs-2019}), they estimate Google's 2019 experiment could be classically simulated on Summit in 15.9 days, whereas their experiment would take 8.24 years.

The experiment is performed on the \textit{Zuchongzhi} quantum computer which consists of a two-dimensional architecture with 66 superconducting transmon qubits and tunable couplers. The average single-qubit gate fidelity is 99.86\% and the average two-qubit gate fidelity is 99.41\%, with an average readout (measurement) fidelity of 95.48\%. Additionally, qubits have an average $T_1$ time of $30.6$ $\mu$s and average $T_2$ time of $5.3$ $\mu$s, both at idle frequency.

\begin{center}
    \hyperref[tab:overview]{\textit{Back to Table I}}
\end{center}

\subsubsection{Description of the second experiment} \label{sec:ustc-rcs-2}

On Sep 08 2021, the same group performed an upgraded experiment with $n = 60$ qubits and $m = 24$ cycles, collecting 70 million bitstrings in 4.2 hours (one million bitstring every 210 seconds) at an estimated fidelity~\eqref{eqn:xeb-fidelity} of $0.0366\%$~\cite{Zhu_Cao_Chen_Chen_Chen_Chung_Deng_Du_Fan_Gong_2022}. The experiment is performed on an updated \textit{Zuchongzhi} computer, dubbed \textit{Zuchongzhi 2.1}. On this device, average single-qubit and two-qubit gate errors were essentially the same as the previous experiment, however the average readout fidelity was improved to 97.74\%. Again building on the latest advances in tensor network simulation (Sec.~\ref{sec:challenges-to-google-rcs-2019}), the authors estimate it would take 48,000 years on Summit to classically simulate their experiment.

\begin{center}
    \hyperref[tab:overview]{\textit{Back to Table I}}
\end{center}

\subsubsection{Challenges, refutations, and loopholes}

Due to the fact that this is again a random circuit sampling experiment, all of the challenges and loopholes from Sec.~\ref{sec:challenges-to-google-rcs-2019} apply to this work. The difference is of course the larger size of the experiment: $n = 53$ qubits and $m = 20$ cycles in Google's 2019 experiment, improved up to $n = 60$ qubits and $m = 24$ cycles here. Recall that the (weak) refutation of Google's 2019 experiment~\cite{Pan_Chen_Zhang_2022} occurred on Nov 04 2021, almost two months after the second RCS experiment on \textit{Zuchongzhi} was published. Hence this work, and the work leading up to it, does not directly comment on the \textit{Zuchongzhi} experiments, but the challenges still apply. One exception is that the previously discussed~\cite{Liu_Guo_Liu_Yang_Song_Gao_Wang_Wu_Peng_Zhao_2021} --- although largely framed around Google's 2019 experiment --- does estimate their methods would take around five years for the \textit{Zuchongzhi} 2.1 experiment. In addition, Ref.~\cite{Kalachev_Panteleev_Zhou_Yung_2021} estimates the $n = 56$, $m = 20$ circuit can be done in four days on a cluster with 4480 GPUs. The loopholes for the cross entropy fidelity in random circuit sampling discussed in Sec.~\ref{sec:loopholes-rcs} also of course apply to these experiments.

\subsection{Gaussian boson sampling on the \textit{Borealis} quantum computer} \label{sec:xanadu-gbs-borealis}

\subsubsection{Description of the experiment} \label{sec:xanadu-gbs-borealis}

On Jun 01 2022, researchers from Xanadu announced a Gaussian boson sampling experiment with $n = 216$ squeezed states and $m = 216$ modes. Up to $N_c = 219$ photons are measured, with a mean photon number of 125. One reason this photon number is larger compared to previous Gaussian boson sampling experiments is that this experiment uses photon number resolving detectors instead of threshold detectors. Also, another important difference is that this experiment is fully programmable. Referring to the time needed to exactly calculate a single probability in Gaussian boson sampling under the best-known classical algorithm, the authors estimate it would take the Fukagu supercomputer 9,000 years to generate one sample, corresponding to 9 billion years to generate the full dataset of one million samples collected by \textit{Borealis}.

To verify the experimental setup, the authors classically simulate the experiment for smaller sizes. In particular, for $3 \le n \le 6$ photons and $m = 16$ modes, the authors calculate the exact output distribution and compare them to the experimental output distribution, finding fidelities greater than 99\% and total variation distances less than or equal to 6.5\%. In the intermediate size regime --- an average photon number of around $n = 21$ with $m = 216$ modes --- a classical computation taking 22 hours is performed which allows the authors to compute the cross entropy of experimental and classically computed samples, finding good agreement. Finally, in the large size regime --- an average photon number of $N_c = 125$ with $m = 216$ modes --- the first and second order cumulants (which can be efficiently computed classically) of the photon-number distributions are calculated from one million samples, for which they again find good agreement. Similar to the previous Gaussian boson sampling experiments, several adversarial distributions are considered --- squashed, thermal, coherent, and distinguishable squeezed light --- and are shown to match poorly with the ground truth. (These distributions are adversarial in the sense that they are easy to classically simulate, presenting a potential challenge to the experiment.) In addition, a fifth adversary is considered, namely the algorithm of Ref.~\cite{Villalonga_Niu_Li_Neven_Platt_Smelyanskiy_Boixo_2022} discussed in Sec.~\ref{sec:loopholes-gbs}. Still in this case the samples from \textit{Borealis} are found to better match the ground truth.

\begin{center}
    \hyperref[tab:overview]{\textit{Back to Table I}}
\end{center}

\subsection{Two new random circuit sampling experiments on the \textit{Sycamore} quantum computer} \label{sec:google-second-experiment}

On Apr 21 2023, the team from Google published a paper describing two new random circuit sampling experiments~\cite{Morvan_2024}. In the first, an $n = 67$ qubit, $m = 32$ cycle (880 two-qubit gates) experiment is performed. Seventy million bitstrings are sampled at an estimated XEB fidelity of 0.1\%. In the second, an $n = 70$ qubit, $m = 24$ cycle random circuit experiment is performed. Seventy million bitstrings are sampled at an estimated XEB fidelity of 0.2\%. In this latter experiment at shorter depth, the Loschmidt echo technique --- which applies the unitary $U^\dagger$ after the random circuit unitary $U$ and counts the number of times the all zero bitstring is measured --- is used to estimate the XEB fidelity, and the results agree with other techniques used.

In addition to presenting the largest random circuit sampling experiments to date, commentary is given as to references challenging, refuting, and/or discussing loopholes in the original 2019 experiment. Indeed, it is shown that the (linear) XEB fidelity~\eqref{eqn:xeb-fidelity} exhibits a phase transition from weak to strong noise, and the authors present a simple experimental protocol to locate this phase transition. Second, the authors estimate the classical runtime in light of the new techniques used to challenge and (weakly) refute the smaller 2019 experiment. Assuming the Frontier cluster, the task of sampling one million uncorrelated bitstrings from their $n = 67$, $m = 32$ experiment is estimated to take ten thousand years. Note that this considers the best case scenario where all RAM is used all bandwidth constraints are ignored. If the memory was expanded to included secondary storage, as previously suggested by~\cite{Pednault_Gunnels_Nannicini_Horesh_Wisnieff_2019} for the 2019 experiment, the authors estimate it would take Frontier 12 years to simulate the experiment.

\begin{center}
    \hyperref[tab:overview]{\textit{Back to Table I}}
\end{center}

\subsection{A third Gaussian Boson sampling experiment on the \Jiuzhang{} quantum computer} \label{sec:ustc-gbs-jiuzhang-3}

On Apr 24 2023, the group from USTC reported a third $n = 50$, $m = 144$ Gaussian boson sampling experiment on an upgraded version \Jiuzhang{} photonic quantum computer, termed \Jiuzhang{} \textit{3.0}~\cite{Deng_Gu_Liu_Gong_Su_Zhang_Tang_Jia_Xu_Chen_2023}. The highest number of detected photons obtained in the experiment is $N_c = 255$, compared to $N_c = 113$ in the previous one. In this third experiment, pseudo photon number resolving detectors are used, as opposed to threshold detectors in previous experiments. Recall that threshold detectors distinguish between two cases --- zero photons or one or more photons. Although easier to work with experimentally, this version of Gaussian boson sampling is easier to classically simulate by exploiting collisions~\cite{Bulmer_Bell_Chadwick_Jones_Moise_Rigazzi_Thorbecke_Haus_Vaerenbergh_Patel_2022}, as previously discussed. The pseudo photon resolving number detectors make the experiment harder to classically simulate, the dominant computational cost being proportional to $m N^3 \sqrt{G}$ where $N := \sum_{i = 1}^{m} n_i$ , $n_i$ being the number of photons resolved in mode $i$, and $G := \prod_{i = 1}^{m} (n_i + 1)$ . Using this cost and runtime estimates from the Frontier supercomputer, the authors estimate it would take at least 600 years to generate a single sample classically, and up to $3.1 \times 10^{10}$ years for the hardest sample, whereas \Jiuzhang{} takes 1.27 $\mu$s per sample.

\begin{center}
    \hyperref[tab:overview]{\textit{Back to Table I}}
\end{center}


\subsubsection{A weak refutation of the experiment and previous Gaussian boson sampling experiments}

On Jun 06 2023, Oh \textit{et al.} presented work using tensor networks to classical simulate Gaussian boson sampling~\cite{Oh_Liu_Alexeev_Fefferman_Jiang_2024}. This builds on previous work (which features the current authors) using matrix product operators to simulate lossy boson sampling~\cite{Oh_Noh_Fefferman_Jiang_2021,Liu_Oh_Liu_Jiang_Alexeev_2023}, and shares some features with previous algorithms exploiting the fact that noise results in an easier distribution to sample from~\cite{Kalai_Kindler_2014}. A key difference in the new algorithm is that thermal states are removed and matrix product states (MPS), instead of matrix product operators, are used. As described, exact sampling algorithms scale exponentially in the number of detected photons, but do not discriminate between which types of photons are detected. The tensor network algorithm is able to significantly reduce the complexity when detected photons are thermal states by removing them. The algorithm is implemented in Python using CuPy and MPI, and the code is executed on a cluster of $m$ GPUs where, as usual, $m$ is the number of modes in Gaussian boson sampling. An MPS with bond dimension $\chi = 10^4$ is used for each of the previous Gaussian boson sampling experiments claiming quantum advantage. The longest time for the MPS construction is 9.5 minutes for the \Jiuzhang{} \textit{2.0} experiment, and the time for generating 10 million samples is 62 minutes, a significant speedup relative to previous classical algorithms. Because the experimental sampling times were a few hundred seconds (e.g., 200 seconds for the \Jiuzhang{} \textit{1.0} and \textit{2.0} experiments), we consider this work to be a weak refutation of these experiments. It is notable that the source code used in this paper is available online (see the Supplementary Material of~\cite{Oh_Liu_Alexeev_Fefferman_Jiang_2024}), and that part of this algorithm directly uses and/or modifies previously mentioned open source software for simulating Gaussian boson sampling~\cite{Quesada_Arrazola_2020}, again highlighting how openly sharing data and code have accelerated --- if not caused --- these developments.

\subsection{A quickly refuted quantum simulation by IBM} \label{sec:ibm-qsim}

On Jun 14 2023, researchers from IBM report quantum simulation of the transverse field Ising model~\cite{Kim_Eddins_Anand_Wei_van_Rosenblatt_Nayfeh_Wu_Zaletel_Temme_2023} defined by $H = -J \sum_{\langle i, j \rangle} Z_i Z_j + h \sum_i X_i$, on an $n = 127$ qubit processor with two-dimensional connectivity $\langle i, j \rangle$. In other words, the connectivity of the Ising model is taken to be that of the quantum processor with qubits arranged in a two-dimensional ``heavy hexagon'' pattern. The quantum simulation is done via Trotterization up to $m = 60$ layers, corresponding to 2880 two-qubit gates. Expectation values of weight 17 observables are computed and taken to be the output of the experiment. (The weight of an observable, taken to be a Pauli string, is the number of non-identity terms in the Pauli string, or the cardinality of the support of the string.) For certain evolution times, the circuit is Clifford and therefore efficiently classically simulable --- this is used to benchmark the performance of the device, analogous to how linear XEB is used to quantify the performance in random circuit sampling experiments. Noise is mitigated by use of zero-noise extrapolation and probabilistic error amplification to produce more accurate expectation values. The authors consider classical simulation by quasi-1D matrix product states and 2D isometric tensor network states, arguing that the size of the experiment would be out of reach for these methods, therefore allowing them to claim quantum advantage.

\begin{center}
    \hyperref[tab:overview]{\textit{Back to Table I}}
\end{center}

\subsubsection{Refutations and loopholes} \label{sec:ibm-refutations}

On Jun 26 2023, just two weeks after the IBM experiment is announced, a classical simulation of the experiment was presented~\cite{Tindall_Fishman_Stoudenmire_Sels_2023}. This classical simulation uses a tensor network ansatz based on the connectivity of the quantum computer and uses the recently-introduced technique of belief propagation~\cite{Tindall_Fishman_2023} to approximately contract it. Belief propagation is a technique for gauging tensor networks and has been shown numerically to be faster than existing methods. In this work, it is used to apply two-qubit gates appearing in the Trotterization to evolve time in the tensor network. The full classical computation simulating IBM's experiment was performed on a laptop in a few minutes. The code to do so is built on open-source libraries and is available online~\cite{Tindall_Fishman_2023}.

Two days later, on Jun 28 2023, 
authors of~\cite{Kechedzhi_Isakov_Mandrà_Villalonga_Mi_Boixo_Smelyanskiy_2024} show the simulation of a smaller $n = 30$ qubit circuit can accurately reproduce the experimental data obtained for the weight 17 observable considered in the IBM experiment. Assuming that $\text{Tr} [ \rho O ] = F \langle O \rangle$ where $O$ is an observable, $|\psi\rangle$ is the noiseless state, and $\rho$ is the noisy state, the authors argue that the effective fidelity $F$ scales as $\exp ( - \epsilon V_O )$ where $\epsilon$ is the error per two-qubit gate and $V_O$, called the ``effective circuit volume'' depending on the observable $O$ and the circuit, is a number of two-qubit gates. Then, the authors argue the classical simulation cost scales as $2^{\alpha \partial V_O}$ where $\alpha$ is a constant and $\partial V_O$ is a boundary or ``cut'' of the effective circuit volume (which again depends on the observable $O$ and the circuit). This demonstrates a tradeoff between high fidelity and high classical simulation cost. These ideas are discussed in the context of random circuit sampling experiments, quantum chaos experiments, and the IBM experiment. For the IBM experiment, it is argued that the effective circuit volume $V_O$ is around 100 two-qubit gates, as opposed to the 2880 two-qubit gates in the experiment. (Note that this is distinct from the lightcone of an observable --- the lightcone of observables used in the IBM experiment covers all 127 qubits.) It is then shown that classically simulating these smaller circuits --- done on a single GPU in less than one second per circuit --- reproduces the experimental results from IBM within one standard deviation. Because the exact circuits are not executed, this computation can be considered a loophole, e.g., ``spoofing'' the expectation values based on the chosen observables and circuit design, just as various vulnerabilities in the XEB fidelity have been exploited (Sec.~\ref{sec:loopholes-rcs}). It is worth noting that these techniques do not lead to efficient spoofing of random circuit sampling experiments as discussed by the authors.

Again on Jun 28 2023, the same day as the previous paper, an approximate simulation technique is used to classically reproduce IBM's experiment~\cite{Begušić_Chan_2023}. The approximate simulation technique is the so-called Clifford perturbation theory --- recently introduced in~\cite{Begušić_Hejazi_Chan_2023} --- which is notable in the context of this review for \textit{not} being a tensor network technique. Rather, the technique updates the assumed Pauli observable in the Heisenberg picture using an expansion into sums of Cliffords. For an actual Clifford gate, the sum is just over one term and the update is trivial. For non-Clifford gates, the sum is over two Cliffords. Thus the cost is exponential in the number of non-Clifford gates. This exponentially growing sum is truncated after a certain number of terms, thus yielding an approximate technique, which is argued to be accurate under certain assumptions. In terms of IBM's experiment, the technique is shown to accurately reproduce the dynamics of the observables, and the overall classical runtime is between one and two minutes on a laptop. An updated and published version of the paper~\cite{Begušić_Gray_Chan_2024} also considers classical simulation via tensor network contraction with projected entangled pair states/operators.

Shortly after, on Aug 06 2023, a classical simulation using projected entangled pair operators (PEPOs) is presented which runs on a single CPU in three seconds~\cite{Liao_Wang_Zhou_Zhang_Xiang_2023}. In some detail, the authors construct the three-dimensional tensor network (two space + one time) $\langle 0 | U^\dagger O U |0\rangle$ and contract it in the Heisenberg picture --- i.e., starting from the center of the network representing the observable $O$ as a PEPO and contracting outward to the boundaries symmetrically. While operations with PEPOs are generally more expensive than with matrix product states (MPS), as considered in the classical simulation discussion of the IBM experiment, the authors note some practical advantages of using PEPOs --- for example, a PEPO reflects the geometry of the underlying hardware and so does not require SWAP operations like in MPS, which increase the bond dimension. Calculations with a PEPO with surprisingly small bond dimension $\chi = 2$ are shown to accurately reproduce IBM's experiment, and an errors with a $\chi = 184$ PEPO are shown to be lower than with an MPO at bond dimension $\chi = 1024$. (Errors with an MPO at bond dimension $\chi = 124$ are comparable to the errors with a PEPO at bond dimension $\chi = 2$.) Running time for $\chi = 184$ is three seconds using a single CPU.

Finally, on Sep 27 2023, a classical simulation of IBM's experiment using graph-based projected entangled pair states (gPEPs) and computing expectation values via mean field theory is presented~\cite{Liao_Wang_Zhou_Zhang_Xiang_2023}. Like others discussed, this classical simulation runs in just a few seconds (two seconds) on a laptop. It is interesting to note that, in addition to the above techniques, the authors considered belief propagation from~\cite{Tindall_Fishman_Stoudenmire_Sels_2023} and found it did not improve accuracy but did increase runtime from two seconds to nine seconds. This paper is particularly notable for not just simulating the $n = 127$ qubit Ising model experiment, but also for classically simulating $n = 433$ and $n = 1121$ qubit experiments (two future milestones on IBM Quantum's roadmap) up to $39$ Trotter steps. 

\subsection{A quantum simulation experiment by D-Wave} \label{sec:dwave-qsim}

On Mar 01 2024, researchers from D-Wave published a paper claiming quantum advantage for the problem of quantum simulation~\cite{dwave2024}. The dynamics simulated are the same as the previously discussed IBM experiment (Sec.~\ref{sec:ibm-qsim}), namely time evolution of the transverse field Ising model defined by $ H(t) = J(t / t_a) \sum_{i < j} J_{ij} Z_i Z_j + h (t / t_a) \sum_i X_i $. Here $t_a$ is the \textit{quench time} and the connectivity (i.e., the sum over $i < j$), is taken to be the connectivity of the quantum processor. Two processors are used in the experiment, termed \textit{Advantage} and \textit{Advantage2} (\textit{ADV1} and \textit{ADV2}). Each device samples 1000 bitstrings per second, taken to be the output of the experiment. Nonzero coupling coefficients $J_{ij}$ are chosen randomly under different topologies including square lattices, diamond lattices, and dimerized biclique graphs. The largest input is a $12 \times 12 \times 16$ diamond lattice requiring $576$ qubits.

The authors consider three classical algorithms to simulate their experiment --- two using tensor networks and one using neural networks (neural quantum states). Both matrix product states and the two-dimensional version of projected entangled pair states are considered for tensor network ansatze, and autoregressive Boltzmann machines as well as transformers and recurrent neural networks are considered for neural quantum states. By studying the numerical scaling of these techniques, the authors conclude that they cannot reach the largest experiment performed on the quantum computer, thereby claiming quantum advantage.

\begin{center}
    \hyperref[tab:overview]{\textit{Back to Table I}}
\end{center}

Due to the recency of this experiment (at the time of writing), there is no literature directly challenging or refuting this claim of quantum advantage, however it seems likely that the techniques of Sec.~\ref{sec:ibm-refutations} as well as others could do so in the future.

\section{Quantum advantage theory} \label{sec:quantum-advantage-theory}

While the preceding discussion focused solely on \textit{experimental} computational advantage and emphasized how this has been classically challenged and even refuted, similar results can be found in \textit{theoretical} computational advantage. We include brief histories for two cases in which this has happened --- computational advantage with respect to approximate optimization and recommendation systems. Last, we discuss prospects for the long-expected (exponential) advantage in simulating quantum systems, in particular quantum chemistry.

\subsection{A brief quantum computational advantage in approximate optimization}

The Quantum Approximate Optimization Algorithm (QAOA)~\cite{Farhi_Goldstone_Gutmann_2014} inputs a classical cost function $C: \{0, 1\}^n \rightarrow \mathbb{R}$ defined on a bitstring $z_1, ..., z_n$ and quantizes it by taking $z_i \mapsto Z_i$. At $p = 1$ layer, the QAOA prepares the state
\begin{equation}
    |\gamma, \beta\rangle := e^{iH_X \beta} e^{-i H_Z \gamma} |+\rangle^{\otimes n}
\end{equation}
where $|+\rangle := (|0\rangle + |1\rangle) / \sqrt{2}$. Thus for $\gamma = \beta = 0$ the QAOA is equivalent to random guessing, and for other parameter values $0 \le \gamma, \beta < 2 \pi$ different performance can be achieved. The \textit{driver} Hamiltonian $H_Z$ embeds the problem information via the cost function $C$, and the \textit{mixer} Hamiltonian $H_X$ is a parameter, commonly taken to be $H_X = \sum_{i = 1}^{n} X_i$. The QAOA can be naturally extended to $p$ layers (or rounds) by repeated application of the driver and mixer Hamiltonians, nominally with different parameters. The $p$ round QAOA thus contains $2p$ parameters $\gamma_1, \beta_1, ..., \gamma_p, \beta_p$. In theory work, $p = 1$ is almost always chosen due to the difficulty of analytical expressions with $p > 1$. By taking $\gamma_j = \beta_j = 0$ for all $j > i$, it is easy to see the additional parameters cannot decrease the performance of the QAOA, and it is hoped that additional rounds can increase the performance. In the original paper~\cite{Farhi_Goldstone_Gutmann_2014}, it was shown that when applied to the MaxCut problem on three-regular graphs, the $p = 1$ QAOA always finds a cut that is at least $0.6924$ times the size of the optimal cut.

On Dec 18 2014, the inventors of the QAOA apply the algorithm to a new problem, Max E3LIN2, and show they achieve an advantage over the best-known classical algorithm.  The Max E3LIN2 problem is a constraint sanctification problem over n bits $z_1, ..., z_n$ where $C(z)$ is a sum of clauses (terms) with \textit{E}xactly \textit{3} (E3) binary variables that sum to zero or one modulo \textit{2} (LIN2). Note that the problem is also known as Max-3XOR and can be easily generalized to Max-$k$XOR. The problem is to maximize the number of satisfied clauses (assuming they cannot all be satisfied, in which case the problem is trivial). A random guess $z$ will satisfy half of the clauses on average, and it is known that for any $\epsilon > 0$ there is no efficient classical algorithm to satisfy $1/2 + \epsilon$ clauses unless P = NP~\cite{Hastad_2000}. With the additional assumption that each variable appears in at most $D$ clauses, a classical algorithm is known that produces an approximation ratio of $1/2 + c / D$ where $c$ is a constant~\cite{Hastad_2001}. 

In~\cite{Farhi_Goldstone_Gutmann_2015}, Farhi, Goldstone, and Gutman applied the $p = 1$ round QAOA to the Max E3LIN2 problem and showed it achieved an approximation ratio of $1/2 + c / D^{3/4}$, thereby demonstrating an advantage over the best-known classical algorithm. Because of this, the result generated significant interest, notably in the form of a blog and discussion~\cite{AaronsonQAOA}. This interest prompted researchers to consider if an equal or better classical algorithm could be found.

Less than five months later on May 13 2015, Barak \textit{et al.} --- directly responding to Farhi, Goldstone, and Gutman ---  presented a randomized classical algorithm able to achieve an approximation ratio of at least $1/2 + c / D^{1/2}$ where $c$ is a constant~\cite{Barak_Moitra_ODonnell_Raghavendra_Regev_Steurer_Trevisan_Vijayaraghavan_Witmer_Wright_2015}. In addition to beating the QAOA, they show that the $1 / D^{1/2}$ dependence is optimal. Thus, the theoretical quantum advantage for Max-E3LIN2 was lost.

It is notable that on Jun 25 2015, Farhi, Goldstone, and Gutman responded to Barak \textit{et al.} and improve their analysis of the QAOA approximation ratio to show it achieves $1/2 + c / D^{1/2} \log D$, and in ``typical'' cases defined by the authors the QAOA will output a string with high probability that satisfies $1/2 + c / D^{1/2}$ clauses. Note that this is not in a separate paper but rather in an updated version of~\cite{Farhi_Goldstone_Gutmann_2015} (v1 and v2 on arXiv). Thus, the QAOA is worse than the best-known classical algorithm by a factor of $\log D$, or (in the best case) achieves equal performance, so a theoretical quantum advantage for Max-E3LIN2 cannot be claimed. It is worthwhile to note that, like most theoretical QAOA analyses, the algorithm is analyzed for $p = 1$ layers (or rounds), and as mentioned the performance of the QAOA cannot decrease for additional layers, and hopefully will increase.

The question of whether the QAOA can achieve a provable computational advantage is still open and highly active. For example, recently Montanaro and Zhou proved that the $p = 1$ QAOA can achieve a success probability of at least $1 / \sqrt{n}$ for particular \textit{symmetric constraint satisfaction problems with planted solutions}~\cite{Montanaro_Zhou_2024}.  Certain classical algorithms are shown to have worse theoretical performance, and other state-of-the-art SAT solvers are studied numerically, suggesting a quantum advantage for the $p = 1$ QAOA on these problems.

Finally, we note there is an interesting connection between the QAOA and Grover's algorithm~\cite{Jiang_Rieffel_Wang_2017}. While Grover's algorithm is a staple quantum algorithm offering a quadratic speedup~\cite{Grover_1997} that is provably optimal~\cite{Bennett_Bernstein_Brassard_Vazirani_1997}, this statement is in the oracle (``black box'') setting, a common model for complexity proofs with both quantum and classical algorithms. While recent work has considered the algorithm in a non-oracle setting~\cite{Stoudenmire_2024} and questioned its advantage, the appropriateness of such an investigation has been called into question~\cite{shtetl}. The principle of amplitude amplification in Grover's algorithm continues to fuel many application-oriented quantum algorithms, notably in quantum machine learning~\cite{Khanal_Orduz_Rivas_Baker_2023,Muser_Zapusek_Belis_Reiter_2024}, while the practicality of quadratic speedups yielding advantage in practice remains a point of debate~\cite{Babbush_McClean_Newman_Gidney_Boixo_Neven_2021}.

\subsection{Dequantization of recommendation systems}

The recommendation systems problem is to produce good products for users based on available data. Specifically, given $m$ users and $n$ products, we can imagine an $m \times n$ preference matrix $T_{ij}$ where $T_{ij} = 1$ is user $i$ likes product $j$, else $T_{ij} = 0$. Given this data in an ``online'' fashion of the form $(i, j, T_{ij})$, the goal is output a small number of new products for a given user. We assume that the preference matrix $T$ has a low rank $k$, specifically taking $k$ to be independent of $m$ and $n$, to make the problem tractable. This low rank assumption arises fairly naturally from social and economic reasons (e.g., people often naturally form groups, and some products will be better than others) and is found to hold well in empirical data.

One classical algorithm to solve this problem first samples entries from $T_{ij}$ to get a random matrix $\hat{T}_{ij}$, computes the singular value decomposition, then truncates to a given rank (which is independent of $m$ and $n$). An input user vector is then projected onto this low-rank subspace, from which we can read out recommendations.

On Mar 29 2016, a quantum algorithm for the recommendation systems problem was presented by Kerenidis and Prakash~\cite{Kerenidis_Prakash_2016}. The algorithm proceeds in line with the previous classical algorithm but uses quantum routines to compute the singular value decomposition and to truncate. The algorithm has the very nice feature that it avoids the ``output problem'' of reading out entries of an exponentially large vector from a quantum computer. This is because we only want to get a small number of recommended products. So, after the quantum singular value decomposition and truncation is performed, simply measuring the final state samples from the desired distribution, and good products are sampled with high probability. The quantum recommendation systems algorithm runs in time $O(\text{poly}(k) \log m n)$ which achieves an exponential speedup over the best-known classical algorithms (which depend polynomially on the dimensions $m$ and $n$). The algorithm is a cornerstone of theoretical advantage in quantum algorithms and quantum machine learning.

However, the quantum algorithm must assume a new data structure --- a so-called binary tree data structure --- to be able to efficiently read in the problem input (the preference matrix $T_{ij}$ and a given user $i$). What if the same data structure was allowed for classical algorithms? Ewin Tang presented the answer to this question on Jul 10 2018, showing that there exists a classical algorithm running in time $O(\text{poly}(k) \log m n)$ using the binary tree data structure~\cite{Tang_2019}. In other words, the quantum recommendation systems algorithm only achieves an (exponential) advantage due to assumptions on the input of the problem. The term \textit{dequantization} is used to refer to this situation of a new classical algorithm inspired by a quantum algorithm with the same (or better) performance.

The dequantized algorithm works as follows. The binary tree data structure leads to efficient sample and query access of a vector: i.e., given $x \in \mathbb{C}^N$, we can efficiently sample $i$ from the distribution $|x_i^2 / ||x||^2$ and query $i \mapsto x_i$. This leads to advantages in certain computations. For example, finding a large, hidden element of a vector takes time $\Omega(N)$, however with sample and query access can be done in constant time --- the large element will have high weight and so will be sampled with high probability in time independent of $n$. In a similar way, sample and query access leads to efficient inner products and thus efficient projections as needed in the recommendation systems algorithm. The dequnatized algorithm then computes the singular value decomposition, projects, and samples entries (products). It is worth noting that the singular value decomposition is performed via the random (classical) algorithm of~\cite{Frieze_Kannan_Vempala_2004}. 

While the runtime of the quantum and dequantized classical algorithms are the same asymptotically, they have different constant factors and different degree polynomials. A paper by Arrazola \textit{et al.}~\cite{Arrazola_Delgado_Bardhan_Lloyd_2020} explored the performance of both in practice, finding that the dequnatized algorithm can perform well in practice despite strong requirements, but also has significant polynomial overhead relative to the quantum algorithm. Assuming the binary tree data structure of~\cite{Kerenidis_Prakash_2016} could be efficiently prepared, then, it follows that the quantum recommendation systems algorithm could be advantageous in practice. However, it could also be possible to improve the constants and polynomials in the dequantized algorithm, negating this advantage.

It is worthwhile to note that the dequantization ideas from Tang~\cite{Tang_2019} were applied to other quantum algorithms, for example linear systems~\cite{Chia_Lin_Wang_2018}, linear regression~\cite{Gilyen_Song_Tang_2022}, and principal component analysis~\cite{Tang_2021}. The question of whether the sample and query access assumption of dequantized algorithms is \textit{too} powerful a model has been considered~\cite{Cotler_Huang_McClean_2021}, motivated by the result that classical algorithms with sample and query access can solve certain learning tasks exponentially faster than quantum algorithms with quantum state inputs. It is expected that, as quantum machine learning continues to develop~\cite{Biamonte_Wittek_Pancotti_Rebentrost_Wiebe_Lloyd_2017,Cerezo_Verdon_Huang_Cincio_Coles_2022}, advantage will move back and forth as new protocols such as quantum support vector machines~\cite{Gentinetta_Thomsen_Sutter_Woerner_2024}, quantum kernel methods~\cite{Liu_Arunachalam_Temme_2021}, and others are developed.


\subsection{Quantum chemistry}

Perhaps the clearest application in which we can expect quantum computers to achieve computational advantage is simulating quantum systems, one of the first motivations for the idea of quantum computers in the 1980s~\cite{Feynman_1982}. Roughly speaking, computational advantage is expected because all known classical algorithms for simulating quantum systems scale exponentially in some parameter ---- e.g. dimensionality, entanglement, or coherence --- whereas quantum computers naturally mimic the behavior of (many-body) quantum systems. 

One formalization of this intuition is the quantum phase estimation (QPE) algorithm~\cite{Kitaev_1995}. The goal of QPE is to find the ground state energy (smallest eigenvalue) of a given a $n$-qubit unitary operator $U$. This energy can be written $e^{2 \pi i \phi}$ where $\phi$ is known as the \textit{phase}, whence the name of the algorithm. Computing $\phi$ to accuracy $\epsilon$ requires a number of gates which scales as $O(\epsilon^{-1} \text{poly}(n) \text{polylog}\left( \epsilon^{-1} \pi_0^{-1}) \right)$, where $\pi_0$ is the overlap of the initial state and the ground state. With respect to general quantum systems, the $\text{poly}(n)$ term is exponentially better than all known classical algorithms. The dependence on $\pi_0$ reflects the fact that QPE must have a relatively good initial state --- measured by overlap with the desired state --- in order to succeed.

Quantum phase estimation is a staple quantum algorithm, and much work has been done to improve the practicality of the algorithm while retaining its complexity --- for some recent developments, see for example~\cite{Lin_2022,Lin_Tong_2022,Clinton_Cubitt_Garcia-Patron_Montanaro_Stanisic_Stroeks_2024}. On Aug 22 2022, however, Lee \textit{et al.} asked if there is evidence for QPE to achieve an exponential speedup on typical problems in quantum chemistry~\cite{Lee_Lee_Zhai_Tong_Dalzell_Kumar_Helms_Gray_Cui_Liu_2023}. These typical problems are taken to be ground state energy calculation --- one of the most common tasks in chemistry --- for which QPE is designed. Highlighting the powerful classical heuristic algorithms which have been developed over many years for these problems, the authors suggest primarily through numerical analyses that evidence for exponential advantage for QPE in ground state energy calculation has yet to be firmly demonstrated (leaving the possibility for polynomial speedups). It is worth noting that, although the scaling of QPE is exponentially better than known algorithms, the overhead is significant despite ongoing work to make the algorithm more practical and amenable to current quantum computers. For this reason QPE has not been experimentally demonstrated on problem sizes which could claim quantum advantage. Further improvements in both the algorithm and in quantum hardware are required to perform such experiments and conclusively demonstrate the advantage of QPE. 

Beyond QPE, heuristic algorithms for quantum chemistry (and other problems) such as variational quantum algorithms~\cite{Cerezo_Arrasmith_Babbush_Benjamin_Endo_Fujii_McClean_Mitarai_Yuan_Cincio__2021}, quantum Krylov algorithms~\cite{Yoshioka_Amico_Kirby_Jurcevic_Dutt_Fuller_Garion_Haas_Hamamura_Ivrii_2025}, and sample-based quantum diagonalization~\cite{Piccinelli_Baiardi_Rossmannek_Vazquez_Tacchino_Mensa_Altamura_Alavi_Motta_RobledoMoreno_2025} may offer some hope for demonstrating quantum advantage as they are (more) amenable to the constraint of current and near-term quantum computers. However, being heuristics, many points remain unclear --- for example, in variational quantum algorithms, training circuits at scale may be computational hard~\cite{Larocca_Thanasilp_Wang_Sharma_Biamonte_Coles_Cincio_McClean_Holmes_Cerezo_2025} --- and much theoretical work stands to be done to prove such claims.


\subsection{Shor's algorithm}

Perhaps the most solid example of an algorithm with theoretical quantum advantage is Shor's algorithm for prime factorization (or discrete logarithms)~\cite{Shor_1997}. Whereas every known classical algorithm takes time exponential in the number of bits in the integer to be factored, Shor's quantum algorithm requires polynomial time. The catch is that the overhead is so large that an experimental demonstration of the algorithm will require quantum error correction. In this sense the final frontier towards experimental quantum advantage in prime factorization is quantum error correction, and we turn our final part of the review towards this topic.

Prior to this point, however, it is worth noting that the overhead of Shor's algorithm has been significantly brought down since the inception of the algorithm in 1995. Shor's algorithm works by reducing the problem of factoring to period finding, which quantum computers are particularly suitable for due to constructive and destructive interference effects. Factoring can be reduced to period finding by considering the modular exponential function $f_x (r) := x^r \mod n$ where $x$ is an integer and $n = p q$ is the number to be factored into unknown primes $p$ and $q$. This function is periodic with period $\phi(n) := (p - 1) (q - 1)$, known as Euler's totient function. Once the period $\phi(n)$ is known, the two equations $n = pq$ and $\phi(n) = (p - 1)(q - 1)$ can be readily solved for $p$ and $q$. (It may turn out that there is a smaller fundamental period of the modular exponential function $f_x(r)$, however in this case the period can still be determined efficiently by Euclid's algorithm for finding the greatest common divisor.)
The quantum part of Shor's algorithm simply computes the modular exponential function on a superposition of all inputs, then implements the quantum Fourier transform to determine it's period with high probability. The quantum Fourier transform is relatively easy, so the bulk of the quantum part of Shor's algorithm is computing $f_x(r)$ through multiplication. Much of the work reducing the overhead of Shor's algorithm has been developing better circuits for this task. Recent proposals with clear cost estimates and summaries of prior techniques can be found in~\cite{Gidney_Ekerå_2021,Gouzien_Sangouard_2021,Proctor_Young_Baczewski_BlumeKohout_2025}. To factor an $n$ bit RSA number, the technique of~\cite{Gidney_Ekerå_2021} requires $3n + 0.002 n \log n$ logical qubits and $0.3 n ^ 3 + 0.0005 n^3 \log n$ Toffoli gates, ignoring the cost of magic state distillation to implement Toffoli gates as well as the cost of routing (due to limited hardware connectivity) and the overhead from error correction. The final estimate is that it would take 20 million noisy qubits and eight hours to factor 2048-bit RSA numbers. Reducing this cost further is still an active an open problem for which new results are still appearing, e.g.~\cite{Regev_2024}.

\section{Towards fault tolerance} \label{sec:qec-experiments}

\begin{table*}
\centering
\begin{tabular}{|c|c|c|c|c|c|}
\hline
Year & Code name                  & Params.          & $n_Q$ & Qubit type        & Ref.                                \\ \hline
1998 & Repetition            & [3,1,3]               & 3     & NMR               & \cite{cory1998}                     \\
2004 & Repetition            & [3, 1, 3]             & 3     & Trapped ion      & \cite{2004}                         \\ 
2005 & Cat                 & N/A                   & 2     & Photonic          & \cite{pittman2005}                  \\ 
2001 & Perfect               & [[5,1,3]]             & 5     & NMR               & \cite{knill2001}                    \\
2011 & Repetition            & [3,1,3]               & 3     & Trapped ion          & \cite{schindler2011}                \\
2011 & Repetition            & [3,1,3]               & 3     & NMR               & \cite{moussa2011}                   \\ 
2011 & Repetition            & [3,1,3]               & 3     & NMR               & \cite{zhang2011}                    \\
2012 & Repetition            & [3,1,3]               & 3     & Superconducting   & \cite{reed2012}                     \\
2012 & Perfect               & [[5,1,3]]             & 5     & NMR               & \cite{zhang2012}                    \\
2014 & Surface               & [[4,1,2]]             & 4     & Photonic           & \cite{bell2014}                     \\
2014 & Repetition            & [5,1,5]               & 9     & Superconducting   & \cite{kelly2014}                    \\
2014 & Color                 & [[7,1,3]]             & 7     & Trapped ion          & \cite{nigg2014}                     \\
2014 & Repetition            & [3,1,3]               & 4     & Diamond         & \cite{waldherr2014}                 \\ 
2015 & Repetition            & [3,1,3]               & 5     & Superconducting   & \cite{riste2015detecting}           \\ 
2015 & Bell                 & [[2,0,2]]             & 4     & Superconducting   & \cite{corcoles2015demonstration}    \\ 
2016 & Repetition            & [3,1,3]               & 4     & Superconducting   & \cite{cramer2016repeated}           \\ 
2016 & Cat                 & N/A                   & 1     & 3D cavity         & \cite{ofek2016extending}            \\ 
2017 & Color                 & [[4,2,2]]             & 5     & Superconducting   & \cite{takita2017experimental}       \\ 
2017 & Color                 & [[4,2,2]]             & 5     & Trapped ion          & \cite{linke2017fault}               \\ 
2017 & Cat                 & N/A                   & 1     & Superconducting   & \cite{heeres2017}                   \\ 
2018 & Repetition            & [8,1,8]               & 15    & Superconducting   & \cite{wootton2018repetition}        \\ 
2019 & Bell                 & [[2,0,2]]             & 3     & Superconducting   & \cite{andersen2019entanglement}     \\ 
2019 & Perfect               & [[5,1,3]]             & 5     & Superconducting   & \cite{gong2019experimental}         \\ 
2019 & Binomial    & N/A                   & 1     & 3D cavity         & \cite{hu2019quantum}                \\ 
2019 & Color                 & [[4,2,2]]             & 4     & Superconducting   & \cite{harper2019}                   \\ 
2020 & Repetition            & [22,1,22]             & 43    & Superconducting   & \cite{Wootton_2020}                 \\ 
2020 & Surface               & [[4,1,2]]             & 7     & Superconducting   & \cite{andersen2020}                 \\ 
2020 & Bacon-Shor            & [[9,1,3]]             & 15    & Trapped ion          & \cite{egan2021fault}                \\ 
2020 & Bacon-Shor            & [[9,1,3]]             & 11    & Photonic           & \cite{luo2020quantum}               \\ 
2020 & GKP                 & N/A                   & 1     & 3D cavity         & \cite{campagne2020quantum}          \\ 
2021 & Repetition            & [11,1,11]             & 21    & Superconducting   & \cite{chen2021}                     \\ 
2021  & Steane       & [[7, 1, 3]]           & 10    & Trapped ion       & \cite{anderson2021realization}      \\
2022 & Surface                    & [[9, 1, 3]]           & 17    & Superconducting   & \cite{krinner2022realizing}         \\
2022 & Five-qubit            & [[5, 1, 3]]           & 7     & Diamond           & \cite{abobeih2022fault}             \\
2022 & Surface               & [[9, 1, 3]]           & 17    & Superconducting   & \cite{zhao2022}                     \\
2023 & GKP                 & N/A                   & 1     & Superconducting   & \cite{sivak2023real}                \\
2023 & Surface               & [[25, 1, 5]]          & 49    & Superconducting   & \cite{ahcarya2023suppresing}        \\
2024 & Carbon      & [[12, 2, 4]]          & 30    & Trapped ion       & \cite{paetznick2024demo}            \\ 
2024 & Surface/Color      & $k \le 48$  & 218   & Neutral atom      & \cite{48}                           \\ 
2024 & Surface               & [[49, 1, 7]]          & 101   & Superconducting   & \cite{acharya2024quantum}           \\ \hline 
\end{tabular}
\caption{A selection of quantum error correction experiments ordered chronologically. The code parameters are the $[[n, k, d]]$ notation (see main text for a definition) and the $n_Q$ column denotes the total number of physical qubits, including any ancilla qubits used for stabilizer measurement or other tasks outside of the $n$ data qubits. Note that if a reference implements multiple error correction experiments we list the parameters for the largest experiment. Many entries prior to 2021 are taken from a similar list in the 2021 experiment of~\cite{chen2021}, and later entries build on this history.}
\label{tab:quantum_error_correction}
\end{table*}

We finish this review article by briefly covering progress in quantum error correction, the final frontier to realizing experimental quantum advantage in Shor's algorithm and others. For readers familiar with classical error correction, quantum error correction can be intuitively understood as using redundancy. For example, the repetition code maps the basis states of a qubit to $|0\rangle \mapsto |000\rangle$ and $|1\rangle \mapsto |111\rangle$. In this way information is encoded redundantly in three physical qubits representing a single \textit{logical} qubit. This particular code is able to correct a single bit flip on any physical qubit by taking a majority vote in the decoding process. For example, if a single bit flip occurs on the last qubit, the error state $|001\rangle$ would be corrected to $|000\rangle$, and so the bit flip can be corrected. However if two or more bit flips happen, the state would get mapped to the other logical basis state, known as a \textit{logical error}. This intuition for how many errors a code can correct is reflected in a parameter known as the \textit{distance} of the code. Thus higher distance is desirable, and a certain physical error rate necessitates a minimum distance for error correction to work. A quantum error correction code is described in the notation $[[n, k, d]]$ where $n$ physical qubits are used to encode $k$ logical qubits and $d$ is the distance of the code. For example, the previously described repetition code is a $[[3, 1, 3]]$ code. (Since this code is only capable of correcting a single type of error (bit flips), it sometimes may be written with single brackets $[3, 1, 3]$ to emphasize it is a classical code.) An error corrected quantum computer is said to be \textit{fault tolerant} when errors do not propagate between logical qubits during operations that are necessary to implement the code and manipulate the logical information to perform a computation.

Table~\ref{tab:quantum_error_correction} shows a chronological history of quantum error correction experiments, from which one can see the remarkable progress in recent years. Notably, Ref.~\cite{schindler2011} constitutes the first repetitive error correction (with only phase flip errors).  The first error correction experiment to achieve a logical qubit lifetime longer than any component physical qubit occurred in 2016~\cite{ofek2016extending}. In this work, a logical qubit is encoded in a superconducting resonator in a so-called \textit{cat state}, exhibiting a lifetime 1.1 times as long as the best physical qubit, demonstrating a fundamental feature of error correction that additional noisy components can combine to create higher fidelity quantum information. Later, in 2020, Ref.~\cite{Wootton_2020} is notable for implementing the largest repetition codes at that time, using up to $43$ physical qubits of a superconducting (IBM) quantum computer. Also of note from 2020 is the first experimental demonstration of the surface code on a superconducting quantum computer~\cite{andersen2020} constituting the first repetitive error detection (while~\cite{anderson2021realization} in 2021 constituted the first repetitive error correction). The surface code is one of the most targeted codes due to its high threshold and relatively low experimental complexity, so although this work involved only four data qubits it generated significant interest. Another work from 2020 showed the first experimental demonstration of fault-tolerant preparation, measurement, and rotation using 13 trapped ion qubits implementing the $[[9, 1, 3]]$ Bacon-Shor code. In 2021, the team from Google published the first error correction experiment on its \textit{Sycamore} computer used in random circuit sampling experiments~\cite{chen2021}. This work scaled the size of repetition codes and showed, although the logical error rates were worse than physical component error rates, that the logical error rate decreased exponentially in the distance of the code. Subsequently, building on~\cite{andersen2020}, in 2022 Ref.~\cite{krinner2022realizing} implemented a distance three surface code using 17 physical qubits on a superconducting quantum computer. The USTC group also implemented a distance three surface code in 2022 on the \Zuchongzhi{} quantum computer used for random circuit sampling experiments. The Google group regained the lead with the largest surface code implementation of distance five (49 physical qubits) in 2023 on the \textit{Sycamore} computer~\cite{ahcarya2023suppresing}. This line of work culminated in the most recent surface code experiment by Google in 2024~\cite{acharya2024quantum}. In this work, using a new 105 qubit quantum computer called \textit{Willow}, surface codes up to distance seven (101 physical qubits) were implemented, and for the first time error rates were \textit{below} the surface code threshold. (Note that Google has claimed quantum advantage in a new random circuit sampling experiment on \textit{Willow} in several press releases~\cite{willow}, however no details about the experiment have been given at the time of writing.) Other experimental work in error correction in 2024 showed up to $218$ physical qubits encoding $k = 48$ logical qubits and implementing logical operations on a neutral atom quantum computer~\cite{48}, and still other work has demonstrated necessary protocols for fault-tolerant quantum computing, e.g. magic state distillation in~\cite{magic2024}.

While the pace of these advancements has been remarkable, there is still a long road to fault-tolerant quantum computation. For example, although the latest surface code experiment from Google is below the threshold, this is a ``memory'' experiment with a single logical qubit, lacking any logical operations. We expect to see an even larger volume of quantum error correction experiments in the coming years to continue reaching milestones on the path towards a large-scale, fault-tolerant quantum computer capable of demonstrating quantum advantage for factoring and other problems.

\section{Conclusion}

In the long run, it seems reasonable to expect quantum computational advantage for certain problems in which quantum algorithms have exponential or strong polynomial speedups. However, while quantum algorithms may be faster, quantum information is inherently sensitive, limiting the size of experiments and opening the door to classical challenges and refutations. The alternative route to using noisy, physical qubits to demonstrate quantum advantage is to develop quantum error correction, for which the engineering challenges are formidable, but also for which there has been a tremendous amount of progress and exciting advances in recent years. With regards theoretical computational advantage, the possibility remains to devise new classical algorithms which are even faster, as we have seen for example problems such as approximate optimization and recommendation systems. Ultimately, any hypothesized computational advantage must be experimentally performed to be substantiated, and we have seen in this review that experiments can be challenged and refuted. It seems at this moment in history we are just on the boundary between quantum and classical computational advantage, and in the near future we expect the status of computational advantage to continue shifting between quantum and classical. 
We hope that this brief history helps to propel readers to the research frontier and develop new ideas which advance both classical and quantum computation.

\subsection{Outlook}

As mentioned in Sec.~\ref{sec:general-remarks-on-quantum-advantage}, a key frontier in future quantum advantage experiments is the formulation of new problems which have solutions that can be efficiently verified. Such problems would avoid the current issue with advantage experiments that exponential classical resources must be used to verify the output from the quantum computer. For example, this occurs in random circuit sampling when evaluating the cross entropy fidelity~\eqref{eqn:xeb-fidelity} --- classical computation must be used to compute $p(z_i)$ for sampled bitstrings $z_i$, and this is classically hard. Perhaps the most obvious candidate for efficiently verifiable quantum advantage is the factoring problem, which will entail fault-tolerant quantum computation as described in Sec.~\ref{sec:qec-experiments}. As a nearer-term alternative to factoring, one recent proposal for an efficiently verifiable quantum advantage experiment is known as peaked circuit sampling~\cite{Aaronson_Zhang_2024}. Intuitively, a peaked circuit is simply one which contains one or more bitstrings that occur with high probability. Formally, an $n$-qubit circuit $C$ is said to be $\delta$ peaked if and only if
\begin{equation}
    \max_{z \in \{0, 1\}^n} | \langle z | C | 0^{\otimes n} \rangle | ^ 2 \ge \delta 
\end{equation}
for $0 \le \delta \le 1$. The problem of peaked circuit sampling is to find the peaked bitstring(s) given the circuit $C$. Given access to a perfect (noiseless) quantum computer, it is clear that one can solve this problem with high probability by collecting $O(1 / \delta_z)$ samples, with $\delta_z$ being the maximum bitstring probability. On the classical side, however, the best approach known at present is to classically simulate the circuit, which will require exponential resources in some parameter --- number of qubits, depth, entanglement, magic, etc. In Ref.~\cite{Aaronson_Zhang_2024}, Aaronson and Zhang propose constructing peaked quantum circuits by adding $\gamma_p$ ``peaking layers'' on top of random quantum circuits that have $\gamma_r$ ``random layers''. Analytically, they prove that at least $\gamma_p = \Omega \left( \left[  \frac{\gamma_r}{n} \right] ^ {0.19} \right) $ peaking layers are required to obtain $\delta = \Omega(1 / \text{poly}(n))$ peakedness for an $n$-qubit circuit with $\gamma_r$ random layers. Numerically, a gradient descent scheme is used to add peaked layers, leading to the finding that relatively small $\gamma_p$ is required relative to $\gamma_r$ to enforce peakedness, though the exact behavior and rigorous understanding is left to future work. Notably, also left to future work is a strategy for efficiently generating such peaked circuits, as the gradient descent scheme used for numerical analysis suffers from barren plateaus~\cite{McClean_Boixo_Smelyanskiy_Babbush_Neven_2018}.

In follow up work building on Aaronson and Zhang's peaked circuit sampling proposal, Ref.~\cite{Gharibyan_Mullath_Sherman_Su_Tepanyan_Zhang_2025} uses tensor network methods in the optimization to create larger peaked circuits, and they perform an experimental implementation on Quantinuum's H2 quantum computer. The largest experimental implementation uses $n = 56$ qubits and $2044$ two-qubit (RZZ) gates with a peak weight of $\delta_z \approx 0.1$. In the experiment, the authors show that the quantum computer can consistently detect the peak bitstring in under two hours with $1000$ shots, while the best classical algorithms (including matrix product state simulators, tensor networks with belief propagation, and Pauli propoagation simulators) are only able to solve the circuits with up to $\sim{}700$ two-qubit gates within ten hours.

In the same spirit of peaked output distributions, Ref.~\cite{Deshpande_Fefferman_Ghosh_Gullans_Hangleiter_2025} proposes a new problem for efficiently verifiable quantum advantage called Hidden Code Sampling. Leaving full details to the paper, in brief this problem relies on principles of quantum error correction to generate and verify a peaked state. One party, Bob, prepares a logical state of a CSS code, applies a coherent error, then measures a number of samples. Another party, Alice, runs a verification protocol, computing a quantity called the relative entropy different (related to the linear XEB for random circuit sampling) which concentrates around a large positive (peaked) value when Bob's samples are from the correct distribution. The authors conjecture that this problem is classically hard, provide evidence that there are no efficient classical algorithms for this task, and explicitly show under mild assumptions that there exists a large gap between simulation and verification.

While current quantum advantage experiments are classically hard and quantumly easy, they require verification that is classically hard. Peaked circuit sampling proposals are a logical next step in quantum advantage experiments to remove the undesirable hard classical verification aspect, and we expect to see additional problem proposals and experimental implementations. Even beyond problems with efficiently verifiable solutions, a next step is the formulation of problems which are useful in a scientific or technological sense. Although utility is a somewhat objective measure --- and we have even discussed potential utility in certified random number generation from the seemingly arbitrary problem of random circuit sampling --- the solution of useful problems constitutes a natural next frontier in the next generation of quantum advantage experiments, building on the first generation of (relatively) arbitrary problems. It is exciting to see rapid progress towards the holy grail of quantum computation --- quantum advantage that is provably classically hard, quantumly easy, efficiently verifiable, and useful for science.


\bibliographystyle{unsrtnat}
\bibliography{refs}

@article{McClean_Boixo_Smelyanskiy_Babbush_Neven_2018, title={Barren plateaus in quantum neural network training landscapes}, url={http://arxiv.org/abs/1803.11173}, abstractNote={Many experimental proposals for noisy intermediate scale quantum devices involve training a parameterized quantum circuit with a classical optimization loop. Such hybrid quantum-classical algorithms are popular for applications in quantum simulation, optimization, and machine learning. Due to its simplicity and hardware efficiency, random circuits are often proposed as initial guesses for exploring the space of quantum states. We show that the exponential dimension of Hilbert space and the gradient estimation complexity make this choice unsuitable for hybrid quantum-classical algorithms run on more than a few qubits. Specifically, we show that for a wide class of reasonable parameterized quantum circuits, the probability that the gradient along any reasonable direction is non-zero to some fixed precision is exponentially small as a function of the number of qubits. We argue that this is related to the 2-design characteristic of random circuits, and that solutions to this problem must be studied.}, note={arXiv: 1803.11173}, journal={arXiv:1803.11173 [physics, physics:quant-ph]}, author={McClean, Jarrod R. and Boixo, Sergio and Smelyanskiy, Vadim N. and Babbush, Ryan and Neven, Hartmut}, year={2018}, month=mar, doi={10.1038/s41467-018-07090-4} }

@article{Deshpande_Fefferman_Ghosh_Gullans_Hangleiter_2025, title={Peaked quantum advantage using error correction}, url={http://arxiv.org/abs/2510.05262}, DOI={10.48550/arXiv.2510.05262}, abstractNote={A key issue of current quantum advantage experiments is that their verification requires a full classical simulation of the ideal computation. This limits the regime in which the experiments can be verified to precisely the regime in which they are also simulatable. An important outstanding question is therefore to find quantum advantage schemes that are also classically verifiable. We make progress on this question by designing a new quantum advantage proposal--Hidden Code Sampling--whose output distribution is conditionally peaked. These peaks enable verification in far less time than it takes for full simulation. At the same time, we show that exactly sampling from the output distribution is classically hard unless the polynomial hierarchy collapses, and we propose a plausible conjecture regarding average-case hardness. Our scheme is based on ideas from quantum error correction. The required quantum computations are closely related to quantum fault-tolerant circuits and can potentially be implemented transversally. Our proposal may thus give rise to a next generation of quantum advantage experiments en route to full quantum fault tolerance.}, note={arXiv:2510.05262 [quant-ph]}, number={arXiv:2510.05262}, publisher={arXiv}, author={Deshpande, Abhinav and Fefferman, Bill and Ghosh, Soumik and Gullans, Michael and Hangleiter, Dominik}, year={2025}, month=oct }

@article{Gharibyan_Mullath_Sherman_Su_Tepanyan_Zhang_2025, title={Heuristic Quantum Advantage with Peaked Circuits}, url={http://arxiv.org/abs/2510.25838}, DOI={10.48550/arXiv.2510.25838}, abstractNote={We design and demonstrate heuristic quantum advantage with peaked circuits (HQAP circuits) on Quantinuum’s System Model H2 quantum processor. Through extensive experimentation with state-of-the-art classical simulation strategies, we identify a clear gap between classical and quantum runtimes. Our largest instance involves all-to-all connectivity with 2000 two-qubit gates, which H2 can produce the target peaked bitstring directly in under 2 hours. Our extrapolations from leading classical simulation techniques such as tensor networks with belief propagation and Pauli path simulators indicate the same instance would take years on exascale systems (Frontier, Summit), suggesting a potentially exponential separation. This work marks an important milestone toward verifiable quantum advantage, as well as providing a useful benchmarking protocol for current utility-scale quantum hardware. We sketch our protocol for designing these circuits and provide extensive numerical results leading to our extrapolation estimates. Separate from our constructed HQAP circuits, we prove hardness on a decision problem involving generic peaked circuits. When both the input and output bitstrings of a peaked circuit are unknown, determining whether the circuit is peaked constitutes a QCMA-complete problem, meaning the problem remains hard even for a quantum polynomial-time machine under commonly accepted complexity assumptions. Inspired by this observation, we propose an application of the peaked circuits as a potentially quantum-safe encryption scheme~cite{chen2016report,kumar2020post,joseph2022transitioning,dam2023survey}. We make our peaked circuits publicly available and invite the community to try additional methods to solve these circuits to see if this gap persists even with novel classical techniques.}, note={arXiv:2510.25838 [quant-ph]}, number={arXiv:2510.25838}, publisher={arXiv}, author={Gharibyan, Hrant and Mullath, Mohammed Zuhair and Sherman, Nicholas E. and Su, Vincent P. and Tepanyan, Hayk and Zhang, Yuxuan}, year={2025}, month=oct }

@article{Gouzien_Sangouard_2021, title={Factoring 2048-bit RSA Integers in 177 Days with 13 436 Qubits and a Multimode Memory}, volume={127}, DOI={10.1103/PhysRevLett.127.140503}, abstractNote={We analyze the performance of a quantum computer architecture combining a small processor and a storage unit. By focusing on integer factorization, we show a reduction by several orders of magnitude of the number of processing qubits compared with a standard architecture using a planar grid of qubits with nearest-neighbor connectivity. This is achieved by taking advantage of a temporally and spatially multiplexed memory to store the qubit states between processing steps. Concretely, for a characteristic physical gate error rate of 10−3, a processor cycle time of 1 microsecond, factoring a 2 048-bit RSA integer is shown to be possible in 177 days with 3D gauge color codes assuming a threshold of 0.75% with a processor made with 13 436 physical qubits and a memory that can store 28 million spatial modes and 45 temporal modes with 2 hours’ storage time. By inserting additional error-correction steps, storage times of 1 second are shown to be sufficient at the cost of increasing the run-time by about 23%. Shorter run-times (and storage times) are achievable by increasing the number of qubits in the processing unit. We suggest realizing such an architecture using a microwave interface between a processor made with superconducting qubits and a multiplexed memory using the principle of photon echo in solids doped with rare-earth ions.}, number={14}, journal={Physical Review Letters}, publisher={American Physical Society}, author={Gouzien, Élie and Sangouard, Nicolas}, year={2021}, month=sep, pages={140503} }

@article{Proctor_Young_Baczewski_BlumeKohout_2025, title={Benchmarking quantum computers}, volume={7}, rights={2025 Springer Nature Limited}, ISSN={2522-5820}, DOI={10.1038/s42254-024-00796-z}, abstractNote={The rapid pace of development in quantum computing technology has sparked a proliferation of benchmarks to assess the performance of quantum computing hardware and software. However, not all benchmarks are of equal merit. Good ones empower scientists, engineers, programmers and users to understand the power of a computing system, whereas bad ones can misdirect research and inhibit progress. In this Perspective, we survey the science of quantum computer benchmarking. We discuss the role of benchmarks and benchmarking and how good benchmarks can drive and measure progress towards the long-term goal of useful quantum computations, known as quantum utility. We explain how different kinds of benchmark quantify the performance of different parts of a quantum computer, discuss existing benchmarks, examine recent trends in benchmarking, and highlight important open research questions in this field.}, number={2}, journal={Nature Reviews Physics}, publisher={Nature Publishing Group}, author={Proctor, Timothy and Young, Kevin and Baczewski, Andrew D. and Blume-Kohout, Robin}, year={2025}, month=feb, pages={105–118}, language={en} }

@article{Piccinelli_Baiardi_Rossmannek_Vazquez_Tacchino_Mensa_Altamura_Alavi_Motta_RobledoMoreno_2025, title={Quantum chemistry with provable convergence via randomized sample-based quantum diagonalization}, url={http://arxiv.org/abs/2508.02578}, DOI={10.48550/arXiv.2508.02578}, abstractNote={Sample-based quantum diagonalization (SQD) is a recently proposed algorithm to approximate the ground-state wave function of many-body quantum systems on near-term and early-fault-tolerant quantum devices. In SQD, the quantum computer acts as a sampling engine that generates the subspace in which the Hamiltonian is classically diagonalized. A recently proposed SQD variant, Sample-based Krylov Quantum Diagonalization (SKQD), uses quantum Krylov states as circuits from which samples are collected. Convergence guarantees can be derived for SKQD under similar assumptions to those of quantum phase estimation, provided that the ground-state wave function is concentrated, i.e., has support on a small subset of the full Hilbert space. Implementations of SKQD on current utility-scale quantum computers are limited by the depth of time-evolution circuits needed to generate Krylov vectors. For many complex many-body Hamiltonians of interest, such as the molecular electronic-structure Hamiltonian, this depth exceeds the capability of state-of-the-art quantum processors. In this work, we introduce a new SQD variant that combines SKQD with the qDRIFT randomized compilation of the Hamiltonian propagator. The resulting algorithm, termed SqDRIFT, enables SQD calculations at the utility scale on chemical Hamiltonians while preserving the convergence guarantees of SKQD. We apply SqDRIFT to calculate the electronic ground-state energy of several polycyclic aromatic hydrocarbons, up to system sizes beyond the reach of exact diagonalization.}, note={arXiv:2508.02578 [quant-ph]}, number={arXiv:2508.02578}, publisher={arXiv}, author={Piccinelli, Samuele and Baiardi, Alberto and Rossmannek, Max and Vazquez, Almudena Carrera and Tacchino, Francesco and Mensa, Stefano and Altamura, Edoardo and Alavi, Ali and Motta, Mario and Robledo-Moreno, Javier and Kirby, William and Sharma, Kunal and Mezzacapo, Antonio and Tavernelli, Ivano}, year={2025}, month=aug }

@article{Yoshioka_Amico_Kirby_Jurcevic_Dutt_Fuller_Garion_Haas_Hamamura_Ivrii_2025, title={Diagonalization of large many-body Hamiltonians on a quantum processor}, volume={16}, ISSN={2041-1723}, DOI={10.1038/s41467-025-59716-z}, abstractNote={The estimation of low energies of many-body systems is a cornerstone of computational quantum sciences. Variational quantum algorithms can be used to prepare ground states on pre-fault-tolerant quantum processors, but their lack of convergence guarantees and impractical number of cost function estimations prevent systematic scaling of experiments to large systems. Alternatives to variational approaches are needed for large-scale experiments on pre-fault-tolerant devices. Here, we use a superconducting quantum processor to compute eigenenergies of quantum many-body systems on two-dimensional lattices of up to 56 sites, using the Krylov quantum diagonalization algorithm, an analog of the well-known classical diagonalization technique. We construct subspaces of the many-body Hilbert space using Trotterized unitary evolutions executed on the quantum processor, and classically diagonalize many-body interacting Hamiltonians within those subspaces. These experiments show that quantum diagonalization algorithms are poised to complement their classical counterpart at the foundation of computational methods for quantum systems.}, note={arXiv:2407.14431 [quant-ph]}, number={1}, journal={Nature Communications}, author={Yoshioka, Nobuyuki and Amico, Mirko and Kirby, William and Jurcevic, Petar and Dutt, Arkopal and Fuller, Bryce and Garion, Shelly and Haas, Holger and Hamamura, Ikko and Ivrii, Alexander and Majumdar, Ritajit and Minev, Zlatko and Motta, Mario and Pokharel, Bibek and Rivero, Pedro and Sharma, Kunal and Wood, Christopher J. and Javadi-Abhari, Ali and Mezzacapo, Antonio}, year={2025}, month=jun, pages={5014} }

@article{Cerezo_Arrasmith_Babbush_Benjamin_Endo_Fujii_McClean_Mitarai_Yuan_Cincio__2021, title={Variational quantum algorithms}, volume={3}, rights={2021 Springer Nature Limited}, ISSN={2522-5820}, DOI={10.1038/s42254-021-00348-9}, abstractNote={Applications such as simulating complicated quantum systems or solving large-scale linear algebra problems are very challenging for classical computers, owing to the extremely high computational cost. Quantum computers promise a solution, although fault-tolerant quantum computers will probably not be available in the near future. Current quantum devices have serious constraints, including limited numbers of qubits and noise processes that limit circuit depth. Variational quantum algorithms (VQAs), which use a classical optimizer to train a parameterized quantum circuit, have emerged as a leading strategy to address these constraints. VQAs have now been proposed for essentially all applications that researchers have envisaged for quantum computers, and they appear to be the best hope for obtaining quantum advantage. Nevertheless, challenges remain, including the trainability, accuracy and efficiency of VQAs. Here we overview the field of VQAs, discuss strategies to overcome their challenges and highlight the exciting prospects for using them to obtain quantum advantage.}, number={9}, journal={Nature Reviews Physics}, publisher={Nature Publishing Group}, author={Cerezo, M. and Arrasmith, Andrew and Babbush, Ryan and Benjamin, Simon C. and Endo, Suguru and Fujii, Keisuke and McClean, Jarrod R. and Mitarai, Kosuke and Yuan, Xiao and Cincio, Lukasz and Coles, Patrick J.}, year={2021}, month=sep, pages={625–644}, language={en} }

@article{Larocca_Thanasilp_Wang_Sharma_Biamonte_Coles_Cincio_McClean_Holmes_Cerezo_2025, title={Barren plateaus in variational quantum computing}, volume={7}, rights={2025 Springer Nature Limited}, ISSN={2522-5820}, DOI={10.1038/s42254-025-00813-9}, abstractNote={Variational quantum computing offers a flexible computational approach with a broad range of applications. However, a key obstacle to realizing their potential is the barren plateau (BP) phenomenon. When a model exhibits a BP, its parameter optimization landscape becomes exponentially flat and featureless as the problem size increases. Importantly, all the moving pieces of an algorithm — choices of ansatz, initial state, observable, loss function and hardware noise — can lead to BPs if they are ill-suited. As BPs strongly impact on trainability, researchers have dedicated considerable effort to develop theoretical and heuristic methods to understand and mitigate their effects. As a result, the study of BPs has become a thriving area of research, influencing and exchanging ideas with other fields such as quantum optimal control, tensor networks and learning theory. This article provides a review of the current understanding of the BP phenomenon.}, number={4}, journal={Nature Reviews Physics}, publisher={Nature Publishing Group}, author={Larocca, Martín and Thanasilp, Supanut and Wang, Samson and Sharma, Kunal and Biamonte, Jacob and Coles, Patrick J. and Cincio, Lukasz and McClean, Jarrod R. and Holmes, Zoë and Cerezo, M.}, year={2025}, month=apr, pages={174–189}, language={en} }

@article{Biamonte_Wittek_Pancotti_Rebentrost_Wiebe_Lloyd_2017, title={Quantum machine learning}, volume={549}, rights={2017 Macmillan Publishers Limited, part of Springer Nature. All rights reserved.}, ISSN={1476-4687}, DOI={10.1038/nature23474}, abstractNote={Fuelled by increasing computer power and algorithmic advances, machine learning techniques have become powerful tools for finding patterns in data. Quantum systems produce atypical patterns that classical systems are thought not to produce efficiently, so it is reasonable to postulate that quantum computers may outperform classical computers on machine learning tasks. The field of quantum machine learning explores how to devise and implement quantum software that could enable machine learning that is faster than that of classical computers. Recent work has produced quantum algorithms that could act as the building blocks of machine learning programs, but the hardware and software challenges are still considerable.}, number={7671}, journal={Nature}, publisher={Nature Publishing Group}, author={Biamonte, Jacob and Wittek, Peter and Pancotti, Nicola and Rebentrost, Patrick and Wiebe, Nathan and Lloyd, Seth}, year={2017}, month=sep, pages={195–202}, language={en} }

@article{Cerezo_Verdon_Huang_Cincio_Coles_2022, title={Challenges and opportunities in quantum machine learning}, volume={2}, rights={2022 Springer Nature America, Inc.}, ISSN={2662-8457}, DOI={10.1038/s43588-022-00311-3}, abstractNote={At the intersection of machine learning and quantum computing, quantum machine learning has the potential of accelerating data analysis, especially for quantum data, with applications for quantum materials, biochemistry and high-energy physics. Nevertheless, challenges remain regarding the trainability of quantum machine learning models. Here we review current methods and applications for quantum machine learning. We highlight differences between quantum and classical machine learning, with a focus on quantum neural networks and quantum deep learning. Finally, we discuss opportunities for quantum advantage with quantum machine learning.}, number={9}, journal={Nature Computational Science}, publisher={Nature Publishing Group}, author={Cerezo, M. and Verdon, Guillaume and Huang, Hsin-Yuan and Cincio, Lukasz and Coles, Patrick J.}, year={2022}, month=sep, pages={567–576}, language={en} }

@article{Gentinetta_Thomsen_Sutter_Woerner_2024, title={The complexity of quantum support vector machines}, volume={8}, DOI={10.22331/q-2024-01-11-1225}, abstractNote={Gian Gentinetta, Arne Thomsen, David Sutter, and Stefan Woerner,
Quantum 8, 1225 (2024).
Quantum support vector machines employ quantum circuits to define the kernel function. It has been shown that this approach offers a provable exponential speedup compared to any known classi…}, journal={Quantum}, publisher={Verein zur Förderung des Open Access Publizierens in den Quantenwissenschaften}, author={Gentinetta, Gian and Thomsen, Arne and Sutter, David and Woerner, Stefan}, year={2024}, month=jan, pages={1225}}

@article{Liu_Arunachalam_Temme_2021, title={A rigorous and robust quantum speed-up in supervised machine learning}, volume={17}, rights={2021 The Author(s), under exclusive licence to Springer Nature Limited}, ISSN={1745-2481}, DOI={10.1038/s41567-021-01287-z}, abstractNote={Recently, several quantum machine learning algorithms have been proposed that may offer quantum speed-ups over their classical counterparts. Most of these algorithms are either heuristic or assume that data can be accessed quantum-mechanically, making it unclear whether a quantum advantage can be proven without resorting to strong assumptions. Here we construct a classification problem with which we can rigorously show that heuristic quantum kernel methods can provide an end-to-end quantum speed-up with only classical access to data. To prove the quantum speed-up, we construct a family of datasets and show that no classical learner can classify the data inverse-polynomially better than random guessing, assuming the widely believed hardness of the discrete logarithm problem. Furthermore, we construct a family of parameterized unitary circuits, which can be efficiently implemented on a fault-tolerant quantum computer, and use them to map the data samples to a quantum feature space and estimate the kernel entries. The resulting quantum classifier achieves high accuracy and is robust against additive errors in the kernel entries that arise from finite sampling statistics.}, number={9}, journal={Nature Physics}, publisher={Nature Publishing Group}, author={Liu, Yunchao and Arunachalam, Srinivasan and Temme, Kristan}, year={2021}, month=sep, pages={1013–1017}, language={en} }

@article{Tang_2021, title={Quantum Principal Component Analysis Only Achieves an Exponential Speedup Because of Its State Preparation Assumptions}, volume={127}, DOI={10.1103/PhysRevLett.127.060503}, abstractNote={A central roadblock to analyzing quantum algorithms on quantum states is the lack of a comparable input model for classical algorithms. Inspired by recent work of the author [E. Tang, STOC 2019.], we introduce such a model, where we assume we can efficiently perform ℓ2-norm samples of input data, a natural analog to quantum algorithms that assume efficient state preparation of classical data. Though this model produces less practical algorithms than the (stronger) standard model of classical computation, it captures versions of many of the features and nuances of quantum linear algebra algorithms. With this model, we describe classical analogs to Lloyd, Mohseni, and Rebentrost’s quantum algorithms for principal component analysis [S. Lloyd, M. Mohseni, and P. Rebentrost, Nat. Phys. 10, 631 (2014).] and nearest-centroid clustering [S. Lloyd, M. Mohseni, and P. Rebentrost, Quantum algorithms for supervised and unsupervised machine learning]. Since they are only polynomially slower, these algorithms suggest that the exponential speedups of their quantum counterparts are simply an artifact of state preparation assumptions.}, number={6}, journal={Physical Review Letters}, publisher={American Physical Society}, author={Tang, Ewin}, year={2021}, month=aug, pages={060503} }

@article{Babbush_McClean_Newman_Gidney_Boixo_Neven_2021, title={Focus beyond Quadratic Speedups for Error-Corrected Quantum Advantage}, volume={2}, DOI={10.1103/PRXQuantum.2.010103}, abstractNote={In this perspective we discuss conditions under which it would be possible for a modest fault-tolerant quantum computer to realize a runtime advantage by executing a quantum algorithm with only a small polynomial speedup over the best classical alternative. The challenge is that the computation must finish within a reasonable amount of time while being difficult enough that the small quantum scaling advantage would compensate for the large constant factor overheads associated with error correction. We compute several examples of such runtimes using state-of-the-art surface code constructions under a variety of assumptions. We conclude that quadratic speedups will not enable quantum advantage on early generations of such fault-tolerant devices unless there is a significant improvement in how we realize quantum error correction. While this conclusion persists even if we were to increase the rate of logical gates in the surface code by more than an order of magnitude, we also repeat this analysis for speedups by other polynomial degrees and find that quartic speedups look significantly more practical.}, number={1}, journal={PRX Quantum}, publisher={American Physical Society}, author={Babbush, Ryan and McClean, Jarrod R. and Newman, Michael and Gidney, Craig and Boixo, Sergio and Neven, Hartmut}, year={2021}, month=mar, pages={010103} }

@article{Khanal_Orduz_Rivas_Baker_2023, title={Supercomputing leverages quantum machine learning and Grover’s algorithm}, volume={79}, ISSN={1573-0484}, DOI={10.1007/s11227-022-04923-4}, abstractNote={The complexity of searching algorithms in classical computing is a classic problem and a research area. Quantum computers and quantum algorithms can efficiently compute some classically hard problems. In addition, quantum machine learning algorithms could be an important avenue to boost existing and new quantum-based technology, reducing the supercomputing requirements for executing such problems. This paper reviews and explores topics such as variational quantum algorithms, kernel methods, and Grover’s algorithm (GA). GA is a quantum search algorithm that achieves a quadratic speed improvement as a quantum classifier. We exploit GA or amplitude amplification to simulate rudimentary classical logical gates into quantum circuits considering AND, XOR, and OR gates. Our experiments in our review suggest that the algorithms discussed can be implemented and verified with relative ease, suggesting that researchers can investigate problems in the areas discussed related to quantum machine learning and more.}, number={6}, journal={The Journal of Supercomputing}, author={Khanal, Bikram and Orduz, Javier and Rivas, Pablo and Baker, Erich}, year={2023}, month=apr, pages={6918–6940}, language={en} }

@article{Muser_Zapusek_Belis_Reiter_2024, title={Provable advantages of kernel-based quantum learners and quantum preprocessing based on Grover’s algorithm}, volume={110}, DOI={10.1103/PhysRevA.110.032434}, abstractNote={There is an ongoing effort to find quantum speedups for learning problems. Recently [Y. Liu et al., Nat. Phys. 17, 1013 (2021)] proved an exponential speedup for quantum support vector machines by leveraging the speedup of Shor’s algorithm. We expand upon this result and identify a speedup utilizing Grover’s algorithm in the kernel of a support vector machine. To show the practicality of the kernel structure we apply it to a problem related to pattern matching, providing a practical yet provable advantage. Moreover, we show that combining quantum computation in a preprocessing step with classical methods for classification further improves classifier performance.}, number={3}, journal={Physical Review A}, publisher={American Physical Society}, author={Muser, T. and Zapusek, E. and Belis, V. and Reiter, F.}, year={2024}, month=sep, pages={032434} }

@misc{shtetl, title={Shtetl-Optimized: The Blog of Scott Aaronson}, author={Aaronson, Scott}, howpublished={https://scottaaronson.blog/?p=7157}, journal={Shtetl-Optimized}, year={2023}, month=mar }

@article{Stoudenmire_2024, title={Opening the Black Box inside Grover’s Algorithm}, volume={14}, DOI={10.1103/PhysRevX.14.041029}, number={4}, journal={Physical Review X}, author={Stoudenmire, E.M. and Waintal, Xavier}, year={2024} }

@article{Bennett_Bernstein_Brassard_Vazirani_1997, title={Strengths and Weaknesses of Quantum Computing}, volume={26}, ISSN={0097-5397, 1095-7111}, DOI={10.1137/S0097539796300933}, abstractNote={Recently a great deal of attention has focused on quantum computation following a sequence of results suggesting that quantum computers are more powerful than classical probabilistic computers. Following Shor’s result that factoring and the extraction of discrete logarithms are both solvable in quantum polynomial time, it is natural to ask whether all of NP can be efficiently solved in quantum polynomial time. In this paper, we address this question by proving that relative to an oracle chosen uniformly at random, with probability 1, the class NP cannot be solved on a quantum Turing machine in time $o(2^{n/2})$. We also show that relative to a permutation oracle chosen uniformly at random, with probability 1, the class $NP cap coNP$ cannot be solved on a quantum Turing machine in time $o(2^{n/3})$. The former bound is tight since recent work of Grover shows how to accept the class NP relative to any oracle on a quantum computer in time $O(2^{n/2})$.}, note={arXiv:quant-ph/9701001}, number={5}, journal={SIAM Journal on Computing}, author={Bennett, Charles H. and Bernstein, Ethan and Brassard, Gilles and Vazirani, Umesh}, year={1997}, month=oct, pages={1510–1523} }

@article{Grover_1997, title={Quantum Mechanics helps in searching for a needle in a haystack}, volume={79}, ISSN={0031-9007, 1079-7114}, DOI={10.1103/PhysRevLett.79.325}, abstractNote={Quantum mechanics can speed up a range of search applications over unsorted data. For example imagine a phone directory containing N names arranged in completely random order. To find someone’s phone number with a probability of 50%, any classical algorithm (whether deterministic or probabilistic) will need to access the database a minimum of O(N) times. Quantum mechanical systems can be in a superposition of states and simultaneously examine multiple names. By properly adjusting the phases of various operations, successful computations reinforce each other while others interfere randomly. As a result, the desired phone number can be obtained in only O(sqrt(N)) accesses to the database.}, note={arXiv: quant-ph/9706033}, number={2}, journal={Physical Review Letters}, author={Grover, Lov K.}, year={1997}, month=jul, pages={325–328} }

@article{Jiang_Rieffel_Wang_2017, title={Near-optimal quantum circuit for Grover’s unstructured search using a transverse field}, volume={95}, DOI={10.1103/PhysRevA.95.062317}, number={6}, journal={Physical Review A}, publisher={American Physical Society}, author={Jiang, Zhang and Rieffel, Eleanor G. and Wang, Zhihui}, year={2017}, month=jun, pages={062317} }

@article{Belkin_Allen_Ghosh_Kang_Lin_Sud_Chong_Fefferman_Clark_2024, title={Approximate t-designs in generic circuit architectures}, url={http://arxiv.org/abs/2310.19783}, DOI={10.48550/arXiv.2310.19783}, abstractNote={Unitary t-designs are distributions on the unitary group whose first t moments appear maximally random. Previous work has established several upper bounds on the depths at which certain specific random quantum circuit ensembles approximate t-designs. Here we show that these bounds can be extended to any fixed architecture of Haar-random two-site gates. This is accomplished by relating the spectral gaps of such architectures to those of 1D brickwork architectures. Our bound depends on the details of the architecture only via the typical number of layers needed for a block of the circuit to form a connected graph over the sites. When this quantity is independent of width, the circuit forms an approximate t-design in linear depth. We also give an implicit bound for nondeterministic architectures in terms of properties of the corresponding distribution over fixed architectures.}, note={arXiv:2310.19783 [quant-ph]}, number={arXiv:2310.19783}, publisher={arXiv}, author={Belkin, Daniel and Allen, James and Ghosh, Soumik and Kang, Christopher and Lin, Sophia and Sud, James and Chong, Fred and Fefferman, Bill and Clark, Bryan K.}, year={2024}, month=may }

@article{Dalzell_Hunter, title={Random quantum circuits anti-concentrate in log depth}, volume={3}, ISSN={2691-3399}, DOI={10.1103/PRXQuantum.3.010333}, abstractNote={We consider quantum circuits consisting of randomly chosen two-local gates and study the number of gates needed for the distribution over measurement outcomes for typical circuit instances to be anti-concentrated, roughly meaning that the probability mass is not too concentrated on a small number of measurement outcomes. Understanding the conditions for anti-concentration is important for determining which quantum circuits are difficult to simulate classically, as anti-concentration has been in some cases an ingredient of mathematical arguments that simulation is hard and in other cases a necessary condition for easy simulation. Our definition of anti-concentration is that the expected collision probability, that is, the probability that two independently drawn outcomes will agree, is only a constant factor larger than if the distribution were uniform. We show that when the 2-local gates are each drawn from the Haar measure (or any two-design), at least $Ω(n log(n))$ gates (and thus $Ω(log(n))$ circuit depth) are needed for this condition to be met on an $n$ qudit circuit. In both the case where the gates are nearest-neighbor on a 1D ring and the case where gates are long-range, we show $O(n log(n))$ gates are also sufficient, and we precisely compute the optimal constant prefactor for the $n log(n)$. The technique we employ relies upon a mapping from the expected collision probability to the partition function of an Ising-like classical statistical mechanical model, which we manage to bound using stochastic and combinatorial techniques.}, note={arXiv:2011.12277 [quant-ph]}, number={1}, journal={PRX Quantum}, author={Dalzell, Alexander M. and Hunter-Jones, Nicholas and Brandão, Fernando G. S. L.}, year={2022}, month=mar, pages={010333} }

@article{Deshpande_Niroula_Shtanko_Gorshkov_Fefferman_Gullans_2022, title={Tight bounds on the convergence of noisy random circuits to the uniform distribution}, volume={3}, ISSN={2691-3399}, DOI={10.1103/PRXQuantum.3.040329}, abstractNote={We study the properties of output distributions of noisy, random circuits. We obtain upper and lower bounds on the expected distance of the output distribution from the “useless” uniform distribution. These bounds are tight with respect to the dependence on circuit depth. Our proof techniques also allow us to make statements about the presence or absence of anticoncentration for both noisy and noiseless circuits. We uncover a number of interesting consequences for hardness proofs of sampling schemes that aim to show a quantum computational advantage over classical computation. Specifically, we discuss recent barrier results for depth-agnostic and/or noise-agnostic proof techniques. We show that in certain depth regimes, noise-agnostic proof techniques might still work in order to prove an often-conjectured claim in the literature on quantum computational advantage, contrary to what was thought prior to this work.}, note={arXiv:2112.00716 [quant-ph]}, number={4}, journal={PRX Quantum}, author={Deshpande, Abhinav and Niroula, Pradeep and Shtanko, Oles and Gorshkov, Alexey V. and Fefferman, Bill and Gullans, Michael J.}, year={2022}, month=dec, pages={040329} }

@article{Schuster_Haferkamp_Huang_2025, title={Random unitaries in extremely low depth}, url={http://arxiv.org/abs/2407.07754}, DOI={10.48550/arXiv.2407.07754}, abstractNote={We prove that random quantum circuits on any geometry, including a 1D line, can form approximate unitary designs over $n$ qubits in $log n$ depth. In a similar manner, we construct pseudorandom unitaries (PRUs) in 1D circuits in $text{poly}(log n)$ depth, and in all-to-all-connected circuits in $text{poly}(log log n)$ depth. In all three cases, the $n$ dependence is optimal and improves exponentially over known results. These shallow quantum circuits have low complexity and create only short-range entanglement, yet are indistinguishable from unitaries with exponential complexity. Our construction glues local random unitaries on $log n$-sized or $text{poly}(log n)$-sized patches of qubits to form a global random unitary on all $n$ qubits. In the case of designs, the local unitaries are drawn from existing constructions of approximate unitary $k$-designs, and hence also inherit an optimal scaling in $k$. In the case of PRUs, the local unitaries are drawn from existing PRU constructions. Applications of our results include proving that classical shadows with 1D log-depth Clifford circuits are as powerful as those with deep circuits, demonstrating superpolynomial quantum advantage in learning low-complexity physical systems, and establishing quantum hardness for recognizing phases of matter with topological order.}, note={arXiv:2407.07754 [quant-ph]}, number={arXiv:2407.07754}, publisher={arXiv}, author={Schuster, Thomas and Haferkamp, Jonas and Huang, Hsin-Yuan}, year={2025}, month=jan }

@article{Choi_Shaw_Madjarov_Xie_Finkelstein_Covey_Cotler_Mark_Huang_Kale_2023, title={Preparing random states and benchmarking with many-body quantum chaos}, volume={613}, ISSN={0028-0836, 1476-4687}, DOI={10.1038/s41586-022-05442-1}, note={arXiv:2103.03535 [quant-ph]}, number={7944}, journal={Nature}, author={Choi, Joonhee and Shaw, Adam L. and Madjarov, Ivaylo S. and Xie, Xin and Finkelstein, Ran and Covey, Jacob P. and Cotler, Jordan S. and Mark, Daniel K. and Huang, Hsin-Yuan and Kale, Anant and Pichler, Hannes and Brandão, Fernando G. S. L. and Choi, Soonwon and Endres, Manuel}, year={2023}, month=jan, pages={468–473} }

@article{Cotler_Mark_Huang_Hernandez_Choi_Shaw_Endres_Choi_2023, title={Emergent quantum state designs from individual many-body wavefunctions}, volume={4}, ISSN={2691-3399}, DOI={10.1103/PRXQuantum.4.010311}, abstractNote={Quantum chaos in many-body systems provides a bridge between statistical and quantum physics with strong predictive power. This framework is valuable for analyzing properties of complex quantum systems such as energy spectra and the dynamics of thermalization. While contemporary methods in quantum chaos often rely on random ensembles of quantum states and Hamiltonians, this is not reflective of most real-world systems. In this paper, we introduce a new perspective: across a wide range of examples, a single non-random quantum state is shown to encode universal and highly random quantum state ensembles. We characterize these ensembles using the notion of quantum state $k$-designs from quantum information theory and investigate their universality using a combination of analytic and numerical techniques. In particular, we establish that $k$-designs arise naturally from generic states as well as individual states associated with strongly interacting, time-independent Hamiltonian dynamics. Our results offer a new approach for studying quantum chaos and provide a practical method for sampling approximately uniformly random states; the latter has wide-ranging applications in quantum information science from tomography to benchmarking.}, note={arXiv:2103.03536 [quant-ph]}, number={1}, journal={PRX Quantum}, author={Cotler, Jordan S. and Mark, Daniel K. and Huang, Hsin-Yuan and Hernandez, Felipe and Choi, Joonhee and Shaw, Adam L. and Endres, Manuel and Choi, Soonwon}, year={2023}, month=jan, pages={010311} }

@inproceedings{Bergamaschi_Chen_Liu_2024, title={Quantum computational advantage with constant-temperature Gibbs sampling}, url={http://arxiv.org/abs/2404.14639}, DOI={10.1109/FOCS61266.2024.00071}, abstractNote={A quantum system coupled to a bath at some fixed, finite temperature converges to its Gibbs state. This thermalization process defines a natural, physically-motivated model of quantum computation. However, whether quantum computational advantage can be achieved within this realistic physical setup has remained open, due to the challenge of finding systems that thermalize quickly, but are classically intractable. Here we consider sampling from the measurement outcome distribution of quantum Gibbs states at constant temperatures, and prove that this task demonstrates quantum computational advantage. We design a family of commuting local Hamiltonians (parent Hamiltonians of shallow quantum circuits) and prove that they rapidly converge to their Gibbs states under the standard physical model of thermalization (as a continuous-time quantum Markov chain). On the other hand, we show that no polynomial time classical algorithm can sample from the measurement outcome distribution by reducing to the classical hardness of sampling from noiseless shallow quantum circuits. The key step in the reduction is constructing a fault-tolerance scheme for shallow IQP circuits against input noise.}, note={arXiv:2404.14639 [quant-ph]}, booktitle={2024 IEEE 65th Annual Symposium on Foundations of Computer Science (FOCS)}, author={Bergamaschi, Thiago and Chen, Chi-Fang and Liu, Yunchao}, year={2024}, month=oct, pages={1063–1085} }

@article{BermejoVega_2018, title={Architectures for Quantum Simulation Showing a Quantum Speedup}, volume={8}, DOI={10.1103/PhysRevX.8.021010}, number={2}, journal={Physical Review X}, author={Bermejo-Vega, Juan}, year={2018} }

@article{Haferkamp_Hangleiter_Bouland_Fefferman_Eisert_Bermejo2020, title={Closing gaps of a quantum advantage with short-time Hamiltonian dynamics}, volume={125}, ISSN={0031-9007, 1079-7114}, DOI={10.1103/PhysRevLett.125.250501}, abstractNote={Demonstrating a quantum computational speedup is a crucial milestone for near-term quantum technology. Recently, quantum simulation architectures have been proposed that have the potential to show such a quantum advantage, based on commonly made assumptions. The key challenge in the theoretical analysis of this scheme - as of other comparable schemes such as boson sampling - is to lessen the assumptions and close the theoretical loopholes, replacing them by rigorous arguments. In this work, we prove two open conjectures for these architectures for Hamiltonian quantum simulators: Anticoncentration of the generated probability distributions and average-case hardness of exactly evaluating those probabilities. The latter is proven building upon recently developed techniques for random circuit sampling. For the former, we develop new techniques that exploit the insight that approximate 2-designs for the unitary group admit anticoncentration. We prove that the 2D translation-invariant, constant depth architectures of quantum simulation form approximate 2-designs in a specific sense, thus obtaining a significantly stronger result. Our work provides the strongest evidence to date that Hamiltonian quantum simulation architectures are classically intractable.}, note={arXiv:1908.08069 [quant-ph]}, number={25}, journal={Physical Review Letters}, author={Haferkamp, Jonas and Hangleiter, Dominik and Bouland, Adam and Fefferman, Bill and Eisert, Jens and Bermejo-Vega, Juani}, year={2020}, month=dec, pages={250501} }

@article{Rajakumar_Watson_2024, title={Gibbs Sampling gives Quantum Advantage at Constant Temperatures with O(1)-Local Hamiltonians}, url={http://arxiv.org/abs/2408.01516}, DOI={10.48550/arXiv.2408.01516}, abstractNote={Sampling from Gibbs states -- states corresponding to system in thermal equilibrium -- has recently been shown to be a task for which quantum computers are expected to achieve super-polynomial speed-up compared to classical computers, provided the locality of the Hamiltonian increases with the system size (Bergamaschi et al., arXiv: 2404.14639). We extend these results to show that this quantum advantage still occurs for Gibbs states of Hamiltonians with O(1)-local interactions at constant temperature by showing classical hardness-of-sampling and demonstrating such Gibbs states can be prepared efficiently using a quantum computer. In particular, we show hardness-of-sampling is maintained even for 5-local Hamiltonians on a 3D lattice. We additionally show that the hardness-of-sampling is robust when we are only able to make imperfect measurements.}, note={arXiv:2408.01516 [quant-ph]}, number={arXiv:2408.01516}, publisher={arXiv}, author={Rajakumar, Joel and Watson, James D.}, year={2024}, month=oct }

@book{Rajakumar_Watson_Liu_2025, title={Polynomial-Time Classical Simulation of Noisy IQP Circuits with Constant Depth}, url={http://arxiv.org/abs/2403.14607}, DOI={10.1137/1.9781611978322.30}, note={arXiv:2403.14607 [quant-ph]}, author={Rajakumar, Joel and Watson, James D. and Liu, Yi-Kai}, year={2025}, month=jan }

@article{Codsi_Wetering_2025, title={Classically Simulating Quantum Supremacy IQP Circuits through a Random Graph Approach}, volume={111}, ISSN={2469-9926, 2469-9934}, DOI={10.1103/PhysRevA.111.012422}, note={arXiv:2212.08609 [quant-ph]}, number={1}, journal={Physical Review A}, author={Codsi, Julien and Wetering, John van de}, year={2025}, month=jan, pages={012422} }

@article{Jozsa_Ghosh_Strelchuk_2025, title={IQP computations with intermediate measurements}, url={http://arxiv.org/abs/2408.10093}, DOI={10.48550/arXiv.2408.10093},  note={arXiv:2408.10093 [quant-ph]}, number={arXiv:2408.10093}, publisher={arXiv}, author={Jozsa, Richard and Ghosh, Soumik and Strelchuk, Sergii}, year={2025}, month=jul }

@article{Bremner_Jozsa_Shepherd_2011, title={Classical simulation of commuting quantum computations implies collapse of the polynomial hierarchy}, volume={467}, ISSN={1364-5021, 1471-2946}, DOI={10.1098/rspa.2010.0301}, note={arXiv:1005.1407 [quant-ph]}, number={2126}, journal={Proceedings of the Royal Society A: Mathematical, Physical and Engineering Sciences}, author={Bremner, Michael J. and Jozsa, Richard and Shepherd, Dan J.}, year={2011}, month=feb, pages={459–472} }

@article{Bremner_Montanaro_Shepherd_2017, title={Achieving quantum supremacy with sparse and noisy commuting quantum computations}, volume={1}, ISSN={2521-327X}, DOI={10.22331/q-2017-04-25-8}, note={arXiv:1610.01808 [quant-ph]}, journal={Quantum}, author={Bremner, Michael J. and Montanaro, Ashley and Shepherd, Dan J.}, year={2017}, month=apr, pages={8} }

@article{Bluvstein_Evered_Geim_Li_Zhou_Manovitz_Ebadi_Cain_Kalinowski_Hangleiter_2024, title={Logical quantum processor based on reconfigurable atom arrays}, volume={626}, rights={2023 The Author(s)}, ISSN={1476-4687}, DOI={10.1038/s41586-023-06927-3}, number={7997}, journal={Nature}, publisher={Nature Publishing Group}, author={Bluvstein, Dolev and Evered, Simon J. and Geim, Alexandra A. and Li, Sophie H. and Zhou, Hengyun and Manovitz, Tom and Ebadi, Sepehr and Cain, Madelyn and Kalinowski, Marcin and Hangleiter, Dominik and Bonilla Ataides, J. Pablo and Maskara, Nishad and Cong, Iris and Gao, Xun and Sales Rodriguez, Pedro and Karolyshyn, Thomas and Semeghini, Giulia and Gullans, Michael J. and Greiner, Markus and Vuletić, Vladan and Lukin, Mikhail D.}, year={2024}, month=feb, pages={58–65} }

@article{Bremner_Montanaro_Shepherd_2016, title={Average-Case Complexity Versus Approximate Simulation of Commuting Quantum Computations}, volume={117}, DOI={10.1103/PhysRevLett.117.080501},  number={8}, journal={Physical Review Letters}, publisher={American Physical Society}, author={Bremner, Michael J. and Montanaro, Ashley and Shepherd, Dan J.}, year={2016}, month=aug, pages={080501} }

@article{Aaronson_Zhang_2024, title={On verifiable quantum advantage with peaked circuit sampling}, url={http://arxiv.org/abs/2404.14493}, DOI={10.48550/arXiv.2404.14493}, note={arXiv:2404.14493 [quant-ph]}, number={arXiv:2404.14493}, publisher={arXiv}, author={Aaronson, Scott and Zhang, Yuxuan}, year={2024}, month=may }

@article{Bentsen_Fefferman_Ghosh_Gullans_Liu_2024, title={On the complexity of sampling from shallow Brownian circuits}, url={http://arxiv.org/abs/2411.04169}, DOI={10.48550/arXiv.2411.04169}, abstractNote={While many statistical properties of deep random quantum circuits can be deduced, often rigorously and other times heuristically, by an approximation to global Haar-random unitaries, the statistics of constant-depth random quantum circuits are generally less well-understood due to a lack of amenable tools and techniques. We circumvent this barrier by considering a related constant-time Brownian circuit model which shares many similarities with constant-depth random quantum circuits but crucially allows for direct calculations of higher order moments of its output distribution. Using mean-field (large-n) techniques, we fully characterize the output distributions of Brownian circuits at shallow depths and show that they follow a Porter-Thomas distribution, just like in the case of deep circuits, but with a truncated Hilbert space. The access to higher order moments allows for studying the expected and typical Linear Cross-entropy (XEB) benchmark scores achieved by an ideal quantum computer versus the state-of-the-art classical spoofers for shallow Brownian circuits. We discover that for these circuits, while the quantum computer typically scores within a constant factor of the expected value, the classical spoofer suffers from an exponentially larger variance. Numerical evidence suggests that the same phenomenon also occurs in constant-depth discrete random quantum circuits, like those defined over the all-to-all architecture. We conjecture that the same phenomenon is also true for random brickwork circuits in high enough spatial dimension.}, note={arXiv:2411.04169 [quant-ph]}, number={arXiv:2411.04169}, publisher={arXiv}, author={Bentsen, Gregory and Fefferman, Bill and Ghosh, Soumik and Gullans, Michael J. and Liu, Yinchen}, year={2024}, month=nov }

@article{Ware_Deshpande_Hangleiter_Niroula_Fefferman_Gorshkov_Gullans_2023, title={A sharp phase transition in linear cross-entropy benchmarking}, url={http://arxiv.org/abs/2305.04954}, DOI={10.48550/arXiv.2305.04954}, abstractNote={Demonstrations of quantum computational advantage and benchmarks of quantum processors via quantum random circuit sampling are based on evaluating the linear cross-entropy benchmark (XEB). A key question in the theory of XEB is whether it approximates the fidelity of the quantum state preparation. Previous works have shown that the XEB generically approximates the fidelity in a regime where the noise rate per qudit $varepsilon$ satisfies $varepsilon N ll 1$ for a system of $N$ qudits and that this approximation breaks down at large noise rates. Here, we show that the breakdown of XEB as a fidelity proxy occurs as a sharp phase transition at a critical value of $varepsilon N$ that depends on the circuit architecture and properties of the two-qubit gates, including in particular their entangling power. We study the phase transition using a mapping of average two-copy quantities to statistical mechanics models in random quantum circuit architectures with full or one-dimensional connectivity. We explain the phase transition behavior in terms of spectral properties of the transfer matrix of the statistical mechanics model and identify two-qubit gate sets that exhibit the largest noise robustness.}, note={arXiv:2305.04954 [quant-ph]}, number={arXiv:2305.04954}, publisher={arXiv}, author={Ware, Brayden and Deshpande, Abhinav and Hangleiter, Dominik and Niroula, Pradeep and Fefferman, Bill and Gorshkov, Alexey V. and Gullans, Michael J.}, year={2023}, month=may }

@article{Angrisani_Mele_Rudolph_Cerezo_Holmes_2025, title={Simulating quantum circuits with arbitrary local noise using Pauli Propagation}, url={http://arxiv.org/abs/2501.13101}, DOI={10.48550/arXiv.2501.13101}, note={arXiv:2501.13101 [quant-ph]}, number={arXiv:2501.13101}, publisher={arXiv}, author={Angrisani, Armando and Mele, Antonio A. and Rudolph, Manuel S. and Cerezo, M. and Holmes, Zoë}, year={2025}, month=may }

@article{Nelson_Rajakumar_Hangleiter_Gullans_2024, title={Polynomial-Time Classical Simulation of Noisy Circuits with Naturally Fault-Tolerant Gates}, url={http://arxiv.org/abs/2411.02535}, DOI={10.48550/arXiv.2411.02535}, note={arXiv:2411.02535 [quant-ph]}, number={arXiv:2411.02535}, publisher={arXiv}, author={Nelson, Jon and Rajakumar, Joel and Hangleiter, Dominik and Gullans, Michael J.}, year={2024}, month=dec }

@article{Singkanipa_Lidar_2025, title={Beyond unital noise in variational quantum algorithms: noise-induced barren plateaus and limit sets}, volume={9}, ISSN={2521-327X}, DOI={10.22331/q-2025-01-30-1617}, note={arXiv:2402.08721 [quant-ph]}, journal={Quantum}, author={Singkanipa, P. and Lidar, D. A.}, year={2025}, month=jan, pages={1617} }

@inproceedings{Aharonov_Gao_Landau_Liu_Vazirani_2023, title={A polynomial-time classical algorithm for noisy random circuit sampling}, url={http://arxiv.org/abs/2211.03999}, DOI={10.1145/3564246.3585234}, booktitle={Proceedings of the 55th Annual ACM Symposium on Theory of Computing}, author={Aharonov, Dorit and Gao, Xun and Landau, Zeph and Liu, Yunchao and Vazirani, Umesh}, year={2023}, month=jun, pages={945–957} }

@article{Angrisani_Schmidhuber_Rudolph_Cerezo_Holmes_Huang_2025, title={Classically estimating observables of noiseless quantum circuits}, volume={135}, ISSN={0031-9007, 1079-7114}, DOI={10.1103/lh6x-7rc3}, note={arXiv:2409.01706 [quant-ph]}, number={17}, journal={Physical Review Letters}, author={Angrisani, Armando and Schmidhuber, Alexander and Rudolph, Manuel S. and Cerezo, M. and Holmes, Zoë and Huang, Hsin-Yuan}, year={2025}, month=oct, pages={170602} }

@article{Fefferman_Ghosh_Gullans_Kuroiwa_Sharma_2023, title={Effect of non-unital noise on random circuit sampling}, url={http://arxiv.org/abs/2306.16659}, DOI={10.48550/arXiv.2306.16659}, note={arXiv:2306.16659 [quant-ph]}, number={arXiv:2306.16659}, publisher={arXiv}, author={Fefferman, Bill and Ghosh, Soumik and Gullans, Michael and Kuroiwa, Kohdai and Sharma, Kunal}, year={2023}, month=jun }

@article{Gao_Duan_2018, title={Efficient classical simulation of noisy quantum computation}, url={http://arxiv.org/abs/1810.03176}, DOI={10.48550/arXiv.1810.03176}, note={arXiv:1810.03176 [quant-ph]}, number={arXiv:1810.03176}, publisher={arXiv}, author={Gao, Xun and Duan, Luming}, year={2018}, month=oct }

@article{Mele_Angrisani_Ghosh_Khatri_Eisert_França_Quek_2024, title={Noise-induced shallow circuits and absence of barren plateaus}, url={http://arxiv.org/abs/2403.13927}, DOI={10.48550/arXiv.2403.13927}, note={arXiv:2403.13927 [quant-ph]}, number={arXiv:2403.13927}, publisher={arXiv}, author={Mele, Antonio Anna and Angrisani, Armando and Ghosh, Soumik and Khatri, Sumeet and Eisert, Jens and França, Daniel Stilck and Quek, Yihui}, year={2024}, month=oct }

@article{Schuster_Yin_Gao_Yao_2024, title={A polynomial-time classical algorithm for noisy quantum circuits}, url={http://arxiv.org/abs/2407.12768}, DOI={10.48550/arXiv.2407.12768},  note={arXiv:2407.12768 [quant-ph]}, number={arXiv:2407.12768}, publisher={arXiv}, author={Schuster, Thomas and Yin, Chao and Gao, Xun and Yao, Norman Y.}, year={2024}, month=oct }

@article{Dalzell_Jones_Brandao_2021, title={Random quantum circuits transform local noise into global white noise}, url={http://arxiv.org/abs/2111.14907}, DOI={10.48550/arXiv.2111.14907},  note={arXiv:2111.14907 [quant-ph]}, number={arXiv:2111.14907}, publisher={arXiv}, author={Dalzell, Alexander M. and Hunter-Jones, Nicholas and Brandão, Fernando G. S. L.}, year={2021}, month=nov }

@misc{willow, title={Meet Willow, our state-of-the-art quantum chip}, howpublished={\url{https://blog.google/technology/research/google-willow-quantum-chip/}}, author={Neven, Hartmut}, journal={Google}, year={2024}, month=dec }

@article{Wootton_2020, title={Benchmarking near-term devices with quantum error correction}, volume={5}, ISSN={2058-9565}, DOI={10.1088/2058-9565/aba038}, abstractNote={Now that ever more sophisticated devices for quantum computing are being developed, we require ever more sophisticated benchmarks. This includes a need to determine how well these devices support the techniques required for quantum error correction. In this paper we introduce the texttt{topological_codes} module of Qiskit-Ignis, which is designed to provide the tools necessary to perform such tests. Specifically, we use the texttt{RepetitionCode} and texttt{GraphDecoder} classes to run tests based on the repetition code and process the results. As an example, data from a 43 qubit code running on IBM’s emph{Rochester} device is presented.}, note={arXiv:2004.11037 [quant-ph]}, number={4}, journal={Quantum Science and Technology}, author={Wootton, James R.}, year={2020}, month=oct, pages={044004} }

@article{paetznick2024demo, title={Demonstration of logical qubits and repeated error correction with better-than-physical error rates}, url={http://arxiv.org/abs/2404.02280}, DOI={10.48550/arXiv.2404.02280}, note={arXiv:2404.02280 [quant-ph]}, number={arXiv:2404.02280}, publisher={arXiv}, author={Paetznick, A. and Silva, M. P. da and Ryan-Anderson, C. and Bello-Rivas, J. M. and III, J. P. Campora and Chernoguzov, A. and Dreiling, J. M. and Foltz, C. and Frachon, F. and Gaebler, J. P. and Gatterman, T. M. and Grans-Samuelsson, L. and Gresh, D. and Hayes, D. and Hewitt, N. and Holliman, C. and Horst, C. V. and Johansen, J. and Lucchetti, D. and Matsuoka, Y. and Mills, M. and Moses, S. A. and Neyenhuis, B. and Paz, A. and Pino, J. and Siegfried, P. and Sundaram, A. and Tom, D. and Wernli, S. J. and Zanner, M. and Stutz, R. P. and Svore, K. M.}, year={2024}, month=nov }

@article{magic2024, title={Encoding a magic state with beyond break-even fidelity}, volume={625}, rights={2024 The Author(s)}, ISSN={1476-4687}, DOI={10.1038/s41586-023-06846-3}, abstractNote={To run large-scale algorithms on a quantum computer, error-correcting codes must be able to perform a fundamental set of operations, called logic gates, while isolating the encoded information from noise1–8. We can complete a universal set of logic gates by producing special resources called magic states9–11. It is therefore important to produce high-fidelity magic states to conduct algorithms while introducing a minimal amount of noise to the computation. Here we propose and implement a scheme to prepare a magic state on a superconducting qubit array using error correction. We find that our scheme produces better magic states than those that can be prepared using the individual qubits of the device. This demonstrates a fundamental principle of fault-tolerant quantum computing12, namely, that we can use error correction to improve the quality of logic gates with noisy qubits. Moreover, we show that the yield of magic states can be increased using adaptive circuits, in which the circuit elements are changed depending on the outcome of mid-circuit measurements. This demonstrates an essential capability needed for many error-correction subroutines. We believe that our prototype will be invaluable in the future as it can reduce the number of physical qubits needed to produce high-fidelity magic states in large-scale quantum-computing architectures.}, number={7994}, journal={Nature}, publisher={Nature Publishing Group}, author={Gupta, Riddhi S. and Sundaresan, Neereja and Alexander, Thomas and Wood, Christopher J. and Merkel, Seth T. and Healy, Michael B. and Hillenbrand, Marius and Jochym-O’Connor, Tomas and Wootton, James R. and Yoder, Theodore J. and Cross, Andrew W. and Takita, Maika and Brown, Benjamin J.}, year={2024}, month=jan, pages={259–263} }

@article{48, title={Logical quantum processor based on reconfigurable atom arrays}, volume={626}, rights={2023 The Author(s)}, ISSN={1476-4687}, DOI={10.1038/s41586-023-06927-3}, number={7997}, journal={Nature}, publisher={Nature Publishing Group}, author={Bluvstein, Dolev and Evered, Simon J. and Geim, Alexandra A. and Li, Sophie H. and Zhou, Hengyun and Manovitz, Tom and Ebadi, Sepehr and Cain, Madelyn and Kalinowski, Marcin and Hangleiter, Dominik and Bonilla Ataides, J. Pablo and Maskara, Nishad and Cong, Iris and Gao, Xun and Sales Rodriguez, Pedro and Karolyshyn, Thomas and Semeghini, Giulia and Gullans, Michael J. and Greiner, Markus and Vuletić, Vladan and Lukin, Mikhail D.}, year={2024}, month=feb, pages={58–65}}

@article{harper2019, title={Fault-Tolerant Logical Gates in the IBM Quantum Experience}, volume={122}, DOI={10.1103/PhysRevLett.122.080504}, abstractNote={Quantum computers will require encoding of quantum information to protect them from noise. Fault-tolerant quantum computing architectures illustrate how this might be done but have not yet shown a conclusive practical advantage. Here we demonstrate that a small but useful error detecting code improves the fidelity of the fault-tolerant gates implemented in the code space as compared to the fidelity of physically equivalent gates implemented on physical qubits. By running a randomized benchmarking protocol in the logical code space of the [4,2,2] code, we observe an order of magnitude improvement in the infidelity of the gates, with the two-qubit infidelity dropping from 5.8(2)% to 0.60(3)%. Our results are consistent with fault-tolerance theory and conclusively demonstrate the benefit of carrying out computation in a code space that can detect errors. Although the fault-tolerant gates offer an impressive improvement in fidelity, the computation as a whole is not below the fault-tolerance threshold because of noise associated with state preparation and measurement on this device.}, number={8}, journal={Physical Review Letters}, publisher={American Physical Society}, author={Harper, Robin and Flammia, Steven T.}, year={2019}, month=feb, pages={080504} }

@article{2004, title={Realization of quantum error correction}, volume={432}, rights={2004 Macmillan Magazines Ltd.}, ISSN={1476-4687}, DOI={10.1038/nature03074}, abstractNote={Scalable quantum computation1 and communication require error control to protect quantum information against unavoidable noise. Quantum error correction2,3 protects information stored in two-level quantum systems (qubits) by rectifying errors with operations conditioned on the measurement outcomes. Error-correction protocols have been implemented in nuclear magnetic resonance experiments4,5,6, but the inherent limitations of this technique7 prevent its application to quantum information processing. Here we experimentally demonstrate quantum error correction using three beryllium atomic-ion qubits confined to a linear, multi-zone trap. An encoded one-qubit state is protected against spin-flip errors by means of a three-qubit quantum error-correcting code. A primary ion qubit is prepared in an initial state, which is then encoded into an entangled state of three physical qubits (the primary and two ancilla qubits). Errors are induced simultaneously in all qubits at various rates. The encoded state is decoded back to the primary ion one-qubit state, making error information available on the ancilla ions, which are separated from the primary ion and measured. Finally, the primary qubit state is corrected on the basis of the ancillae measurement outcome. We verify error correction by comparing the corrected final state to the uncorrected state and to the initial state. In principle, the approach enables a quantum state to be maintained by means of repeated error correction, an important step towards scalable fault-tolerant quantum computation using trapped ions.}, number={7017}, journal={Nature}, publisher={Nature Publishing Group}, author={Chiaverini, J. and Leibfried, D. and Schaetz, T. and Barrett, M. D. and Blakestad, R. B. and Britton, J. and Itano, W. M. and Jost, J. D. and Knill, E. and Langer, C. and Ozeri, R. and Wineland, D. J.}, year={2004}, month=dec, pages={602–605} }

@article{pittman2005, title={Demonstration of Quantum Error Correction using Linear Optics}, volume={71}, ISSN={1050-2947, 1094-1622}, DOI={10.1103/PhysRevA.71.052332}, abstractNote={We describe a laboratory demonstration of a quantum error correction procedure that can correct intrinsic measurement errors in linear-optics quantum gates. The procedure involves a two-qubit encoding and fast feed-forward-controlled single-qubit operations. In our demonstration the qubits were represented by the polarization states of two single-photons from a parametric down-conversion source, and the real-time feed-forward control was implemented using an electro-optic device triggered by the output of single-photon detectors.}, note={arXiv:quant-ph/0502042}, number={5}, journal={Physical Review A}, author={Pittman, T. B. and Jacobs, B. C. and Franson, J. D.}, year={2005}, month=may, pages={052332} }

@article{zhao2022, title={Realization of an Error-Correcting Surface Code with Superconducting Qubits}, volume={129}, DOI={10.1103/PhysRevLett.129.030501}, number={3}, journal={Physical Review Letters}, publisher={American Physical Society}, author={Zhao, Youwei and Ye, Yangsen and Huang, He-Liang and Zhang, Yiming and Wu, Dachao and Guan, Huijie and Zhu, Qingling and Wei, Zuolin and He, Tan and Cao, Sirui and Chen, Fusheng and Chung, Tung-Hsun and Deng, Hui and Fan, Daojin and Gong, Ming and Guo, Cheng and Guo, Shaojun and Han, Lianchen and Li, Na and Li, Shaowei and Li, Yuan and Liang, Futian and Lin, Jin and Qian, Haoran and Rong, Hao and Su, Hong and Sun, Lihua and Wang, Shiyu and Wu, Yulin and Xu, Yu and Ying, Chong and Yu, Jiale and Zha, Chen and Zhang, Kaili and Huo, Yong-Heng and Lu, Chao-Yang and Peng, Cheng-Zhi and Zhu, Xiaobo and Pan, Jian-Wei}, year={2022}, month=jul, pages={030501} }

@article{abobeih2022fault, title={Fault-tolerant operation of a logical qubit in a diamond quantum processor}, volume={606}, rights={2022 The Author(s)}, ISSN={1476-4687}, DOI={10.1038/s41586-022-04819-6},  number={7916}, journal={Nature}, publisher={Nature Publishing Group}, author={Abobeih, M. H. and Wang, Y. and Randall, J. and Loenen, S. J. H. and Bradley, C. E. and Markham, M. and Twitchen, D. J. and Terhal, B. M. and Taminiau, T. H.}, year={2022}, month=jun, pages={884–889} }

@article{egan2021fault, title={Fault-tolerant control of an error-corrected qubit}, volume={598}, rights={2021 The Author(s), under exclusive licence to Springer Nature Limited}, ISSN={1476-4687}, DOI={10.1038/s41586-021-03928-y}, number={7880}, journal={Nature}, publisher={Nature Publishing Group}, author={Egan, Laird and Debroy, Dripto M. and Noel, Crystal and Risinger, Andrew and Zhu, Daiwei and Biswas, Debopriyo and Newman, Michael and Li, Muyuan and Brown, Kenneth R. and Cetina, Marko and Monroe, Christopher}, year={2021}, month=oct, pages={281–286} }

@article{acharya2024quantum, title={Quantum error correction below the surface code threshold}, rights={2024 The Author(s), under exclusive licence to Springer Nature Limited}, ISSN={1476-4687}, DOI={10.1038/s41586-024-08449-y},  journal={Nature}, publisher={Nature Publishing Group}, author={Acharya, Rajeev and others}, year={2024}, month=dec, pages={1–3} }

@article{ahcarya2023suppresing, title={Suppressing quantum errors by scaling a surface code logical qubit}, volume={614}, rights={2023 The Author(s)}, ISSN={1476-4687}, DOI={10.1038/s41586-022-05434-1},  number={7949}, journal={Nature}, publisher={Nature Publishing Group}, author={Acharya, Rajeev and others}, year={2023}, month=feb, pages={676–681} 
}

@article{sivak2023real, title={Real-time quantum error correction beyond break-even}, volume={616}, rights={2023 The Author(s), under exclusive licence to Springer Nature Limited}, ISSN={1476-4687}, DOI={10.1038/s41586-023-05782-6}, number={7955}, journal={Nature}, publisher={Nature Publishing Group}, author={Sivak, V. V. and Eickbusch, A. and Royer, B. and Singh, S. and Tsioutsios, I. and Ganjam, S. and Miano, A. and Brock, B. L. and Ding, A. Z. and Frunzio, L. and Girvin, S. M. and Schoelkopf, R. J. and Devoret, M. H.}, year={2023}, month=apr, pages={50–55}}

@article{krinner2022realizing, title={Realizing repeated quantum error correction in a distance-three surface code}, volume={605}, rights={2022 The Author(s), under exclusive licence to Springer Nature Limited}, ISSN={1476-4687}, DOI={10.1038/s41586-022-04566-8}, author={Krinner, Sebastian and Lacroix, Nathan and Remm, Ants and Di Paolo, Agustin and Genois, Elie and Leroux, Catherine and Hellings, Christoph and Lazar, Stefania and Swiadek, Francois and Herrmann, Johannes and Norris, Graham J. and Andersen, Christian Kraglund and Müller, Markus and Blais, Alexandre and Eichler, Christopher and Wallraff, Andreas}, year={2022}, month=may, pages={669–674} }

@article{anderson2021realization, title={Realization of Real-Time Fault-Tolerant Quantum Error Correction}, volume={11}, DOI={10.1103/PhysRevX.11.041058}, number={4}, journal={Physical Review X}, publisher={American Physical Society}, author={Ryan-Anderson, C. and others}, year={2021}, month=dec, pages={041058} }

@article{Kalai_Kindler_2014, title={Gaussian Noise Sensitivity and BosonSampling}, url={http://arxiv.org/abs/1409.3093}, DOI={10.48550/arXiv.1409.3093}, abstractNote={We study the sensitivity to noise of |permanent(X)|^2 for random real and complex n x n Gaussian matrices X, and show that asymptotically the correlation between the noisy and noiseless outcomes tends to zero when the noise level is {omega}(1)/n. This suggests that, under certain reasonable noise models, the probability distributions produced by noisy BosonSampling are very sensitive to noise. We also show that when the amount of noise is constant the noisy value of |permanent(X)|^2 can be approximated efficiently on a classical computer. These results seem to weaken the possibility of demonstrating quantum-speedup via BosonSampling without quantum fault-tolerance.}, note={arXiv:1409.3093 [quant-ph]}, number={arXiv:1409.3093}, publisher={arXiv}, author={Kalai, Gil and Kindler, Guy}, year={2014}, month=nov }

@article{Liu_Oh_Liu_Jiang_Alexeev_2023, title={Simulating lossy Gaussian boson sampling with matrix-product operators}, volume={108}, DOI={10.1103/PhysRevA.108.052604}, abstractNote={Gaussian boson sampling, a computational model that is widely believed to admit quantum supremacy, has already been experimentally demonstrated and is claimed to surpass the classical simulation capabilities of even the most powerful supercomputers today. However, whether the current approach limited by photon loss and noise in such experiments prescribes a scalable path to quantum advantage is an open question. To understand the effect of photon loss on the scalability of Gaussian boson sampling, we analytically derive the asymptotic operator entanglement entropy scaling, which relates to the simulation complexity. As a result, we observe that efficient tensor network simulations are likely possible under the ��out∝√�� scaling of the number of surviving photons ��out in the number of input photons ��. We numerically verify this result using a tensor network algorithm with U⁡(1) symmetry, and we overcome previous challenges due to the large local Hilbert-space dimensions in Gaussian boson sampling with hardware acceleration. Additionally, we observe that increasing the photon number through larger squeezing does not increase the entanglement entropy significantly. Finally, we numerically find the bond dimension necessary for fixed accuracy simulations, providing more direct evidence for the complexity of tensor networks.}, number={5}, journal={Physical Review A}, publisher={American Physical Society}, author={Liu, Minzhao and Oh, Changhun and Liu, Junyu and Jiang, Liang and Alexeev, Yuri}, year={2023}, month=nov, pages={052604} }

@article{Oh_Noh_Fefferman_Jiang_2021, title={Classical simulation of lossy boson sampling using matrix product operators}, volume={104}, DOI={10.1103/PhysRevA.104.022407}, abstractNote={Characterizing the computational advantage from noisy intermediate-scale quantum (NISQ) devices is an important task from theoretical and practical perspectives. Here, we numerically investigate the computational power of NISQ devices focusing on boson sampling, one of the well-known promising problems which can exhibit quantum supremacy. We study hardness of lossy boson sampling using matrix product operator (MPO) simulation to address the effect of photon loss on classical simulability using MPO entanglement entropy (EE), which characterizes a running time of an MPO algorithm. An advantage of MPO simulation over other classical algorithms proposed to date is that its simulation accuracy can be efficiently controlled by increasing an MPO’s bond dimension. Notably, we show by simulating lossy boson sampling using an MPO that as an input photon number grows, its computational cost, or MPO EE, behaves differently depending on a loss scaling, exhibiting a different feature from that of lossless boson sampling. Especially when an output photon number scales faster than the square root of an input photon number, our study shows an exponential scaling of time complexity for MPO simulation. On the contrary, when an output photon number scales slower than the square root of an input photon number, MPO EE may decrease, indicating that an exponential time complexity might not be necessary.}, number={2}, journal={Physical Review A}, publisher={American Physical Society}, author={Oh, Changhun and Noh, Kyungjoo and Fefferman, Bill and Jiang, Liang}, year={2021}, month=aug, pages={022407} }

@inproceedings{Bravyi_Gosset_Liu_2024, address={New York, NY, USA}, series={STOC 2024}, title={Classical Simulation of Peaked Shallow Quantum Circuits}, ISBN={9798400703836}, url={https://doi.org/10.1145/3618260.3649638}, DOI={10.1145/3618260.3649638}, abstractNote={An n-qubit quantum circuit is said to be peaked if it has an output probability that is at least inverse-polynomially large as a function of n. We describe a classical algorithm with quasipolynomial runtime nO(logn) that approximately samples from the output distribution of a peaked constant-depth circuit. We give even faster algorithms for circuits composed of nearest-neighbor gates on a D-dimensional grid of qubits, with polynomial runtime nO(1) if D=2 and almost-polynomial runtime nO(loglogn) for D&gt;2. Our sampling algorithms can be used to estimate output probabilities of shallow circuits to within a given inverse-polynomial additive error, improving previously known methods. As a simple application, we obtain a quasipolynomial algorithm to estimate the magnitude of the expected value of any Pauli observable in the output state of a shallow circuit (which may or may not be peaked). This is a dramatic improvement over the prior state-of-the-art algorithm which had an exponential scaling in √n.}, booktitle={Proceedings of the 56th Annual ACM Symposium on Theory of Computing}, publisher={Association for Computing Machinery}, author={Bravyi, Sergey and Gosset, David and Liu, Yinchen}, year={2024}, month=jun, pages={561–572}, collection={STOC 2024} }

@article{Oh_Liu_Alexeev_Fefferman_Jiang_2024, title={Classical algorithm for simulating experimental Gaussian boson sampling}, volume={20}, rights={2024 The Author(s), under exclusive licence to Springer Nature Limited}, ISSN={1745-2481}, DOI={10.1038/s41567-024-02535-8}, abstractNote={Gaussian boson sampling is a form of non-universal quantum computing that has been considered a promising candidate for showing experimental quantum advantage. While there is evidence that noiseless Gaussian boson sampling is hard to efficiently simulate using a classical computer, current Gaussian boson sampling experiments inevitably suffer from high photon loss rates and other noise sources. Nevertheless, they are currently claimed to be hard to classically simulate. Here we present a classical tensor-network algorithm that simulates Gaussian boson sampling and whose complexity can be significantly reduced when the photon loss rate is high. Our algorithm enables us to simulate the largest-scale Gaussian boson sampling experiment so far using relatively modest computational resources. We exhibit evidence that our classical sampler can simulate the ideal distribution better than the experiment can, which calls into question the claims of experimental quantum advantage.}, number={9}, journal={Nature Physics}, publisher={Nature Publishing Group}, author={Oh, Changhun and Liu, Minzhao and Alexeev, Yuri and Fefferman, Bill and Jiang, Liang}, year={2024}, month=sep, pages={1461–1468}, language={en} }

@article{Zhao_Zhong_Pan_Chen_Fu_Su_Xie_Zhao_Zhang_Ouyang__2024, title={Leapfrogging Sycamore: Harnessing 1432 GPUs for 7$\times$ Faster Quantum Random Circuit Sampling}, url={http://arxiv.org/abs/2406.18889}, DOI={10.48550/arXiv.2406.18889}, abstractNote={Random quantum circuit sampling serves as a benchmark to demonstrate quantum computational advantage. Recent progress in classical algorithms, especially those based on tensor network methods, has significantly reduced the classical simulation time and challenged the claim of the first-generation quantum advantage experiments. However, in terms of generating uncorrelated samples, time-to-solution, and energy consumption, previous classical simulation experiments still underperform the textit{Sycamore} processor. Here we report an energy-efficient classical simulation algorithm, using 1432 GPUs to simulate quantum random circuit sampling which generates uncorrelated samples with higher linear cross entropy score and is 7 times faster than textit{Sycamore} 53 qubits experiment. We propose a post-processing algorithm to reduce the overall complexity, and integrated state-of-the-art high-performance general-purpose GPU to achieve two orders of lower energy consumption compared to previous works. Our work provides the first unambiguous experimental evidence to refute textit{Sycamore}’s claim of quantum advantage, and redefines the boundary of quantum computational advantage using random circuit sampling.}, note={arXiv:2406.18889 [quant-ph]}, number={arXiv:2406.18889}, publisher={arXiv}, author={Zhao, Xian-He and Zhong, Han-Sen and Pan, Feng and Chen, Zi-Han and Fu, Rong and Su, Zhongling and Xie, Xiaotong and Zhao, Chaoxing and Zhang, Pan and Ouyang, Wanli and Lu, Chao-Yang and Pan, Jian-Wei and Chen, Ming-Cheng}, year={2024}, month=jun }

@article{Clinton_Cubitt_Garcia-Patron_Montanaro_Stanisic_Stroeks_2024, title={Quantum Phase Estimation without Controlled Unitaries}, url={http://arxiv.org/abs/2410.21517}, DOI={10.48550/arXiv.2410.21517}, abstractNote={In this work we demonstrate the use of adapted classical phase retrieval algorithms to perform control-free quantum phase estimation. We eliminate the costly controlled time evolution and Hadamard test commonly required to access the complex time-series needed to reconstruct the spectrum. This significant reduction of the number of coherent controlled-operations lowers the circuit depth and considerably simplifies the implementation of statistical quantum phase estimation in near-term devices. This seemingly impossible task can be achieved by extending the problem that one wishes to solve to one with a larger set of input signals while exploiting natural constraints on the signal and/or the spectrum. We leverage well-established algorithms that are widely used in the context of classical signal processing, demonstrating two complementary methods to do this, vectorial phase retrieval and two-dimensional phase retrieval. We numerically investigate the feasibility of both approaches for estimating the spectrum of the Fermi-Hubbard model and discuss their resilience to inherent statistical noise.}, note={arXiv:2410.21517 [quant-ph]}, number={arXiv:2410.21517}, publisher={arXiv}, author={Clinton, Laura and Cubitt, Toby S. and Garcia-Patron, Raul and Montanaro, Ashley and Stanisic, Stasja and Stroeks, Maarten}, year={2024}, month=oct }

@article{Lin_2022, title={Lecture Notes on Quantum Algorithms for Scientific Computation}, url={http://arxiv.org/abs/2201.08309}, DOI={10.48550/arXiv.2201.08309}, abstractNote={This is a set of lecture notes used in a graduate topic class in applied mathematics called ``Quantum Algorithms for Scientific Computation’’ at the Department of Mathematics, UC Berkeley during the fall semester of 2021. These lecture notes focus only on quantum algorithms closely related to scientific computation, and in particular, matrix computation. The main purpose of the lecture notes is to introduce quantum phase estimation (QPE) and ``post-QPE’’ methods such as block encoding, quantum signal processing, and quantum singular value transformation, and to demonstrate their applications in solving eigenvalue problems, linear systems of equations, and differential equations. The intended audience is the broad computational science and engineering (CSE) community interested in using fault-tolerant quantum computers to solve challenging scientific computing problems.}, note={arXiv:2201.08309 [quant-ph]}, number={arXiv:2201.08309}, publisher={arXiv}, author={Lin, Lin}, year={2022}, month=jan }

@article{Lin_Tong_2022, title={Heisenberg-Limited Ground-State Energy Estimation for Early Fault-Tolerant Quantum Computers}, volume={3}, DOI={10.1103/PRXQuantum.3.010318}, abstractNote={Under suitable assumptions, the quantum-phase-estimation (QPE) algorithm is able to achieve Heisenberg-limited precision scaling in estimating the ground-state energy. However, QPE requires a large number of ancilla qubits and a large circuit depth, as well as the ability to perform inverse quantum Fourier transform, making it expensive to implement on an early fault-tolerant quantum computer. We propose an alternative method to estimate the ground-state energy of a Hamiltonian with Heisenberg-limited precision scaling, which employs a simple quantum circuit with one ancilla qubit, and a classical postprocessing procedure. Besides the ground-state energy, our algorithm also produces an approximate cumulative distribution function of the spectral measure, which can be used to compute other spectral properties of the Hamiltonian.}, number={1}, journal={PRX Quantum}, publisher={American Physical Society}, author={Lin, Lin and Tong, Yu}, year={2022}, month=feb, pages={010318} }

@article{Regev_2024, title={An Efficient Quantum Factoring Algorithm}, url={http://arxiv.org/abs/2308.06572}, DOI={10.48550/arXiv.2308.06572}, abstractNote={We show that $n$-bit integers can be factorized by independently running a quantum circuit with $tilde{O}(n^{3/2})$ gates for $sqrt{n}+4$ times, and then using polynomial-time classical post-processing. The correctness of the algorithm relies on a number-theoretic heuristic assumption reminiscent of those used in subexponential classical factorization algorithms. It is currently not clear if the algorithm can lead to improved physical implementations in practice.}, note={arXiv:2308.06572 [quant-ph]}, number={arXiv:2308.06572}, publisher={arXiv}, author={Regev, Oded}, year={2024}, month=jan }

@article{Lee_Lee_Zhai_Tong_Dalzell_Kumar_Helms_Gray_Cui_Liu_2023,  number={1}, journal={Nature Communications}, publisher={Nature Publishing Group}, author={Lee, Seunghoon and Lee, Joonho and Zhai, Huanchen and Tong, Yu and Dalzell, Alexander M. and Kumar, Ashutosh and Helms, Phillip and Gray, Johnnie and Cui, Zhi-Hao and Liu, Wenyuan and Kastoryano, Michael and Babbush, Ryan and Preskill, John and Reichman, David R. and Campbell, Earl T. and Valeev, Edward F. and Lin, Lin and Chan, Garnet Kin-Lic}, year={2023}, month=apr, pages={1952}, language={en}, doi={10.1038/s41467-023-37587-6} }

@article{Kitaev_1995, title={Quantum measurements and the Abelian Stabilizer Problem}, url={http://arxiv.org/abs/quant-ph/9511026}, DOI={10.48550/arXiv.quant-ph/9511026}, abstractNote={We present a polynomial quantum algorithm for the Abelian stabilizer problem which includes both factoring and the discrete logarithm. Thus we extend famous Shor’s results. Our method is based on a procedure for measuring an eigenvalue of a unitary operator. Another application of this procedure is a polynomial quantum Fourier transform algorithm for an arbitrary finite Abelian group. The paper also contains a rather detailed introduction to the theory of quantum computation.}, note={arXiv:quant-ph/9511026}, number={arXiv:quant-ph/9511026}, publisher={arXiv}, author={Kitaev, A. Yu}, year={1995}, month=nov }

@article{Feynman_1982, title={Simulating physics with computers}, volume={21}, ISSN={0020-7748, 1572-9575}, DOI={10.1007/BF02650179}, number={6–7}, journal={International Journal of Theoretical Physics}, author={Feynman, Richard P.}, year={1982}, month=jun, pages={467–488}, language={en} }

@article{Cotler_Huang_McClean_2021, title={Revisiting dequantization and quantum advantage in learning tasks}, url={http://arxiv.org/abs/2112.00811}, DOI={10.48550/arXiv.2112.00811}, note={arXiv:2112.00811 [quant-ph]}, number={arXiv:2112.00811}, publisher={arXiv}, author={Cotler, Jordan and Huang, Hsin-Yuan and McClean, Jarrod R.}, year={2021}, month=dec }

@article{Qi_2020, title={Regimes of Classical Simulability for Noisy Gaussian Boson Sampling}, volume={124}, DOI={10.1103/PhysRevLett.124.100502}, abstractNote={As a promising candidate for exhibiting quantum computational supremacy, Gaussian boson sampling (GBS) is designed to exploit the ease of experimental preparation of Gaussian states. However, sufficiently large and inevitable experimental noise might render GBS classically simulable. In this work, we formalize this intuition by establishing a sufficient condition for approximate polynomial-time classical simulation of noisy GBS—in the form of an inequality between the input squeezing parameter, the overall transmission rate, and the quality of photon detectors. Our result serves as a nonclassicality test that must be passed by any quantum computational supremacy demonstration based on GBS. We show that, for most linear-optical architectures, where photon loss increases exponentially with the circuit depth, noisy GBS loses its quantum advantage in the asymptotic limit. Our results thus delineate intermediate-sized regimes where GBS devices might considerably outperform classical computers for modest noise levels. Finally, we find that increasing the amount of input squeezing is helpful to evade our classical simulation algorithm, which suggests a potential route to mitigate photon loss.}, number={10}, journal={Physical Review Letters}, publisher={American Physical Society}, author={Qi, Haoyu and Brod, Daniel J. and Quesada, Nicolás and Garcia-Patron, Raul}, year={2020}, month=mar, pages={100502} }

@article{Renema_2020, title={Simulability of partially distinguishable superposition and Gaussian boson sampling}, volume={101}, DOI={10.1103/PhysRevA.101.063840}, abstractNote={We study the hardness of classically simulating boson sampling with superposition and Gaussian input states at nonzero photon indistinguishability. We find that, similar to regular boson sampling, distinguishability causes exponential attenuation of the many-photon interference terms in both these boson sampling variants. For superposition sampling, we find that it is not simulable with out method at zero indistinguishability, which is evidence for the computational hardness of this problem, and we find that it is simulable at any level of particle distinguishability, similar to regular boson sampling. If an efficient classical algorithm to approximate a given sum over permanents is found, this approach also leads to an efficient classical algorithm to simulate Gaussian boson sampling in the presence of distinguishability.}, number={6}, journal={Physical Review A}, publisher={American Physical Society}, author={Renema, Jelmer J.}, year={2020}, month=jun, pages={063840} }

@article{Shi_Byrnes_2022, title={Effect of partial distinguishability on quantum supremacy in Gaussian Boson sampling}, volume={8}, rights={2022 The Author(s)}, ISSN={2056-6387}, DOI={10.1038/s41534-022-00557-9}, abstractNote={Gaussian boson sampling (GBS) allows for a way to demonstrate quantum supremacy with the relatively modest experimental resources of squeezed light sources, linear optics, and photon detection. In a realistic experimental setting, numerous effects can modify the complexity of the sampling, in particular loss, partial distinguishability of the photons, and the use of threshold detectors rather than photon counting detectors. In this paper, we investigate GBS with partial distinguishability using an approach based on virtual modes and indistinguishability efficiency. We develop a model using these concepts and derive the probabilities of measuring a specific output pattern from partially distinguishable and lossy GBS for both types of detectors. In the case of threshold detectors, the probability as calculated by the Torontonian is a special case under our framework. By analyzing the expressions of these probabilities we propose an efficient classical simulation algorithm which can be used to calculate the probabilities. Our model and algorithm provide foundations for an approximate method for calculating probabilities. It also allows for a way to design sampling algorithms that are not only compatible with existing algorithms for ideal GBS, but can also reduce their complexity exponentially, depending on the indistinguishability efficiency. Using this we show how the boundary of quantum supremacy in GBS can be affected by partial distinguishability.}, number={1}, journal={npj Quantum Information}, publisher={Nature Publishing Group}, author={Shi, Junheng and Byrnes, Tim}, year={2022}, month=may, pages={1–11}, language={en} }

@article{Quesada_Arrazola_2020, title={Exact simulation of Gaussian Boson Sampling in polynomial space and exponential time}, volume={2}, ISSN={2643-1564}, DOI={10.1103/PhysRevResearch.2.023005}, abstractNote={We introduce an exact classical algorithm for simulating Gaussian Boson Sampling (GBS). The complexity of the algorithm is exponential in the number of photons detected, which is itself a random variable. For a fixed number of modes, the complexity is in fact equivalent to that of calculating output probabilities, up to constant prefactors. The simulation algorithm can be extended to other models such as GBS with threshold detectors, GBS with displacements, and sampling linear combinations of Gaussian states. In the specific case of encoding non-negative matrices into a GBS device, our method leads to an approximate sampling algorithm with polynomial runtime. We implement the algorithm, making the code publicly available as part of Xanadu’s The Walrus library, and benchmark its performance on GBS with random Haar interferometers and with encoded ErdH{o}s-Renyi graphs.}, note={arXiv:1908.08068 [quant-ph]}, number={2}, journal={Physical Review Research}, author={Quesada, Nicolás and Arrazola, Juan Miguel}, year={2020}, month=apr, pages={023005} }

@article{Kaposi_Kolarovszki_Kozsik_Zimborás_Rakyta_2022, title={Polynomial speedup in Torontonian calculation by a scalable recursive algorithm}, url={http://arxiv.org/abs/2109.04528}, DOI={10.48550/arXiv.2109.04528}, abstractNote={Evaluating the Torontonian function is a central computational challenge in the simulation of Gaussian Boson Sampling (GBS) with threshold detection. In this work, we propose a recursive algorithm providing a polynomial speedup in the exact calculation of the Torontonian compared to state-of-the-art algorithms. According to our numerical analysis the complexity of the algorithm is proportional to $N^{1.0691}2^{N/2}$ with $N$ being the size of the problem. We also show that the recursive algorithm can be scaled up to HPC use cases making feasible the simulation of threshold GBS up to $35-40$ photon clicks without the needs of large-scale computational capacities.}, note={arXiv:2109.04528 [quant-ph]}, number={arXiv:2109.04528}, publisher={arXiv}, author={Kaposi, Ágoston and Kolarovszki, Zoltán and Kozsik, Tamás and Zimborás, Zoltán and Rakyta, Péter}, year={2022}, month=nov }

@article{Quesada_2022, title={Quadratic Speed-Up for Simulating Gaussian Boson Sampling}, volume={3}, DOI={10.1103/PRXQuantum.3.010306}, number={1}, journal={PRX Quantum}, publisher={American Physical Society}, author={Quesada, Nicolas and Chadwick, Rahel S. and Bell, Bryn A. and Arrazola, Juan Miguel and Vincent, Trevor and Qi, Haoyu and Garcia-Patron, Raul}, year={2022}, month=jan, pages={010306} }

@article{Shor_1997, title={Polynomial-Time Algorithms for Prime Factorization and Discrete Logarithms on a Quantum Computer}, volume={26}, ISSN={0097-5397, 1095-7111}, DOI={10.1137/S0097539795293172}, abstractNote={A digital computer is generally believed to be an eﬃcient universal computing device; that is, it is believed able to simulate any physical computing device with an increase in computation time by at most a polynomial factor. This may not be true when quantum mechanics is taken into consideration. This paper considers factoring integers and ﬁnding discrete logarithms, two problems which are generally thought to be hard on a classical computer and which have been used as the basis of several proposed cryptosystems. Eﬃcient randomized algorithms are given for these two problems on a hypothetical quantum computer. These algorithms take a number of steps polynomial in the input size, e.g., the number of digits of the integer to be factored.}, note={arXiv: quant-ph/9508027}, number={5}, journal={SIAM Journal on Computing}, author={Shor, Peter W.}, year={1997}, month=oct, pages={1484–1509}, language={en} }

@article{heeres2017, title={Implementing a universal gate set on a logical qubit encoded in an oscillator}, volume={8}, rights={2017 The Author(s)}, ISSN={2041-1723}, DOI={10.1038/s41467-017-00045-1}, abstractNote={A logical qubit is a two-dimensional subspace of a higher dimensional system, chosen such that it is possible to detect and correct the occurrence of certain errors. Manipulation of the encoded information generally requires arbitrary and precise control over the entire system. Whether based on multiple physical qubits or larger dimensional modes such as oscillators, the individual elements in realistic devices will always have residual interactions, which must be accounted for when designing logical operations. Here we demonstrate a holistic control strategy which exploits accurate knowledge of the Hamiltonian to manipulate a coupled oscillator-transmon system. We use this approach to realize high-fidelity (98.5%, inferred), decoherence-limited operations on a logical qubit encoded in a superconducting cavity resonator using four-component cat states. Our results show the power of applying numerical techniques to control linear oscillators and pave the way for utilizing their large Hilbert space as a resource in quantum information processing.}, number={1}, journal={Nature Communications}, publisher={Nature Publishing Group}, author={Heeres, Reinier W. and Reinhold, Philip and Ofek, Nissim and Frunzio, Luigi and Jiang, Liang and Devoret, Michel H. and Schoelkopf, Robert J.}, year={2017}, month=jul, pages={94}, language={en} }

@article{riste2015detecting,
  title={Detecting bit-flip errors in a logical qubit using stabilizer measurements},
  author={Riste, Diego and Poletto, Stefano and Huang, M-Z and Bruno, Alessandro and Vesterinen, Visa and Saira, O-P and DiCarlo, Leonardo},
  journal={Nature communications},
  volume={6},
  number={1},
  pages={1--6},
  year={2015},
  publisher={Nature Publishing Group}, doi={10.1038/ncomms7983}
}

@article{corcoles2015demonstration,
  title={Demonstration of a quantum error detection code using a square lattice of four superconducting qubits},
  author={C{\'o}rcoles, Antonio D and Magesan, Easwar and Srinivasan, Srikanth J and Cross, Andrew W and Steffen, Matthias and Gambetta, Jay M and Chow, Jerry M},
  journal={Nature communications},
  volume={6},
  number={1},
  pages={1--10},
  year={2015},
  publisher={Nature Publishing Group}, doi={10.1038/ncomms7979}
}

@article{cramer2016repeated,
  title={Repeated quantum error correction on a continuously encoded qubit by real-time feedback},
  author={Cramer, Julia and Kalb, Norbert and Rol, M Adriaan and Hensen, Bas and Blok, Machiel S and Markham, Matthew and Twitchen, Daniel J and Hanson, Ronald and Taminiau, Tim H},
  journal={Nature communications},
  volume={7},
  number={1},
  pages={1--7},
  year={2016},
  publisher={Nature Publishing Group}, doi={10.1038/ncomms11526}
}

@article{ofek2016extending,
  title={Extending the lifetime of a quantum bit with error correction in superconducting circuits},
  author={Ofek, Nissim and Petrenko, Andrei and Heeres, Reinier and Reinhold, Philip and Leghtas, Zaki and Vlastakis, Brian and Liu, Yehan and Frunzio, Luigi and Girvin, SM and Jiang, Liang and others},
  journal={Nature},
  volume={536},
  number={7617},
  pages={441--445},
  year={2016},
  publisher={Nature Publishing Group}, doi={10.1038/nature18949}
}

@article{takita2017experimental,
  title={Experimental demonstration of fault-tolerant state preparation with superconducting qubits},
  author={Takita, Maika and Cross, Andrew W and C{\'o}rcoles, AD and Chow, Jerry M and Gambetta, Jay M},
  journal={Physical review letters},
  volume={119},
  number={18},
  pages={180501},
  year={2017},
  publisher={APS}, doi={10.1103/PhysRevLett.119.180501}
}

@article{linke2017fault,
  title={Fault-tolerant quantum error detection},
  author={Linke, Norbert M and Gutierrez, Mauricio and Landsman, Kevin A and Figgatt, Caroline and Debnath, Shantanu and Brown, Kenneth R and Monroe, Christopher},
  journal={Science advances},
  volume={3},
  number={10},
  pages={e1701074},
  year={2017},
  publisher={American Association for the Advancement of Science}, doi={10.1126/sciadv.1701074}
}

@article{wootton2018repetition,
  title={Repetition code of 15 qubits},
  author={Wootton, James R and Loss, Daniel},
  journal={Physical Review A},
  volume={97},
  number={5},
  pages={052313},
  year={2018},
  publisher={APS}, doi={10.1103/PhysRevA.97.052313}
}

@article{andersen2019entanglement,
  title={Entanglement stabilization using ancilla-based parity detection and real-time feedback in superconducting circuits},
  author={Andersen, Christian Kraglund and Remm, Ants and Lazar, Stefania and Krinner, Sebastian and Heinsoo, Johannes and Besse, Jean-Claude and Gabureac, Mihai and Wallraff, Andreas and Eichler, Christopher},
  journal={npj Quantum Information},
  volume={5},
  number={1},
  pages={1--7},
  year={2019},
  publisher={Nature Publishing Group}, doi={10.1038/s41534-019-0185-4}
}

@article{gong2019experimental,
  title={Experimental verification of five-qubit quantum error correction with superconducting qubits},
  author={Gong, Ming and Yuan, Xiao and Wang, Shiyu and Wu, Yulin and Zhao, Youwei and Zha, Chen and Li, Shaowei and Zhang, Zhen and Zhao, Qi and Liu, Yunchao and others},
  journal={arXiv preprint arXiv:1907.04507},
  year={2019}, doi={10.1093/nsr/nwab011}
}

@article{hu2019quantum,
  title={Quantum error correction and universal gate set operation on a binomial bosonic logical qubit},
  author={Hu, Ling and Ma, Yuwei and Cai, Weizhou and Mu, Xianghao and Xu, Yuan and Wang, Weiting and Wu, Yukai and Wang, Haiyan and Song, YP and Zou, C-L and others},
  journal={Nature Physics},
  volume={15},
  number={5},
  pages={503--508},
  year={2019},
  publisher={Nature Publishing Group}, doi={10.1038/s41567-018-0414-3}
}

@article{andersen2020,
  title={Repeated quantum error detection in a surface code},
  author={Andersen, Christian Kraglund and Remm, Ants and Lazar, Stefania and Krinner, Sebastian and Lacroix, Nathan and Norris, Graham J and Gabureac, Mihai and Eichler, Christopher and Wallraff, Andreas},
  journal={Nature Physics},
  volume={16},
  pages={875--880},
  year={2020},
  publisher={Nature Publishing Group}, doi={10.1038/s41567-020-0920-y}
}

@article{luo2020quantum,
  title={Quantum teleportation of physical qubits into logical code-spaces},
  author={Luo, Yi-Han and Chen, Ming-Cheng and Erhard, Manuel and Zhong, Han-Sen and Wu, Dian and Tang, Hao-Yang and Zhao, Qi and Wang, Xi-Lin and Fujii, Keisuke and Li, Li and others},
  journal={arXiv preprint arXiv:2009.06242},
  year={2020}, doi={10.1073/pnas.2026250118}
}

@article{campagne2020quantum,
  title={Quantum error correction of a qubit encoded in grid states of an oscillator},
  author={Campagne-Ibarcq, P and Eickbusch, A and Touzard, S and Zalys-Geller, E and Frattini, NE and Sivak, VV and Reinhold, P and Puri, S and Shankar, S and Schoelkopf, RJ and others},
  journal={Nature},
  volume={584},
  number={7821},
  pages={368--372},
  year={2020},
  publisher={Nature Publishing Group}, doi={10.1038/s41586-020-2603-3}
}

@article{cory1998,
  author    = {D. G. Cory and M. Price and W. Maas and E. Knill and R. Laflamme and W. H. Zurek and T. F. Havel and S. S. Somaroo},
  title     = {Experimental quantum error correction},
  journal   = {Physical Review Letters},
  volume    = {81},
  pages     = {2152},
  year      = {1998}, doi={10.1103/PhysRevLett.81.2152}
}

@article{knill2001,
  author    = {E. Knill and R. Laflamme and R. Martinez and C. Negrevergne},
  title     = {Benchmarking quantum computers: the five-qubit error correcting code},
  journal   = {Physical Review Letters},
  volume    = {86},
  pages     = {5811},
  year      = {2001}, doi={10.1103/PhysRevLett.86.5811}
}

@article{schindler2011,
  author    = {P. Schindler and J. T. Barreiro and T. Monz and V. Nebendahl and D. Nigg and M. Chwalla and M. Hennrich and R. Blatt},
  title     = {Experimental repetitive quantum error correction},
  journal   = {Science},
  volume    = {332},
  pages     = {1059},
  year      = {2011}, doi={10.1126/science.1203329}
}

@article{moussa2011,
  author    = {O. Moussa and J. Baugh and C. A. Ryan and R. Laflamme},
  title     = {Demonstration of sufficient control for two rounds of quantum error correction in a solid-state ensemble quantum information processor},
  journal   = {Physical Review Letters},
  volume    = {107},
  pages     = {160501},
  year      = {2011}, doi={10.1103/PhysRevLett.107.160501}
}

@article{zhang2011,
  author    = {J. Zhang and T. Laflamme and D. Cory},
  title     = {Experimental implementation of encoded logical qubits in a 3-qubit decoherence-free subspace},
  journal   = {Physical Review A},
  volume    = {82},
  pages     = {022304},
  year      = {2011},
}

@article{reed2012,
  author    = {M. D. Reed and L. DiCarlo and S. E. Nigg and L. Sun and L. Frunzio and S. M. Girvin and R. J. Schoelkopf},
  title     = {Realization of three-qubit quantum error correction with superconducting circuits},
  journal   = {Nature},
  volume    = {482},
  pages     = {382},
  year      = {2012}, doi={10.1038/nature10786}
}

@article{zhang2012,
  author    = {J. Zhang and R. Laflamme and D. G. Cory},
  title     = {Experimental quantum error correction with large block codes},
  journal   = {Physical Review Letters},
  volume    = {108},
  pages     = {110501},
  year      = {2012},
}

@article{bell2014,
  author    = {B. A. Bell and D. A. Herrera-Martí and M. S. Tame and D. Markham and W. J. Wadsworth and J. G. Rarity},
  title     = {Experimental demonstration of a graph state quantum error-correction code},
  journal   = {Nature Communications},
  volume    = {5},
  pages     = {5480},
  year      = {2014}, doi={10.1038/ncomms4658}
}

@article{kelly2014,
  author    = {J. Kelly and R. Barends and A. G. Fowler and A. Megrant and E. Jeffrey and T. C. White and D. Sank and J. Y. Mutus and B. Campbell and Y. Chen and Z. Chen and B. Chiaro and A. Dunsworth and I. C. Hoi and C. Neill and P. J. J. O’Malley and C. Quintana and P. Roushan and A. Vainsencher and J. Wenner and A. N. Cleland and J. M. Martinis},
  title     = {State preservation by repetitive error detection in a superconducting quantum circuit},
  journal   = {Nature},
  volume    = {519},
  pages     = {66},
  year      = {2015}, doi={10.1038/nature14270}
}

@article{nigg2014,
  author    = {D. Nigg and M. Müller and E. A. Martinez and P. Schindler and M. Hennrich and T. Monz and M. A. Martin-Delgado and R. Blatt},
  title     = {Quantum computations on a topologically encoded qubit},
  journal   = {Science},
  volume    = {345},
  pages     = {302},
  year      = {2014}, doi={10.1126/science.1253742}
}

@article{waldherr2014,
  author    = {G. Waldherr and Y. Wang and S. Zaiser and M. Jamali and T. Schulte-Herbrüggen and H. Abe and T. Ohshima and J. Isoya and J. F. Du and P. Neumann and J. Wrachtrup},
  title     = {Quantum error correction in a solid-state hybrid spin register},
  journal   = {Nature},
  volume    = {506},
  pages     = {204},
  year      = {2014}, doi={10.1038/nature12919}
}

@article{Deng_Gu_Liu_Gong_Su_Zhang_Tang_Jia_Xu_Chen_2023, title={Gaussian Boson Sampling with Pseudo-Photon-Number-Resolving Detectors and Quantum Computational Advantage}, volume={131}, DOI={10.1103/PhysRevLett.131.150601}, abstractNote={We report new Gaussian boson sampling experiments with pseudo-photon-number-resolving detection, which register up to 255 photon-click events. We consider partial photon distinguishability and develop a more complete model for the characterization of the noisy Gaussian boson sampling. In the quantum computational advantage regime, we use Bayesian tests and correlation function analysis to validate the samples against all current classical spoofing mockups. Estimating with the best classical algorithms to date, generating a single ideal sample from the same distribution on the supercomputer Frontier would take ∼600 yr using exact methods, whereas our quantum computer, Jiǔzhāng 3.0, takes only 1.27 μ⁢s to produce a sample. Generating the hardest sample from the experiment using an exact algorithm would take Frontier∼3.1×1010 yr.}, number={15}, journal={Physical Review Letters}, publisher={American Physical Society}, author={Deng, Yu-Hao and Gu, Yi-Chao and Liu, Hua-Liang and Gong, Si-Qiu and Su, Hao and Zhang, Zhi-Jiong and Tang, Hao-Yang and Jia, Meng-Hao and Xu, Jia-Min and Chen, Ming-Cheng and Qin, Jian and Peng, Li-Chao and Yan, Jiarong and Hu, Yi and Huang, Jia and Li, Hao and Li, Yuxuan and Chen, Yaojian and Jiang, Xiao and Gan, Lin and Yang, Guangwen and You, Lixing and Li, Li and Zhong, Han-Sen and Wang, Hui and Liu, Nai-Le and Renema, Jelmer J. and Lu, Chao-Yang and Pan, Jian-Wei}, year={2023}, month=oct, pages={150601} }

@article{Deshpande_Mehta_Vincent_Quesada_Hinsche_Ioannou_Madsen_Lavoie_Qi_Eisert_2022, title={Quantum computational advantage via high-dimensional Gaussian boson sampling}, volume={8}, DOI={10.1126/sciadv.abi7894}, abstractNote={Photonics is a promising platform for demonstrating a quantum computational advantage (QCA) by outperforming the most powerful classical supercomputers on a well-defined computational task. Despite this promise, existing proposals and demonstrations face challenges. Experimentally, current implementations of Gaussian boson sampling (GBS) lack programmability or have prohibitive loss rates. Theoretically, there is a comparative lack of rigorous evidence for the classical hardness of GBS. In this work, we make progress in improving both the theoretical evidence and experimental prospects. We provide evidence for the hardness of GBS, comparable to the strongest theoretical proposals for QCA. We also propose a QCA architecture we call high-dimensional GBS, which is programmable and can be implemented with low loss using few optical components. We show that particular algorithms for simulating GBS are outperformed by high-dimensional GBS experiments at modest system sizes. This work thus opens the path to demonstrating QCA with programmable photonic processors.}, number={1}, journal={Science Advances}, publisher={American Association for the Advancement of Science}, author={Deshpande, Abhinav and Mehta, Arthur and Vincent, Trevor and Quesada, Nicolás and Hinsche, Marcel and Ioannou, Marios and Madsen, Lars and Lavoie, Jonathan and Qi, Haoyu and Eisert, Jens and Hangleiter, Dominik and Fefferman, Bill and Dhand, Ish}, year={2022}, month=jan, pages={eabi7894} }

@phdthesis{Balasubramanian_1980, type={Thesis}, title={Combinatorics and diagonals of matrices}, url={http://localhost:8080/xmlui/handle/10263/3603}, abstractNote={Thesis under supervision of Dr.K R Parthasarthy}, note={Accepted: 2012-04-29T18:32:22Z}, school={Indian Statistical Institute,Calcutta}, author={Balasubramanian, K.}, year={1980}, month=dec,}

@phdthesis{Bax_1998, type={phd}, title={Finite-difference algorithms for counting problems}, rights={other}, url={https://resolver.caltech.edu/CaltechETD:etd-01182008-113319}, DOI={10.7907/fdc3-5s46}, school={California Institute of Technology}, author={Bax, Eric}, year={1998} }

@book{Ryser_1963, title={Combinatorial Mathematics}, ISBN={978-1-61444-014-7}, abstractNote={Herbert J. Ryser is widely regarded as one of the major figures in combinatorics in the 20th century. His Combinatorial Mathematics is a classic which has enticed many young mathematics students into this area.}, publisher={American Mathematical Soc.}, author={Ryser, Herbert John}, year={1963}, month=dec, language={en}, doi={10.5948/UPO9781614440147} }

@misc{Clifford_Clifford_boson_sampling_r_package, title={BosonSampling: Classical Boson Sampling}, rights={GPL-2}, howpublished={\url{https://cran.r-project.org/web/packages/BosonSampling/index.html}}, abstractNote={Classical Boson Sampling using the algorithm of Clifford and Clifford (2017) <doi:10.48550/arXiv.1706.01260>. Also provides functions for generating random unitary matrices, evaluation of matrix permanents (both real and complex) and evaluation of complex permanent minors.}, author={Clifford, Peter and Clifford, Raphaël}, year={2017}, month=oct }

@article{Wu_Liu_Zhang_Jin_Wang_Wang_Yang_2018, title={A benchmark test of boson sampling on Tianhe-2 supercomputer}, volume={5}, ISSN={2095-5138}, DOI={10.1093/nsr/nwy079}, abstractNote={Boson sampling, thought to be intractable classically, can be solved by a quantum machine composed of merely generation, linear evolution and detection of single photons. Such an analog quantum computer for this specific problem provides a shortcut to boost the absolute computing power of quantum computers to beat classical ones. However, the capacity bound of classical computers for simulating boson sampling has not yet been identified. Here we simulate boson sampling on the Tianhe-2 supercomputer, which occupied the first place in the world ranking six times from 2013 to 2016. We computed the permanent of the largest matrix using up to 312 000 CPU cores of Tianhe-2, and inferred from the current most efficient permanent-computing algorithms that an upper bound on the performance of Tianhe-2 is one 50-photon sample per ∼100 min. In addition, we found a precision issue with one of two permanent-computing algorithms.}, number={5}, journal={National Science Review}, author={Wu, Junjie and Liu, Yong and Zhang, Baida and Jin, Xianmin and Wang, Yang and Wang, Huiquan and Yang, Xuejun}, year={2018}, month=sep, pages={715–720} }

@inbook{Clifford_Clifford_2018, series={Proceedings}, title={The Classical Complexity of Boson Sampling}, url={https://epubs.siam.org/doi/10.1137/1.9781611975031.10}, DOI={10.1137/1.9781611975031.10}, abstractNote={We study the classical complexity of the exact Boson Sampling problem where the objective is to produce provably correct random samples from a particular quantum mechanical distribution. The computational framework was proposed in STOC ’11 by Aaronson and Arkhipov in 2011 as an attainable demonstration of ‘quantum supremacy’, that is a practical quantum computing experiment able to produce output at a speed beyond the reach of classical (that is non-quantum) computer hardware. Since its introduction Boson Sampling has been the subject of intense international research in the world of quantum computing. On the face of it, the problem is challenging for classical computation. Aaronson and Arkhipov show that exact Boson Sampling is not efficiently solvable by a classical computer unless P#P = BPPNP and the polynomial hierarchy collapses to the third level.The fastest known exact classical algorithm for the standard Boson Sampling problem requires  time to produce samples for a system with input size n and m output modes, making it infeasible for anything but the smallest values of n and m. We give an algorithm that is much faster, running in  time and  additional space. The algorithm is simple to implement and has low constant factor overheads. As a consequence our classical algorithm is able to solve the exact Boson Sampling problem for system sizes far beyond current photonic quantum computing experimentation, thereby significantly reducing the likelihood of achieving near-term quantum supremacy in the context of Boson Sampling.}, booktitle={Proceedings of the 2018 Annual ACM-SIAM Symposium on Discrete Algorithms (SODA)}, publisher={Society for Industrial and Applied Mathematics}, author={Clifford, Peter and Clifford, Raphaël}, year={2018}, month=jan, pages={146–155}, collection={Proceedings} }

@article{Björklund_Gupt_Quesada_2019, title={A Faster Hafnian Formula for Complex Matrices and Its Benchmarking on a Supercomputer}, volume={24}, ISSN={1084-6654}, DOI={10.1145/3325111}, abstractNote={We introduce new and simple algorithms for the calculation of the number of perfect matchings of complex weighted, undirected graphs with and without loops. Our compact formulas for the hafnian and loop hafnian of n × n complex matrices run in O(n3 2n/2) time, are embarrassingly parallelizable and, to the best of our knowledge, are the fastest exact algorithms to compute these quantities. Despite our highly optimized algorithm, numerical benchmarks on the Titan supercomputer with matrices up to size 56 × 56 indicate that one would require the 288,000 CPUs of this machine for about 6 weeks to compute the hafnian of a 100 × 100 matrix.}, journal={ACM J. Exp. Algorithmics}, author={Björklund, Andreas and Gupt, Brajesh and Quesada, Nicolás}, year={2019}, month=jun, pages={1.11:1-1.11:17} }

@article{Gupt_Arrazola_Quesada_Bromley_2020, title={Classical benchmarking of Gaussian Boson Sampling on the Titan supercomputer}, volume={19}, ISSN={1573-1332}, DOI={10.1007/s11128-020-02713-6}, abstractNote={Gaussian Boson Sampling (GBS) is a model of photonic quantum computing where single-mode squeezed states are sent through linear-optical interferometers and measured using single-photon detectors. In this work, we employ a recent exact sampling algorithm for GBS with threshold detectors to perform classical simulations on the Titan supercomputer. We determine the time and memory resources as well as the amount of computational nodes required to produce samples for different numbers of modes and detector clicks. It is possible to simulate a system with 800 optical modes postselected on outputs with 20 detector clicks, producing a single sample in roughly 2 h using 40% of the available nodes of Titan. Additionally, we benchmark the performance of GBS when applied to dense subgraph identification, even in the presence of photon loss. We perform sampling for several graphs containing as many as 200 vertices. Our findings indicate that large losses can be tolerated and that the use of threshold detectors is preferable over using photon-number-resolving detectors postselected on collision-free outputs.}, number={8}, journal={Quantum Information Processing}, author={Gupt, Brajesh and Arrazola, Juan Miguel and Quesada, Nicolás and Bromley, Thomas R.}, year={2020}, month=jul, pages={249}, language={en} }

@article{Wu_Cheng_Jia_Zhang_Yung_Sun_2020, title={Speedup in classical simulation of Gaussian boson sampling}, volume={65}, ISSN={2095-9273}, DOI={10.1016/j.scib.2020.02.012}, abstractNote={Gaussian boson sampling is an alternative model for demonstrating quantum computational supremacy, where squeezed states are injected into every input mode, instead of applying single photons as in the case of standard boson sampling. Here by analyzing numerically the computational costs, we establish a lower bound for achieving quantum computational supremacy for a class of Gaussian boson-sampling problems. Specifically, we propose a more efficient method for calculating the transition probabilities, leading to a significant reduction of the simulation costs. Particularly, our numerical results indicate that one can simulate up to 18 photons for Gaussian boson sampling at the output subspace on a normal laptop, 20 photons on a commercial workstation with 256 cores, and about 30 photons for supercomputers. These numbers are significantly smaller than those in standard boson sampling, suggesting that Gaussian boson sampling could be experimentally-friendly for demonstrating quantum computational supremacy.}, number={10}, journal={Science Bulletin}, author={Wu, Bujiao and Cheng, Bin and Jia, Fei and Zhang, Jialin and Yung, Man-Hong and Sun, Xiaoming}, year={2020}, month=may, pages={832–841} }

@article{Li_Gan_Chen_Chen_Lu_Lu_Pan_Fu_Yang_2022, title={Benchmarking 50-Photon Gaussian Boson Sampling on the Sunway TaihuLight}, volume={33}, ISSN={1558-2183}, DOI={10.1109/TPDS.2021.3111185}, abstractNote={Boson sampling is expected to be an important milestone that will demonstrate quantum computational advantage (or quantum supremacy). This work establishes the benchmarking of Gaussian boson sampling (GBS) with threshold detection based on the Sunway TaihuLight supercomputer. To achieve the best performance and provide a competitive scenario for future quantum computing studies, the selected simulation algorithm is fully optimized based on a set of innovative approaches, including a parallel framework with almost perfect load balance and an instruction-level optimizing scheme based on a shortest-path-based instruction scheduling. In addition, data precision is carefully processed by an integer-instruction-based and multiple-precision fixed-point implementation, including 128- and 256-bit precison mode, which can be appropriately selected based on an adaptive precision optimizing scheme. Based on these methods, a highly efficient parallel quantum sampling algorithm is designed. The largest run enables us to obtain one Torontonian function of a 100times 100100×100 submatrix from 50-photon GBS within 20 hours in 128-bit precision and 2 days in 256-bit precision. To our knowledge, this was the largest quantum computing simulation based on Boson Sampling by using modern supercomputers.}, number={6}, journal={IEEE Transactions on Parallel and Distributed Systems}, author={Li, Yuxuan and Gan, Lin and Chen, Mingcheng and Chen, Yaojian and Lu, Haitian and Lu, Chaoyang and Pan, Jianwei and Fu, Haohuan and Yang, Guangwen}, year={2022}, month=jun, pages={1357–1372} }

@article{Bentivegna_Spagnolo_Vitelli_Flamini_Viggianiello_Latmiral_Mataloni_Brod_Galvão_Crespi_2015, title={Experimental scattershot boson sampling}, volume={1}, DOI={10.1126/sciadv.1400255}, abstractNote={Boson sampling is a computational task strongly believed to be hard for classical computers, but efficiently solvable by orchestrated bosonic interference in a specialized quantum computer. Current experimental schemes, however, are still insufficient for a convincing demonstration of the advantage of quantum over classical computation. A new variation of this task, scattershot boson sampling, leads to an exponential increase in speed of the quantum device, using a larger number of photon sources based on parametric down-conversion. This is achieved by having multiple heralded single photons being sent, shot by shot, into different random input ports of the interferometer. We report the first scattershot boson sampling experiments, where six different photon-pair sources are coupled to integrated photonic circuits. We use recently proposed statistical tools to analyze our experimental data, providing strong evidence that our photonic quantum simulator works as expected. This approach represents an important leap toward a convincing experimental demonstration of the quantum computational supremacy.}, number={3}, journal={Science Advances}, publisher={American Association for the Advancement of Science}, author={Bentivegna, Marco and Spagnolo, Nicolò and Vitelli, Chiara and Flamini, Fulvio and Viggianiello, Niko and Latmiral, Ludovico and Mataloni, Paolo and Brod, Daniel J. and Galvão, Ernesto F. and Crespi, Andrea and Ramponi, Roberta and Osellame, Roberto and Sciarrino, Fabio}, year={2015}, month=apr, pages={e1400255} }

@article{Broome_Fedrizzi_Rahimi_Keshari_Dove_Aaronson_Ralph_White_2013, title={Photonic Boson Sampling in a Tunable Circuit}, volume={339}, DOI={10.1126/science.1231440}, abstractNote={Quantum computers are unnecessary for exponentially efficient computation or simulation if the Extended Church-Turing thesis is correct. The thesis would be strongly contradicted by physical devices that efficiently perform tasks believed to be intractable for classical computers. Such a task is boson sampling: sampling the output distributions of n bosons scattered by some passive, linear unitary process. We tested the central premise of boson sampling, experimentally verifying that three-photon scattering amplitudes are given by the permanents of submatrices generated from a unitary describing a six-mode integrated optical circuit. We find the protocol to be robust, working even with the unavoidable effects of photon loss, non-ideal sources, and imperfect detection. Scaling this to large numbers of photons should be a much simpler task than building a universal quantum computer.}, number={6121}, journal={Science}, publisher={American Association for the Advancement of Science}, author={Broome, Matthew A. and Fedrizzi, Alessandro and Rahimi-Keshari, Saleh and Dove, Justin and Aaronson, Scott and Ralph, Timothy C. and White, Andrew G.}, year={2013}, month=feb, pages={794–798} }

@article{Carolan_Harrold_Sparrow_Martín_Lopez_Russell_Silverstone_Shadbolt_Matsuda_Oguma_Itoh_2015, title={Universal linear optics}, volume={349}, DOI={10.1126/science.aab3642}, abstractNote={Linear optics underpins fundamental tests of quantum mechanics and quantum technologies. We demonstrate a single reprogrammable optical circuit that is sufficient to implement all possible linear optical protocols up to the size of that circuit. Our six-mode universal system consists of a cascade of 15 Mach-Zehnder interferometers with 30 thermo-optic phase shifters integrated into a single photonic chip that is electrically and optically interfaced for arbitrary setting of all phase shifters, input of up to six photons, and their measurement with a 12-single-photon detector system. We programmed this system to implement heralded quantum logic and entangling gates, boson sampling with verification tests, and six-dimensional complex Hadamards. We implemented 100 Haar random unitaries with an average fidelity of 0.999 ± 0.001. Our system can be rapidly reprogrammed to implement these and any other linear optical protocol, pointing the way to applications across fundamental science and quantum technologies.}, number={6249}, journal={Science}, publisher={American Association for the Advancement of Science}, author={Carolan, Jacques and Harrold, Christopher and Sparrow, Chris and Martín-López, Enrique and Russell, Nicholas J. and Silverstone, Joshua W. and Shadbolt, Peter J. and Matsuda, Nobuyuki and Oguma, Manabu and Itoh, Mikitaka and Marshall, Graham D. and Thompson, Mark G. and Matthews, Jonathan C. F. and Hashimoto, Toshikazu and O’Brien, Jeremy L. and Laing, Anthony}, year={2015}, month=aug, pages={711–716} }

@article{Carolan_Meinecke_Shadbolt_Russell_Ismail_Wörhoff_Rudolph_Thompson_OBrien_Matthews_2014, title={On the experimental verification of quantum complexity in linear optics}, volume={8}, rights={2014 Springer Nature Limited}, ISSN={1749-4893}, DOI={10.1038/nphoton.2014.152}, abstractNote={Quantum computers promise to solve certain problems that are forever intractable to classical computers. The first of these devices are likely to tackle bespoke problems suited to their own particular physical capabilities. Sampling the probability distribution from many bosons interfering quantum-mechanically is conjectured to be intractable to a classical computer but solvable with photons in linear optics. However, the complexity of this type of problem means its solution is mathematically unverifiable, so the task of establishing successful operation becomes one of gathering sufficiently convincing circumstantial or experimental evidence. Here, we develop scalable methods to experimentally establish correct operation for this class of computation, which we implement for three, four and five photons in integrated optical circuits, on Hilbert spaces of up to 50,000 dimensions. Our broad approach is practical for all quantum computational architectures where formal verification methods for quantum algorithms are either intractable or unknown.}, number={8}, journal={Nature Photonics}, publisher={Nature Publishing Group}, author={Carolan, Jacques and Meinecke, Jasmin D. A. and Shadbolt, Peter J. and Russell, Nicholas J. and Ismail, Nur and Wörhoff, Kerstin and Rudolph, Terry and Thompson, Mark G. and O’Brien, Jeremy L. and Matthews, Jonathan C. F. and Laing, Anthony}, year={2014}, month=aug, pages={621–626}, language={en} }

@article{Crespi_Osellame_Ramponi_Brod_Galvão_Spagnolo_Vitelli_Maiorino_Mataloni_Sciarrino_2013, title={Integrated multimode interferometers with arbitrary designs for photonic boson sampling}, volume={7}, rights={2013 Springer Nature Limited}, ISSN={1749-4893}, DOI={10.1038/nphoton.2013.112}, abstractNote={The evolution of bosons undergoing arbitrary linear unitary transformations quickly becomes hard to predict using classical computers as we increase the number of particles and modes. Photons propagating in a multiport interferometer naturally solve this so-called boson sampling problem1, thereby motivating the development of technologies that enable precise control of multiphoton interference in large interferometers2,3,4. Here, we use novel three-dimensional manufacturing techniques to achieve simultaneous control of all the parameters describing an arbitrary interferometer. We implement a small instance of the boson sampling problem by studying three-photon interference in a five-mode integrated interferometer, confirming the quantum-mechanical predictions. Scaled-up versions of this set-up are a promising way to demonstrate the computational advantage of quantum systems over classical computers. The possibility of implementing arbitrary linear-optical interferometers may also find applications in high-precision measurements and quantum communication5.}, number={7}, journal={Nature Photonics}, publisher={Nature Publishing Group}, author={Crespi, Andrea and Osellame, Roberto and Ramponi, Roberta and Brod, Daniel J. and Galvão, Ernesto F. and Spagnolo, Nicolò and Vitelli, Chiara and Maiorino, Enrico and Mataloni, Paolo and Sciarrino, Fabio}, year={2013}, month=jul, pages={545–549}, language={en} }

@article{Loredo_Broome_Hilaire_Gazzano_Sagnes_Lemaitre_Almeida_Senellart_White_2017, title={Boson Sampling with Single-Photon Fock States from a Bright Solid-State Source}, volume={118}, DOI={10.1103/PhysRevLett.118.130503}, abstractNote={A boson-sampling device is a quantum machine expected to perform tasks intractable for a classical computer, yet requiring minimal nonclassical resources as compared to full-scale quantum computers. Photonic implementations to date employed sources based on inefficient processes that only simulate heralded single-photon statistics when strongly reducing emission probabilities. Boson sampling with only single-photon input has thus never been realized. Here, we report on a boson-sampling device operated with a bright solid-state source of single-photon Fock states with high photon-number purity: the emission from an efficient and deterministic quantum dot-micropillar system is demultiplexed into three partially indistinguishable single photons, with a single-photon purity 1 −��(2)⁡(0) of 0.990±0.001, interfering in a linear optics network. Our demultiplexed source is between 1 and 2 orders of magnitude more efficient than current heralded multiphoton sources based on spontaneous parametric down-conversion, allowing us to complete the boson-sampling experiment faster than previous equivalent implementations.}, number={13}, journal={Physical Review Letters}, publisher={American Physical Society}, author={Loredo, J. C. and Broome, M. A. and Hilaire, P. and Gazzano, O. and Sagnes, I. and Lemaitre, A. and Almeida, M. P. and Senellart, P. and White, A. G.}, year={2017}, month=mar, pages={130503} }

@article{Spagnolo_Vitelli_Bentivegna_Brod_Crespi_Flamini_Giacomini_Milani_Ramponi_Mataloni_2014, title={Experimental validation of photonic boson sampling}, volume={8}, rights={2014 Springer Nature Limited}, ISSN={1749-4893}, DOI={10.1038/nphoton.2014.135}, abstractNote={To address the controversy regarding the validation of an experiment that is hard to simulate, boson-sampling experiments are implemented with three photons in randomly designed integrated chips with up to 13 modes. It is experimentally demonstrated that the Aaronson–Arkhipov test allows boson-sampling experiments to be distinguished from uniformly drawn samples.}, number={8}, journal={Nature Photonics}, publisher={Nature Publishing Group}, author={Spagnolo, Nicolò and Vitelli, Chiara and Bentivegna, Marco and Brod, Daniel J. and Crespi, Andrea and Flamini, Fulvio and Giacomini, Sandro and Milani, Giorgio and Ramponi, Roberta and Mataloni, Paolo and Osellame, Roberto and Galvão, Ernesto F. and Sciarrino, Fabio}, year={2014}, month=aug, pages={615–620}, language={en} }

@article{Spring_Metcalf_Humphreys_Kolthammer_Jin_Barbieri_Datta_Thomas_Peter_Langford_Kundys_2013, title={Boson Sampling on a Photonic Chip}, volume={339}, DOI={10.1126/science.1231692}, abstractNote={Although universal quantum computers ideally solve problems such as factoring integers exponentially more efficiently than classical machines, the formidable challenges in building such devices motivate the demonstration of simpler, problem-specific algorithms that still promise a quantum speedup. We constructed a quantum boson-sampling machine (QBSM) to sample the output distribution resulting from the nonclassical interference of photons in an integrated photonic circuit, a problem thought to be exponentially hard to solve classically. Unlike universal quantum computation, boson sampling merely requires indistinguishable photons, linear state evolution, and detectors. We benchmarked our QBSM with three and four photons and analyzed sources of sampling inaccuracy. Scaling up to larger devices could offer the first definitive quantum-enhanced computation.}, number={6121}, journal={Science}, publisher={American Association for the Advancement of Science}, author={Spring, Justin B. and Metcalf, Benjamin J. and Humphreys, Peter C. and Kolthammer, W. Steven and Jin, Xian-Min and Barbieri, Marco and Datta, Animesh and Thomas-Peter, Nicholas and Langford, Nathan K. and Kundys, Dmytro and Gates, James C. and Smith, Brian J. and Smith, Peter G. R. and Walmsley, Ian A.}, year={2013}, month=feb, pages={798–801} }

@article{Tillmann_Dakić_Heilmann_Nolte_Szameit_Walther_2013, title={Experimental boson sampling}, volume={7}, rights={2013 Springer Nature Limited}, ISSN={1749-4893}, DOI={10.1038/nphoton.2013.102}, abstractNote={Universal quantum computers1 promise a dramatic increase in speed over classical computers, but their full-size realization remains challenging2. However, intermediate quantum computational models3,4,5 have been proposed that are not universal but can solve problems that are believed to be classically hard. Aaronson and Arkhipov6 have shown that interference of single photons in random optical networks can solve the hard problem of sampling the bosonic output distribution. Remarkably, this computation does not require measurement-based interactions7,8 or adaptive feed-forward techniques9. Here, we demonstrate this model of computation using laser-written integrated quantum networks that were designed to implement unitary matrix transformations. We characterize the integrated devices using an in situ reconstruction method and observe three-photon interference10,11,12 that leads to the boson-sampling output distribution. Our results set a benchmark for a type of quantum computer with the potential to outperform a conventional computer through the use of only a few photons and linear-optical elements13.}, number={7}, journal={Nature Photonics}, publisher={Nature Publishing Group}, author={Tillmann, Max and Dakić, Borivoje and Heilmann, René and Nolte, Stefan and Szameit, Alexander and Walther, Philip}, year={2013}, month=jul, pages={540–544}, language={en} }

@article{Neville_Sparrow_Clifford_Johnston_Birchall_Montanaro_Laing_2017, title={Classical boson sampling algorithms with superior performance to near-term experiments}, volume={13}, rights={2017 Springer Nature Limited}, ISSN={1745-2481}, DOI={10.1038/nphys4270}, abstractNote={A classical algorithm solves the boson sampling problem for 30 bosons with standard computing hardware, suggesting that a much larger experimental effort will be needed to reach a regime where quantum hardware outperforms classical methods.}, number={12}, journal={Nature Physics}, publisher={Nature Publishing Group}, author={Neville, Alex and Sparrow, Chris and Clifford, Raphaël and Johnston, Eric and Birchall, Patrick M. and Montanaro, Ashley and Laing, Anthony}, year={2017}, month=dec, pages={1153–1157}, language={en} }

@article{Quesada_Arrazola_Killoran_2018, title={Gaussian boson sampling using threshold detectors}, volume={98}, DOI={10.1103/PhysRevA.98.062322}, abstractNote={We study what is arguably the most experimentally appealing boson sampling architecture: Gaussian states sampled with threshold detectors. We show that, in this setting, the probability of observing a given outcome is related to a matrix function that we name the Torontonian, which plays an analogous role to the permanent or the Hafnian in other models. We also prove that, provided that the probability of observing two or more photons in a single output mode is sufficiently small, our model remains intractable to simulate classically under standard complexity-theoretic conjectures. Finally, we leverage the mathematical simplicity of the model to introduce a physically motivated, exact sampling algorithm for all boson sampling models that employ Gaussian states and threshold detectors.}, number={6}, journal={Physical Review A}, publisher={American Physical Society}, author={Quesada, Nicolás and Arrazola, Juan Miguel and Killoran, Nathan}, year={2018}, month=dec, pages={062322} }

@article{Bulmer_Bell_Chadwick_Jones_Moise_Rigazzi_Thorbecke_Haus_Vaerenbergh_Patel_2022, title={The Boundary for Quantum Advantage in Gaussian Boson Sampling}, volume={8}, ISSN={2375-2548}, DOI={10.1126/sciadv.abl9236}, abstractNote={Identifying the boundary beyond which quantum machines provide a computational advantage over their classical counterparts is a crucial step in charting their usefulness. Gaussian Boson Sampling (GBS), in which photons are measured from a highly entangled Gaussian state, is a leading approach in pursuing quantum advantage. State-of-the-art quantum photonics experiments that, once programmed, run in minutes, would require 600 million years to simulate using the best pre-existing classical algorithms. Here, we present substantially faster classical GBS simulation methods, including speed and accuracy improvements to the calculation of loop hafnians, the matrix function at the heart of GBS. We test these on a $sim ! 100,000$ core supercomputer to emulate a range of different GBS experiments with up to 100 modes and up to 92 photons. This reduces the run-time of classically simulating state-of-the-art GBS experiments to several months -- a nine orders of magnitude improvement over previous estimates. Finally, we introduce a distribution that is efficient to sample from classically and that passes a variety of GBS validation methods, providing an important adversary for future experiments to test against.}, note={arXiv:2108.01622 [quant-ph]}, number={4}, journal={Science Advances}, author={Bulmer, Jacob F. F. and Bell, Bryn A. and Chadwick, Rachel S. and Jones, Alex E. and Moise, Diana and Rigazzi, Alessandro and Thorbecke, Jan and Haus, Utz-Uwe and Vaerenbergh, Thomas Van and Patel, Raj B. and Walmsley, Ian A. and Laing, Anthony}, year={2022}, month=jan, pages={eabl9236} }

@article{Wang_Qin_Ding_Chen_Chen_You_He_Jiang_Wang_You_2019, title={Boson sampling with 20 input photons in 60-mode interferometers at $10^{14}$ state spaces}, volume={123}, ISSN={0031-9007, 1079-7114}, DOI={10.1103/PhysRevLett.123.250503}, abstractNote={Quantum computing experiments are moving into a new realm of increasing size and complexity, with the short-term goal of demonstrating an advantage over classical computers. Boson sampling is a promising platform for such a goal, however, the number of involved single photons was up to five so far, limiting these small-scale implementations to a proof-of-principle stage. Here, we develop solid-state sources of highly efficient, pure and indistinguishable single photons, and 3D integration of ultra-low-loss optical circuits. We perform an experiment with 20 single photons fed into a 60-mode interferometer, and, in its output, sample over Hilbert spaces with a size of $10^{14}$ $-$over ten orders of magnitude larger than all previous experiments. The results are validated against distinguishable samplers and uniform samplers with a confidence level of 99.9%.}, note={arXiv:1910.09930 [quant-ph]}, number={25}, journal={Physical Review Letters}, author={Wang, Hui and Qin, Jian and Ding, Xing and Chen, Ming-Cheng and Chen, Si and You, Xiang and He, Yu-Ming and Jiang, Xiao and Wang, Z. and You, L. and Renema, J. J. and Hoefling, Sven and Lu, Chao-Yang and Pan, Jian-Wei}, year={2019}, month=dec, pages={250503} }

@article{Knill_Laflamme_Milburn_2001, title={A scheme for efficient quantum computation with linear optics}, volume={409}, rights={2001 Macmillan Magazines Ltd.}, ISSN={1476-4687}, DOI={10.1038/35051009}, abstractNote={Quantum computers promise to increase greatly the efficiency of solving problems such as factoring large integers, combinatorial optimization and quantum physics simulation. One of the greatest challenges now is to implement the basic quantum-computational elements in a physical system and to demonstrate that they can be reliably and scalably controlled. One of the earliest proposals for quantum computation is based on implementing a quantum bit with two optical modes containing one photon. The proposal is appealing because of the ease with which photon interference can be observed. Until now, it suffered from the requirement for non-linear couplings between optical modes containing few photons. Here we show that efficient quantum computation is possible using only beam splitters, phase shifters, single photon sources and photo-detectors. Our methods exploit feedback from photo-detectors and are robust against errors from photon loss and detector inefficiency. The basic elements are accessible to experimental investigation with current technology.}, number={6816}, journal={Nature}, publisher={Nature Publishing Group}, author={Knill, E. and Laflamme, R. and Milburn, G. J.}, year={2001}, month=jan, pages={46–52}, language={en} }

@article{Tindall_Fishman_2023, title={Gauging tensor networks with belief propagation}, volume={15}, ISSN={2542-4653}, DOI={10.21468/SciPostPhys.15.6.222}, abstractNote={SciPost Journals Publication Detail SciPost Phys. 15, 222 (2023) Gauging tensor networks with belief propagation}, number={6}, journal={SciPost Physics}, author={Tindall, Joseph and Fishman, Matthew}, year={2023}, month=dec, pages={222}, language={en} }

\end{document}